\documentclass[a4paper,11pt]{article}
\pdfoutput=1 

\usepackage{jheppub} 
\usepackage{algorithmic,algorithm}
\usepackage{longtable,booktabs}
\usepackage{doi}
\usepackage{siunitx}
\usepackage{verbatim}

\usepackage[T1]{fontenc} 

\title{Numerical stochastic perturbation theory applied to the twisted Eguchi-Kawai model}

\author[a,b]{Antonio~Gonz\'{a}lez-Arroyo,}
\author[c,d]{Issaku~Kanamori,}
\author[c,e]{Ken-Ichi~Ishikawa,}
\author[c]{Kanata~Miyahana,}
\author[c,e]{Masanori~Okawa,}
\author[c]{and Ryoichiro~Ueno}

\newcommand{\Nc}{N}                       
\newcommand{\SUNc}{\mathrm{SU}(\Nc)}      
\newcommand{\Nptmax}{N_{\mathrm{trunc}}}  
\newcommand{\NMD}{N_{\mathrm{MD}}}        
\newcommand{\RE}{\mathit{Re}}             
\newcommand{\IM}{\mathit{Im}}             
\newcommand{\Lbox}{\hat{L}}               

\newlength\myindent
\newcommand\bindent{%
  \begingroup
  \setlength{\itemindent}{\myindent}
  \addtolength{\algorithmicindent}{\myindent}
}
\newcommand\eindent{\endgroup}

\affiliation[a]{Instituto    de F\'{i}sica Te\'{o}rica UAM/CSIC, Nicol\'{a}s Cabrera 13-15, E-28049 Universidad Aut\'{o}noma de Madrid, Madrid, Spain}
\affiliation[b]{Departamento de F\'{i}sica Te\'{o}rica, M\'{ó}dulo 15, Cantoblanco,         E-28049 Universidad Aut\'{o}noma de Madrid, Madrid, Spain}
\affiliation[c]{Graduate School of Science, Hiroshima University, Higashi-Hiroshima 739-8526, Japan}
\affiliation[d]{\footnote{Present affiliation}RIKEN Center for Computational Science, Kobe 650-0047, Japan}
\affiliation[e]{Core of Research for the Energetic Universe, Hiroshima University, Higashi-Hiroshima 739-8526, Japan}

\emailAdd{antonio.gonzalez-arroyo@uam.es}
\emailAdd{kanamori-i@riken.jp}
\emailAdd{ishikawa@theo.phys.sci.hiroshima-u.ac.jp}
\emailAdd{m180230@hiroshima-u.ac.jp}
\emailAdd{okawa@hiroshima-u.ac.jp}
\emailAdd{ryoichiro-ueno@hiroshima-u.ac.jp}

\abstract{
We present the results of an exploratory study of the
numerical stochastic perturbation theory (NSPT) applied to the four
dimensional twisted Eguchi-Kawai  (TEK) model. 
We employ a Kramers type algorithm based on the Generalized Hybrid Molecular Dynamics
(GHMD) algorithm.
We have computed the
perturbative expansion of square Wilson loops up to $O(g^8)$. The results of
the first two coefficients (up to $O(g^4)$)  have a high precision and
match well with the exact values. The next two coefficients can be
determined and even extrapolated to large $\Nc$, where they should
coincide with the corresponding coefficients for ordinary Yang-Mills
theory on an infinite lattice. Our analysis shows the behaviour of the
probability distribution for each coefficient tending to Gaussian for
larger $\Nc$. 
The results allow us to establish the requirements to extend
this analysis to much higher order.
}

\DeclareMathOperator{\Tr}{Tr} 

\preprint{%
{\flushright
IFT-UAM/CSIC-19-18\\
FTUAM-19-4\\
HUPD-1903\\
}}

\begin{document} 
\maketitle
\flushbottom

\section{Introduction}
\label{sec:intro}

Gauge theories in the limit of infinite number of colours
($\Nc \longrightarrow \infty$) are very
interesting theoretically~\cite{tHooft:1973alw}. They are simpler than their
finite $\Nc$ counterparts, but share most of  the fascinating properties  
of the latter. Nonetheless, their understanding remains a challenge.
The relevance  of this goal is also given by the fact that they seem a
point of contact with other approaches~\cite{Maldacena:1997re}. A good deal of
their difficulty lies in the non-perturbative character of most of its
properties. The standard first principles approach to this kind of
problems is the lattice formulation of quantum field theory~\cite{Wilson:1974sk}. However,
in contrast with what happens in perturbation theory, the large $\Nc$
limit involves an extrapolation and seems harder than the finite $\Nc$
study. Only recently this line of approach has led to
trustworthy computations (for a review see ~\cite{Lucini:2012gg}),  pioneered by
the works of Teper and collaborators~\cite{Lucini:2003zr,Lucini:2004my,Lucini:2005vg}.

Fortunately, there
is a very specific simplification which emerges when studying large
$\Nc$ gauge theories on the lattice. This is the so-called Eguchi-Kawai
(EK) reduction~\cite{Eguchi:1982nm}. According to this result  finite volume
corrections are subleading in the large $\Nc$ limit. Hence, there is the
possibility that the dynamics of the large $\Nc$ gauge theories is
captured by a matrix model. There are several proposals that have been
put forward  to transform this possibility into a
reality~\cite{Bhanot:1982sh,Gross:1982at,Narayanan:2003fc,Kovtun:2007py,Unsal:2008ch}.
Here, we will be concerned with one
of the early proposals called Twisted Eguchi-Kawai model (TEK for
short)~\cite{GonzalezArroyo:1982ub, GonzalezArroyo:1982hz}, and introduced by two of the present authors. 
Indeed, it was used shortly after EK
proposal to compute the string tension at large
$\Nc$~\cite{GonzalezArroyo:1983pw,Fabricius:1984un}. Lately this has led to a calculation
of this quantity with at least compatible  precision to other methods~\cite{GonzalezArroyo:2012fx}. 
Several  precise tests of the  reduction mechanism have been
obtained recently for other quantities~\cite{Gonzalez-Arroyo:2014dua}, 
providing  a strong verification of the validity and usefulness of this approach.

The previous paragraph justifies our interest in the TEK model.
The finite $\Nc$ corrections are different for the matrix model and for
the ordinary gauge theory, and their size and nature are very important
from the practical point of view, in order to make this approach
competitive computationally. Curiously, these corrections have a
theoretically interesting interpretation in terms of the so-called
non-commutative gauge theories~\cite{Douglas:2001ba}. Their
Lagrangian and Feynman rules first appeared in the
literature~\cite{GonzalezArroyo:1983ac} when looking for a continuous generalization
of the TEK model, and before they emerged from the mathematical
construction of non-commutative geometries~\cite{Connes:1987ue}. A
more direct connection appears as a result of Morita duality on the
non-commutative torus. This shows that ordinary gauge theories with
twisted boundary conditions (TBC) a la 't Hooft~\cite{tHooft:1979rtg} are particular
cases of non-commutative field theories. The TEK model is nothing but
the volume reduced version of a gauge theory with twisted boundary
conditions  on the lattice. TBC are characterized by a collection of 
integer-valued fluxes, and their appropriate choice has been found to
have a crucial impact upon the size of the corrections and the absence of phase
transitions.  

The last ingredient entering  in this work is perturbation theory.
Although our major interest when using  the lattice approach was in studying the 
non-perturbative aspects of the theory, there is also interest in
understanding the theory from the perturbative side. At the least it
offers us a method to analytically determine certain observables and
estimating the size and nature of its $\Nc$ dependence. In this spirit
a recent perturbative calculation of Wilson loops in lattice gauge
theories with twisted boundary conditions (including the TEK model)
was addressed~\cite{Perez:2017jyq}. 

In addition, a new set of ideas has focused in explaining  the usefulness of
perturbation theory in understanding also non-perturbative contributions. 
This takes its most extreme form in the concept of {\em resurgence}~(see for example \cite{Dunne:2015eaa,Cherman:2014ofa} 
and references therein). 
At the least, as was already known,  the large order behaviour of the
perturbative coefficients is associated with
non-perturbative aspects such as the action of other saddles.
Very interesting results have been obtained in this spirit for 
large $\Nc$ matrix theories~\cite{Cherman:2014ofa,Marino:2012zq}.
For example, it has been identified that the large $\Nc$ phase transition 
of the Gross-Witten-Wadia unitary matrix model~\cite{Gross:1980he,Wadia:2012fr} is governed by 
non-trivial saddles and the action can be reconstructed via the resurgence on the asymptotic expansion 
about the vacuum~\cite{Buividovich:2015oju}.

For large $\Nc$ gauge theories in 4 dimensions there are many
interesting aspects  to be studied. Perturbation theory is dominated
by planar diagrams, whose number does not grow factorially. This could
suggest that  the perturbative series is convergent within a
given radius. Furthermore,  the instanton action at fixed value of
't Hooft coupling diverges. This  implies that the corresponding
singularity in the Borel plane moves to infinity, suggesting at least
Borel summability. However, the expected singularity associated to
infrared renormalons~\cite{tHooft:1977xjm,Beneke:1998ui} does not move away and
would induce  a factorial growth of the coefficients similar to the
finite $\Nc$ case. An analysis of the perturbative expansion at high orders should
settle this point. Another issue to be studied  is the interplay between
the order of perturbation theory and the value of $\Nc$. The higher rate
of growth in the number of non-planar diagrams would suggest that
their contribution could overcome the $1/\Nc^2$ suppression at
sufficiently high orders. Something of this kind emerges when
analyzing the simultaneous expansion in powers of 't Hooft coupling and
$1/\Nc^2$ as seen in Ref.~\cite{Marino:2012zq}. The behaviour of the
reduced model could however be quite different.

To close the circle, recently a new method has arisen that allows the
numerical computation of high order coefficients in the perturbative expansion.
It goes under the name of Numerical Stochastic perturbation theory
(NSPT) and was pioneered by the works of Di~Renzo and
collaborators~\cite{DiRenzo:1993hs,DiRenzo:1994av,DiRenzo:1994sy,DiRenzo:1995qc,Burgio:1997hc,DiRenzo:2004hhl}
(for earlier developments on stochastic quantization~\cite{Parisi:1980ys} 
 and stochastic perturbation theory, see also~\cite{Damgaard:1987rr,BOOK_SQ1,BOOK_SQ2}). 
Recently, this methodology has led to
remarkable results about high order coefficients in SU(3) gauge 
theories~\cite{DiRenzo:2000ua,Rakow:2005yn,Bauer:2011ws,Horsley:2012ra,Bali:2013pla,Horsley:2013pra,Bali:2014fea,DelDebbio:2018ftu,DelDebbio:2018vhr}.
As the memory size required  is proportional to the highest perturbation order 
involved in the computation in NSPT and to the volume,
the authors of~\cite{Bali:2013pla,Bali:2014fea,DelDebbio:2018ftu}
have used twisted boundary conditions aware that they lead to reduced finite volume dependence. 

Stochastic quantization is based on the Langevin equation, and the
first NSPT studies used the  perturbative expansion of this equation.
However, for lattice theories various other Monte Carlo algorithms,
such as Kramers, hybrid molecular dynamics (HMD), and hybrid Monte Carlo (HMC) algorithms,
have been developed and used. Recently, extensive studies have been carried out  
using NSPT algorithms~\cite{DallaBrida:2017pex,DallaBrida:2017tru}
based on them. In particular, the HMD based algorithm for NSPT is easy to implement
by  perturbatively expanding  the  HMD/HMC codes used for non-perturbative simulations.
In this case,  one can  benefit from various numerical integration schemes for the molecular dynamics part 
to reduce the finite step size error of the integration. 
Thus, the NSPT algorithm based on the HMD/HMC algorithm could open the way to 
efficiently estimating very high order coefficients.

Our work is a first attempt to apply this methodology to
matrix models. For the time being, our work is  mostly exploratory but
a necessary step before any attempt of a higher order and larger $\Nc$
study. Nonetheless, apart from showing that the method works with an
incredibly high precision for the low-order coefficients, we also present 
results which extend to higher order the previous analytical results~\cite{Perez:2017jyq}. 
In addition, our work provides interesting  information  about the 
probability distribution of the perturbative coefficients. In particular, 
we have studied the
dependence of the cumulants of these distributions with respect to the
different parameters: the value of $\Nc$ and of the fluxes, the size of
the loops and the order of perturbation theory. These results allow
us  to determine the necessary computational requirements for any
further extension of these  studies. 
 
The layout of the paper is the following.
In section~\ref{sec:TEKNSPT}, after introducing the TEK model,  we
explain the  application of the NSPT algorithm based on the HMD
algorithm to the TEK model.
In Section~\ref{sec:NumericalResults} we present  the numerical
results for the perturbative coefficients of 
the Wilson loops. We employ $(N,k)=(16,1),(49,2),(121,3)$ for $\SUNc$ and flux parameter $k$ of the TEK model.
The coefficients are computed up to four-loop ($O(g^8)$) and the first two-loop coefficients are compared to 
the analytic values. 
The probability  distribution of these estimates is investigated and
used to explore the feasibility for extending our results to 
higher orders.
We estimate the numerical computational cost of NSPT for the TEK model
at  large values of $\Nc$ and  
large perturbative orders in section~\ref{sec:costestimate}. 
Possible improvements on the algorithm are also discussed.
In the last section we summarize the main results of the paper. Two
technical points are included in appendices.

\section{TEK model and NSPT}
\label{sec:TEKNSPT}

In this section we start by  briefly introducing the TEK model together with
the gauge fixing method. Then we recall   the Hybrid Molecular
Dynamics (HMD)  algorithm for nonperturbative simulations of  the TEK
model.  This  algorithm is then perturbatively expanded to derive the equation of motion for the NSPT algorithm.

The computational  cost is estimated in terms of the highest
order  $\Nptmax$ in the perturbative expansion involved in the algorithm 
and the matrix size $\Nc$ of $\SUNc$.

\subsection{TEK model}

The partition function of the TEK model in four-dimensions is defined by
\begin{align}
    Z&= \int \prod_{\mu=1}^{4}\mathrm{d}U_{\mu}\,  e^{-S[U]},
\label{eq:PartTEK}
    \\
 S[U]&= \beta \sum_{\mu,\nu=1,\mu\ne\nu}^{4}
\Tr\left[I - z_{\mu\nu} U_{\mu}U_{\nu}U_{\mu}^{\dag}U_{\nu}^{\dag} \right],
\label{eq:ActionTEK}
\end{align}
where $U_{\mu}$ are $\SUNc$ matrices.  We choose the symmetric twist
characterized by $\Nc=\Lbox^2$ and $z_{\mu\nu}$ given by
\begin{align}
  z_{\mu\nu} &= \exp\left[  \dfrac{2\pi i}{\Nc} n_{\mu\nu}\right], \notag
\\
  n_{\mu\nu} &= \epsilon_{\mu\nu}k \Lbox, \notag
\\ 
\epsilon_{\mu\nu} &= \left\{
      \begin{matrix}
          +1 & \quad \mbox{for $\mu<\nu$}\\
           0 & \quad \mbox{for $\mu=\nu$}\\
          -1 & \quad \mbox{for $\mu>\nu$}
      \end{matrix}\right.,
\label{eq:twistphase}
\end{align}
where $k$ is an integer coprime with $\Lbox$ ($=\sqrt{\Nc}$). 
The bare coupling constant $g$ is defined through $\beta=1/g^2$.

The Wilson loop operator for an  $R\times T$ rectangle in the
$\mu$--$\nu$ plane is defined by
\begin{align}
 W_{\mu\nu}(R,T)&= \dfrac{1}{\Nc}\left(z_{\mu\nu}\right)^{RT}  \Tr\left[ \left(U_{\mu}\right)^{R} 
 \left(U_{\nu}\right)^{T}
 \left(U_{\mu}^{\dag}\right)^{R}\left(U_{\nu}^{\dag}\right)^{T}\right].
\label{eq:WLOOPOP}
\end{align}
In this paper we restrict ourselves to square loops $R=T$.

The analytic perturbative expansion of the observables proceeds by expanding the link matrices
around the classical vacuum as
\begin{align}
    U_{\mu}=e^{-ig A_{\mu}}\Gamma_\mu,
\end{align}
where the matrices $\Gamma_\mu$ define the classical vacuum of
eq.~\eqref{eq:ActionTEK} and have the following property
\begin{align}
\Gamma_{\mu}\Gamma_{\nu}  =
\Gamma_{\nu}\Gamma_{\mu} z_{\mu\nu}^{*}.
\label{eq:TwistEater}
\end{align}
A particular solution (up to multiplication by a phase) is singled out
by an appropriately chosen gauge fixing condition, accompanied by the
corresponding ghost term.

In order to stabilize the runaway trajectories,  NSPT also requires
the use of the  stochastic gauge fixing method~\cite{Zwanziger:1981kg}.
Here we are  using the  functional for the Landau gauge condition, from which we can extract 
the  stochastic gauge fixing contribution in NSPT~\cite{DiRenzo:1993hs,DiRenzo:1994av}.
The gauge fixing functional $F[G]$ is defined by
\begin{align}
    F[G]&= \sum_{\mu=1}^{4}\mathrm{Re}\left[\Tr\left[ G U_{\mu} G^{\dag} \Gamma_\mu^{\dag}\right]\right],
\end{align}
where $G \in \SUNc$ is the gauge transformation matrix. The
Landau gauge condition is achieved by maximizing the functional, by
application of several  iterations
of the form:
\begin{align}
   U_{\mu} \to  U_{\mu}'=G U_\mu G^{\dag},
\end{align}
with 
\begin{align}
G &= \exp\left[ i \alpha \Theta \right],\\
 \Theta &= i \left[ \left(Y - Y^\dag\right) - \dfrac{1}{\Nc}
\Tr\left(Y - Y^\dag\right)\right],
\label{eq:GFForce}
\\
Y &= \sum_{\mu=1}^{4} \left[ U_{\mu}\Gamma_{\mu}^{\dag}- \Gamma_{\mu}^{\dag}U_{\mu}\right],
\label{eq:GFForce2}
\end{align}
where $\alpha$ is a parameter. 
When this iteration process is transformed  into a continuous evolution equation with a fictitious time $t$, 
we obtain the following differential equation:
\begin{align}
            G(t) &= \exp\left[i  w(t) \right], 
\label{eq:transMat}
\\
 \dfrac{d w}{dt} &\equiv \alpha  \Theta[U(t)], 
\label{eq:timedepsgfixtrans}
\\
  U_{\mu}(t)     &=  G(t) U_{\mu} G(t)^{\dag},
\label{eq:Gtrans}
\end{align}
where $w(t)$ is an auxiliary Hermitian-traceless matrix tracing the steepest descent trajectory.
This evolution equation can be combined with   the NSPT process to
implement  the stochastic gauge fixing.

\subsection{NSPT for the TEK model}

NSPT is based on stochastic quantization~\cite{Parisi:1980ys}. 
This amounts to writing down a stochastic differential equation of
the Langevin type for the field  variables of a target system, which 
relaxes to the
equilibrium probability density at large times~\cite{Damgaard:1987rr,BOOK_SQ1,BOOK_SQ2}.
The NSPT method is obtained by  expanding these field variables in
powers of the coupling constant and casting the original Langevin
equation  into a tower of equations, one for each power. 
 Observables that are analytic
functions of the field variables acquire a corresponding perturbative expansion.
Each term in the expansion becomes an stochastic variable whose mean
value at large times gives us the corresponding coefficient that  we are
after.

In this paper we employ the HMD based NSPT, which has been introduced in refs.~\cite{DallaBrida:2017tru,DallaBrida:2017pex}.
This is easy  to implement through modifications of existing
codes  of nonperturbative HMD or HMC algorithms, and  is preferable to systematically improve the molecular 
dynamics (MD) part, reducing the error arising from the finite MD step size.

To explain the method in our case, we will begin by reviewing  the generalized HMD (GHMD) algorithm for nonperturbative simulations.
Then we will apply the perturbative expansion to the algorithm, to
derive  the HMD based NSPT algorithm. 

The HMD partition function is introduced by adding the canonical momentum variable $P_{\mu}$ 
conjugate to $U_{\mu}$ to the original partition function eq.~\eqref{eq:PartTEK}.
\begin{align}
    Z_{\mathrm{HMD}}&=
    \int \prod_{\mu=1}^{4}\mathrm{d}U_{\mu}
    \prod_{\mu=1}^{4}\mathrm{d}P_{\mu}\, e^{-H[P,U]},
\\
H[P,U]&= \dfrac{1}{2}\sum_{\mu=1}^{4}\Tr\left[P_{\mu}P_{\mu}\right] + S[U].
\label{eq:HMDHamil}
\end{align}
The classical dynamics for $(P_\mu,U_\mu)$ with the Hamiltonian $H[P,U]$ reproduces 
the microcanonical ensemble at fixed energy $E=H[P,U]$. Subsequent
refreshment of the variable  $P_\mu$ with the Gaussian distributions
generates the corresponding canonical ensemble $e^{-H[P,U]}$. 
The marginal distribution for $U_{\mu}$ becomes the  $e^{-S[U]}$
distribution that we are looking for.

\begin{algorithm}[t]
\caption{HMD algorithm for nonperturbative simulations}
\label{eq:HMDALG}
\begin{algorithmic}[0]
\STATE \textbf{Step 0}: Set initial state $U_{0,\mu}$ arbitrary.
\STATE \textbf{Step 1}: Set initial momentum $P_{0,\mu}$ from the Gaussian distribution as
    \begin{align}
      P_{0,\mu} &= \sum_{a=1}^{\Nc^2-1}\eta_\mu^a T^a \equiv \eta_\mu,
      \label{eq:MomUpdation}
      \\
      \mathrm{Prob}(\eta_\mu^a)d\eta^a & \propto \exp(-(\eta_\mu^a)^2/4)d\eta^a ,
      \label{eq:MomDistribution}
     \end{align}
    where the matrices $T^a$ are $\SUNc$ generators in the fundamental
    representation normalized as $\mathrm{Tr}(T^aT^b)=\frac{1}{2}\delta_{a b}$.
\STATE \textbf{Step 2}: Solve the MD equation defined by 
\begin{align}
\dot{U}_{\mu} & = i P_{\mu}U_{\mu},\\
\dot{P}_{\mu} & = F_{\mu},\\
F_{\mu}&\equiv i \beta\left[ \left[S_{\mu} - S_{\mu}^{\dag} \right] - \dfrac{1}{\Nc}\Tr\left[S_{\mu} - S_{\mu}^{\dag} \right]\right],
\label{eq:MDforce}
\\
S_{\mu}&\equiv 
U_{\mu}\left[ \sum_{\nu=1,\nu\ne\mu}^{4} \left(z_{\mu\nu} U_{\nu} U_{\mu}^{\dag} U_{\nu}^{\dag}+ z_{\mu\nu}^* U_{\nu}^{\dag} U_{\mu}^{\dag} U_{\nu}\right)\right],
\label{eq:eom}
\end{align}
from the initial state $(P_{0,\mu},U_{0,\mu})$ for a fixed interval in
fictitious time. 
The dot $\dot{\ }$ represents the left-wise derivative with respect to  the
fictitious time. These evolution equations preserve the energy value $E=H[P,U]$.
\STATE \textbf{Step 3}: Store $U_\mu$ as the configuration and set $U_{0,\mu}=U_{\mu}$, then return to Step 1.
\end{algorithmic}
\end{algorithm}

The HMD algorithm proceeds as described in Algorithm~\ref{eq:HMDALG},
where the equations in Step 2 are discretized in time. This
discretization violates  the exact energy conservation,  which
distorts the distribution shape. In order to ensure that  the
discretized  Markov chain leads to the correct probability
distribution the Monte Carlo process must
satisfy time reversibility and area preservation in the MD evolution, 
for which the leapfrog type scheme is normally used.
For nonperturbative simulations, the Metropolis test can be inserted between 
 Step 1 and 2 to compensate the violation of energy conservation, yielding the HMC algorithm.

The full momentum refreshment in  Step 1 will cause random walking in  phase space so that 
the autocorrelation times becomes longer for the ensemble.
To relax random walking behaviour, the generalized HMC algorithm (GHMC) has been introduced in ref.~\cite{Kennedy:2000ju}.
For the GHMC algorithm the momentum is partially refreshed in Step 1 as
\begin{align}
    P_{0,\mu} = c_1 P_{\mu} + \sqrt{1-c_1^2}\, \eta_{\mu},
\label{eq:momupdate}
\end{align}
instead of eq.~\eqref{eq:MomUpdation}, 
where $c_1$ is a mixing parameter. $P_\mu$ in the right hand-side is the solution of  Step 2.
A momentum reflection step is added after the rejection of the Metropolis test in the GHMC algorithm, 
while this step is absent in the GHMD algorithm.
We can further include the gauge fixing process eq.~\eqref{eq:timedepsgfixtrans} into the MD evolution equation.

With the leapfrog algorithm and  the partial momentum refreshment, the
updating algorithm for one step having a small  $\Delta t$ interval,
is given by 
\begin{align}
& \left\{\ 
    \begin{aligned}
 P_{\mu} &= c_1  P_{0,\mu} + \sqrt{1-c_1^2} \eta_{\mu}, \\
 w       &= c_1  w_{0}, \\
 U_{\mu} &= U_{0,\mu}, 
    \end{aligned}
\right.
\label{eq:MDinitailze}
\\
& \left\{\ 
    \begin{aligned}
  \bar{P}_{\mu}  &= P_{\mu} + (\Delta t/2) F_{\mu}[U], \\
        U_{\mu}' &= \exp\left[ i \bar{P}_{\mu} \Delta t\right] U_{\mu},\\
        P_{\mu}' &= \bar{P}_{\mu} + (\Delta t/2) F_{\mu}[U'],
    \end{aligned}
\right.
\label{eq:simpleLeapFrog}
\\
&
\left\{\ 
    \begin{aligned}
        w_1 &= w + \Delta t \alpha \Theta[U'], \\
  P_{1,\mu} &= \exp\left[ i w_1 \Delta t \right] P'_{\mu} \exp\left[ -i  w_1\Delta t \right], \\
  U_{1,\mu} &= \exp\left[ i w_1 \Delta t \right] U'_{\mu} \exp\left[ -i  w_1\Delta t \right],
    \end{aligned}
\right.
\label{eq:GF}
\end{align}
where $(P_{0,\mu},U_{0,\mu},w_{0})$ is the initial state and 
      $(P_{1,\mu},U_{1,\mu},w_{1})$ is the final state.  
Eq.~\eqref{eq:simpleLeapFrog} corresponds to the leapfrog evolution for $\Delta t$, 
for which various higher-order schemes are available. For simplicity
we will explain  the second order leapfrog scheme only.
The gauge fixing evolution \eqref{eq:transMat}--\eqref{eq:Gtrans} is interleaved into 
the MD evolution~\cite{Rossi:1987hv,Davies:1987vs}
and the transformation in eq.~\eqref{eq:GF} does not affect the gauge invariant observables, however, 
we introduce this to explain the stochastic gauge fixing for NSPT~\cite{Zwanziger:1981kg,DiRenzo:1993hs,DiRenzo:1994av}.
Taking the $\Delta t \to 0$ limit with $c_1=e^{-\gamma \Delta t}$, as
shown in ref.~\cite{DallaBrida:2017tru}, 
this evolution reduces to 
\begin{align}
\dot{U}_{\mu} &= i \left[ P_{\mu} - D_{\mu}w \right] U_{\mu},\notag \\
\dot{P}_{\mu} &= F_{\mu} - \gamma P_{\mu} + i \left[ w,P_{\mu}\right] + \zeta_\mu,\notag \\
      \dot{w} &= -\gamma w + \alpha \Theta,
\label{eq:LCMC}
\end{align}
where $D_{\mu}w$ is defined as
\begin{align}
D_{\mu}w &= U_{\mu}w U_{\mu}^{\dag} -w,
\end{align}
and $\zeta_\mu$ is a random noise satisfying
\begin{align}
 \langle \zeta_\mu^a(t) \zeta_\nu^b(s) \rangle &= 4\gamma \delta(t-s)\delta^{a,b}\delta_{\mu,\nu}.
\end{align}
The terms with $w$ act as the gauge damping force.
Eq.~\eqref{eq:LCMC} corresponds to the Kramers equation and in the limit
$\gamma\to +\infty$ ($c_1=0$) to the Langevin equation with gauge damping force.

In the GHMD algorithm Step 2 is obtained by repeating eqs.~\eqref{eq:simpleLeapFrog} and \eqref{eq:GF}  until 
the total evolution time becomes a fixed value, and then applying   
eq.~\eqref{eq:MDinitailze} to implement Step 1.
When the evolution time in Step 2 reduces to  $\Delta t$, the method becomes Langevin ($c_1=0$) or Kramers ($c_1\ne 0$) algorithm.

Having explained the GHMD algorithm, we now describe the corresponding
NSPT algorithm. This follows  by expanding $(P_\mu, U_\mu, w)$ and 
the MD equation~\eqref{eq:eom} or 
eqs.~\eqref{eq:MDinitailze}--\eqref{eq:GF}  in a power series in $g$. 
We will now describe the  perturbative expansion for eqs.~\eqref{eq:MDinitailze}--\eqref{eq:GF},
because it is sufficient to write the simulation program.
In order to simplify notation, we first define the $\star$-product as the
convolution product of two perturbative series, given by 
\begin{align}
C &= A\star B,\\
A  = \sum_{k=0}^{\infty} g^k A^{(k)},\quad
B &= \sum_{k=0}^{\infty} g^k B^{(k)},\quad
C  = \sum_{k=0}^{\infty} g^k C^{(k)},\\
C^{(k)}=(A\star B)^{(k)} &\equiv \sum_{j=0}^{k} A^{(j)} B^{(k-j)},
\end{align}
where $A,B,C$ are matrices and $A^{(k)},B^{(k)},C^{(k)}$ are the coefficient matrices.

The perturbative expansion for $(P_{\mu}, U_{\mu}, w)$ is defined by
\begin{align}
   P_{\mu} &= \beta^{1/2}\sum_{k=1}^{\infty}\beta^{-k/2} P_{\mu}^{(k)}
 = P_{\mu}^{(1)}+g P_{\mu}^{(2)}+g^2 P_{\mu}^{(3)}+\dots,
\label{eq:PTp}
\\
   U_{\mu} &= \sum_{k=0}^{\infty}\beta^{-k/2} U_{\mu}^{(k)}
 = U_{\mu}^{(0)}+g U_{\mu}^{(1)}+g^2 U_{\mu}^{(2)}+g^3 U_{\mu}^{(3)}+\dots,
\label{eq:PTu}
\\
         w &= \beta^{1/2}\sum_{k=1}^{\infty}\beta^{-k/2} w^{(k)}
 = w^{(1)}+g w^{(2)}+g^2 w^{(3)}+\dots,
\label{eq:PTg}
\end{align}
where $U_{\mu}^{(0)}=\Gamma_{\mu}$ is kept fixed in NSPT as it is the perturbative vacuum.
Rescaling the fictitious time and the gauge fixing parameter as $t'=t/g$ and $\alpha'=g^2 \alpha$, 
and substituting eqs.~\eqref{eq:PTp}--\eqref{eq:PTg} into eqs.~\eqref{eq:MDinitailze}--\eqref{eq:GF}, 
we obtain
\begin{align}
& \left\{\ 
    \begin{aligned}
    {P}^{(k)}_{\mu} &= c_1 P^{(k)}_{0,\mu} + \sqrt{1-c_1^2} \eta_{\mu}\delta_{k,1},\\
          {w}^{(k)} &= c_1 w_0^{(k)},\\
      U^{(k)}_{\mu} &= U^{(k)}_{0,\mu}, 
    \end{aligned}
\right.
\label{eq:NSPTMDinitailze}
\\
& \left\{\ 
    \begin{aligned}
  \bar{P}_{\mu}^{(k)}   &= P^{(k)}_{\mu} + (\Delta t'/2) F^{(k)}_{\mu}[U], \\
       {U^{(k)}_{\mu}}' &= \left(\exp\left[ i \bar{P}_{\mu} \Delta t' \right] \star U_{\mu}\right)^{(k)},\\
       {P^{(k)}_{\mu}}' &= \bar{P}^{(k)}_{\mu} + (\Delta t'/2) F^{(k)}_{\mu}[U'],
    \end{aligned}
\right.
\label{eq:NSPTsimpleLeapFrog}
\\
&
\left\{\ 
    \begin{aligned}
        w^{(k)}_1 &= w^{(k)} + \Delta t' \alpha' \Theta^{(k)}[U'], \\
  P^{(k)}_{1,\mu} &= \left(\exp\left[ i w_1 \Delta t' \right] \star P'_{\mu} \star \exp\left[ -i  w_1\Delta t' \right]\right)^{(k)}, \\
  U^{(k)}_{1,\mu} &= \left(\exp\left[ i w_1 \Delta t' \right] \star U'_{\mu} \star \exp\left[ -i  w_1\Delta t' \right]\right)^{(k)},
    \end{aligned}
\right.
\label{eq:NSPTGF}
\end{align}
for each perturbative order $k=1,2,\dots$. The details of the perturbative expansion of the matrix exponential 
$\exp[A]=I+g \left(\exp[A]\right)^{(1)}+g^2 \left(\exp[A]\right)^{(2)}+\dots$ with $A=g A^{(1)} + g^2 A^{(2)} +\dots$ 
are explained in Appendix~\ref{sec:AppendixA}. 
The perturbative expressions for $F^{(k)}_{\mu}$ and $\Theta^{(k)}$ are extracted from
eqs.~\eqref{eq:MDforce}--\eqref{eq:eom}, 
 and \eqref{eq:GFForce}--\eqref{eq:GFForce2}, respectively:
\begin{align}
F_{\mu}&= \beta \sum_{k=1}^{\infty} \beta^{-k/2} F^{(k)}_{\mu},
\quad 
\Theta  = \sum_{k=1}^{\infty}\beta^{-k/2}\Theta^{(k)},
\\
F^{(k)}_{\mu}&=
i \left[S^{(k)}_{\mu}-{S^{(k)}}_{\mu}^{\dag}-\dfrac{1}{\Nc}\Tr\left[S^{(k)}_{\mu}-{S^{(k)}}_{\mu}^{\dag}\right]\right],
\\
 S^{(k)}_{\mu} &= \left( U_{\mu}\star 
\sum_{\nu\ne\mu}\left[ 
    z_{\mu\nu} U_{\nu}        \star U^{\dag}_{\mu} \star U^{\dag}_{\nu}
+ z^*_{\mu\nu} U^{\dag}_{\nu} \star U^{\dag}_{\mu} \star U_{\nu}
\right]\right)^{(k)},\\
\Theta^{(k)}&= i \left[ \left(Y^{(k)} - {Y^{(k)}}^\dag\right) - \dfrac{1}{\Nc}\Tr\left(Y^{(k)} - {Y^{(k)}}^\dag\right)\right],\\
Y^{(k)} &= \sum_{\mu=1}^{4} \left[ U^{(k)}_{\mu}\Gamma_{\mu}^{\dag}- \Gamma_{\mu}^{\dag}U^{(k)}_{\mu}\right].
\end{align}
In the limit  $\Delta t'\to 0$, the equation of motion should conserve
the energy eq.~\eqref{eq:HMDHamil} order by order in perturbation theory.
This energy conservation can be monitored during the simulation.

The Wilson loop operator eq.~\eqref{eq:WLOOPOP} is expanded similarly:
\begin{align}
 W^{(k)}_{\mu\nu}(R,T)&= \dfrac{\left(z_{\mu\nu}\right)^{RT}}{\Nc}\Tr\left[ \left( \left(U_{\mu}\right)^{\star R} \star
 \left(U_{\nu}\right)^{\star T} \star \left(U_{\mu}^{\dag}\right)^{\star R} \star \left(U_{\nu}^{\dag}\right)^{\star T}
\right)^{(k)}
\right],
\label{eq:NSPTWLOOPOP}
\end{align}
where we used a shorthand notation for matrix power with the $\star$-product such as $(A\star A \star A)\equiv A^{\star 3}$.
The expectation values at large times of our $W^{(k)}_{\mu\nu}(R,T)$
yield the coefficients of the perturbative expansion of Wilson loops.
For the latter we take the notation given in ref.~\cite{Perez:2017jyq}
where the first two coefficients have been computed analytically. 
If we consider for example an $R\times T$ Wilson loop in the $\mu$--$\nu$ plane, the
relation is as follows 
\begin{align}
\hat{W}_{\ell}^{(R\times T)} &= -\Nc^{-\ell} \left\langle W^{(2\ell)}_{\mu \nu}(R,T) \right\rangle,
 \notag \\
0 &= \left\langle W^{(2\ell+1)}_{\mu \nu}(R,T) \right\rangle.
\end{align}
Notice that, since the perturbative expansion is in powers of 't Hooft
coupling $\lambda=g^2 N$,  the expectation value of $W^{(k)}_{\mu\nu}$ for  odd
values of $k$ has to vanish.

A technical point is the necessity  to correct for deviations from
the unitarity constraint induced by the numerical round-off error from
the finite precision of computer arithmetic. While imposing the
traceless-Hermitian character  of  $(P_{\mu}^{(k)},w^{(k)})$ is easily
done, the conditions on $\{U_{\mu}^{(k)}\}$ following from the
unitarity of the link matrices can be imposed by the matrix logarithm
scheme for  reunitarization~\cite{DiRenzo:2004hhl}.
The details of the group projection is described in Appendix~\ref{sec:AppendixB}.

\begin{algorithm}[t]
\caption{NSPT algorithm based on GHMD. 
The MD time step size $\Delta t'=1/\NMD$, momentum mixing parameter $\gamma$, and gauge fixing parameter $\alpha'$ are given.}
\label{eq:NSPTHMDALG}
\begin{algorithmic}[0]
\STATE \textbf{Step 0}: Set an initial state $P^{(k)}_{0,\mu}=0,U^{(k)}_{0,\mu}=0, w^{(k)}_{0}=0$ for $k=1,\cdots$.
\STATE \textbf{Step 1}: Repeat the MD evolution and gauge transformation as
\bindent
\FOR{$j=0,\NMD-1$}
\STATE \textbf{1-1:}  Compute eq.~\eqref{eq:NSPTMDinitailze} with partial momentum refreshment parameter $c_1=e^{-\gamma \Delta t'}$.
\STATE \textbf{1-2:}  Evolve state with the discretized MD equation eq.~\eqref{eq:NSPTsimpleLeapFrog} or higher order scheme for $\Delta t'$.
\STATE \textbf{1-3:}  Transform state with eq.~\eqref{eq:NSPTGF} for gauge fixing.
\STATE \textbf{1-4:}  Set $(P^{(k)}_{0,\mu},U^{(k)}_{0,\mu},w^{(k)}_{0})=(P^{(k)}_{1,\mu},U^{(k)}_{1,\mu},w^{(k)}_{1})$.
\ENDFOR
\eindent
\STATE \textbf{Step 3}: Project the state $(P^{(k)}_{0,\mu},U^{(k)}_{0,\mu},w^{(k)}_{0})$ out to $\SUNc$ group and algebra.
\STATE \textbf{Step 4}: Compute observables with ${U^{(k)}_{0,\mu}}$ and return to Step 1. 
\end{algorithmic}
\end{algorithm}

Let us conclude by summarizing our NSPT algorithm. As explained
earlier, the finite time step induces a distortion in the probability
distribution, which in NSPT cannot be corrected by a Metropolis test.
Hence, it is preferable to employ a higher order integration scheme
in the MD evolution. Although, for simplicity, we used the simple
(second order) leapfrog scheme to explain the HMD based NSPT algorithm,
in practice we  employed the fourth order leapfrog scheme for
eq.~\eqref{eq:NSPTsimpleLeapFrog},
while the evolution of $w$ is not changed.
The properties of second and fourth order Omelyan-Mryglod-Folk
schemes~\cite{OMF,Takaishi:2005tz}, 
and second order Runge-Kutta scheme in NSPT have 
been investigated in~\cite{DallaBrida:2017pex,DallaBrida:2017tru}. 
We summarize the NSPT algorithm used in this paper in
algorithm~\ref{eq:NSPTHMDALG}.
This corresponds to the Kramers type NSPT algorithm (called KSPT
in~\cite{DallaBrida:2017pex}).
The specific parameters are fixed to $\gamma=0.5$ for the momentum
refreshing parameter $c_1=\mathrm{e}^{-\gamma \Delta t'}$ and
$\alpha'=2$ for  the gauge damping parameter. The trajectories are
performed over a fixed time $t'=1$, with a perturbative
reunitarization step at the end. The Wilson loop coefficients are 
computed at the end of  every trajectory.

\subsubsection{Computational estimates}
The maximum power of $g$ that is studied $\Nptmax$ is limited by the total
computer memory available. For the $\SUNc$ TEK model, the total memory
requirement is of $O(\Nptmax \Nc^2)$.  
Most of the computational time is spent in matrix multiplication, 
whose cost is of $O(\Nc^3)$. The cost of the convolutional $\star$-product is 
of $O(\Nptmax^2)$ with a naive implementation. 
The evaluation of the perturbative matrix function requires one more factor of $\Nptmax$,
yielding the cost of $O(\Nptmax^3)$ (see Appendix~\ref{sec:AppendixA}).
The cost of the reunitarization we implemented is of $O(\Nptmax^4)$ (see Appendix~\ref{sec:AppendixB}).

The energy difference in a trajectory $\Delta H$ is non-zero for a
finite MD step size $\Delta t$. The relation between the energy
conservation violation $\Delta H$ and $\Delta t$ depends on the MD
integration scheme. We assume that algorithm~\ref{eq:NSPTHMDALG} with 4th order leapfrog
scheme yields
$\Delta H^{(k)} \sim \Delta t^4$ at each perturbation
order~\cite{DallaBrida:2017pex}, where
$\Delta H^{(k)}$ is the energy conservation violation for
the perturbative coefficient of the total energy in NSPT. Since the
energy is proportional to $\Nc^2$, to achieve a constant energy
conservation violation, the number of steps $\NMD$, therefore, should
scale as $\NMD=\sqrt{\Nc}$.

The total computational cost of the NSPT algorithm truncated at $\Nptmax$ for the $\SUNc$ TEK model, 
then, scales as $O(\Nptmax^3 \Nc^3 \NMD)=O(\Nptmax^3 \Nc^{7/2})$ for the MD part and 
$O(\Nptmax^4 \Nc^3)$ for the reunitarization part. Our estimates of
total CPU time do not take into account  autocorrelation times. 
For that we need experimental studies on the statistical properties of
the Monte Carlo data, which  depends on models and algorithms.
As far as the  parameters we investigated, no sizable autocorrelation and parameter dependence are observed
and the autocorrelation length is of $O(1)$ in units of the trajectory length $t'=1$.

\section{Numerical results}
\label{sec:NumericalResults}

In this section we show the results for the perturbative coefficients
of the Wilson loops with NSPT, and 
investigate the statistical properties of the distribution. 
We have implemented the NSPT algorithm as explained in the previous section. 
We accumulate the statistics for the perturbative coefficients of
Wilson loops up to order $g^8$  ($\Nptmax=8$).
The mean value gives us the corresponding perturbative quantity, but
the variance and higher cumulants allow us to estimate the required
statistics to achieve a given error in these coefficients.
Fortunately, we have  analytic results for the one and two-loop
coefficients to test our results at these orders, but we can extend
these results two more orders in powers of $\lambda$. Furthermore,  we monitor 
the coefficients  at odd-orders $O(g^{2\ell+1})$ which should be  zero for Wilson loops.
This provides a test of  the cancellation of the non-loop effect in
NSPT. Indeed,  our results are  consistent with zero within the two standard deviation.
Therefore, we will concentrate in giving the results of  even-order coefficients.

The parameters and statistics are shown in table~\ref{tab:simpara}. $\NMD$ is the number of MD steps for unit trajectory $t'=1$. 
Statistical errors are estimated with the jackknife method after binning in  1000 trajectory samples. 
We discard the first $\ge 1000$ trajectories to account for thermalization.
Since we employ the 4th order leapfrog scheme for the MD integrator, a ${\Delta t'}^4=1/\NMD^4$ dependence is expected in observables~\cite{DallaBrida:2017pex}.
We also study the limit of vanishing step size for some representative
cases. 

Given the pilot nature of our study we have concentrated in studying
cases which have been analyzed in detail in the analytic calculations
of ref.~\cite{Perez:2017jyq}. Thus we concentrated in the three values of
$\Nc=16,49,121$. The lowest values are not so interesting from the
point of view of approximating large $\Nc$ Yang-Mills theory at
infinite volume. Corrections are expected to be large. However,
only for a value as low as $N=16$ can one see clearly two effects of
the twisted boundary conditions: the breakdown of CP and of
cubic-rotational invariance. The first phenomenon manifests itself in
non-vanishing imaginary parts of Wilson loops. We included this case
to test this phenomenon in our NSPT determinations. On the other hand
having $\Nc=49$ and $121$ allows us to test the dependence on $\Nc$ of
 both the physical and computational parameters. We also tested several
values of the flux parameter $k$. For sufficiently large values of
$k/\Lbox$ the dependence is quite small. For large values of $\Nc$ and
small values of $k$ the dependence can be quite strong even in the
perturbative calculation. Non-perturbatively this restriction is even
more important to preserve a remnant of  center
symmetry~\cite{GonzalezArroyo:2010ss} that validates reduction in the
limit $\Nc\to \infty$.

\begin{table}[t]
    \centering
    \begin{tabular}{ccccS[table-format=8.1]}
\toprule
  $\Nc$ & $\Lbox$ & $k$ & $\NMD$ & {Statistics} \\
\midrule
 121    & 11  &  3  & 40 &   216000  \\ 
        &     &     & 32 &   175000  \\ 
        &     &     & 28 &   182000  \\ \midrule
  49    &  7  &  1  & 32 &   516000  \\ \midrule
  49    &  7  &  3  & 32 &   520000  \\ \midrule
  49    &  7  &  2  & 32 &   510000  \\
        &     &     & 24 &   528000  \\
        &     &     & 20 &   504000  \\ \midrule
  16    &  4  &  1  & 32 &  1140000  \\
        &     &     & 20 &  1170000  \\
        &     &     & 16 &  1120000  \\ 
\bottomrule
    \end{tabular}
    \caption{Simulation parameters and statistics.}
    \label{tab:simpara}
\end{table}

\newcommand{\figscale}{0.55}
\graphicspath{{FIGS/}}

\begin{figure}[t]
    \centering
\includegraphics[scale=\figscale,clip]{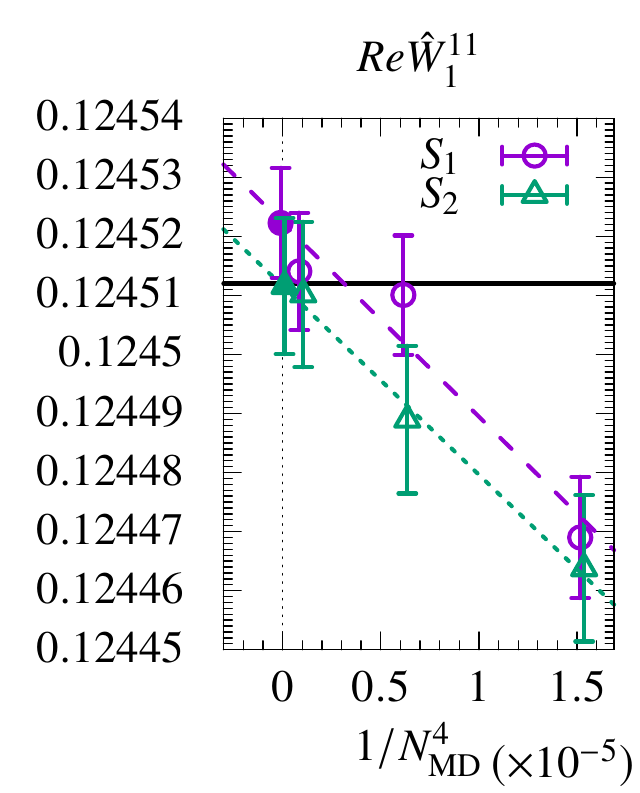}
\includegraphics[scale=\figscale,clip]{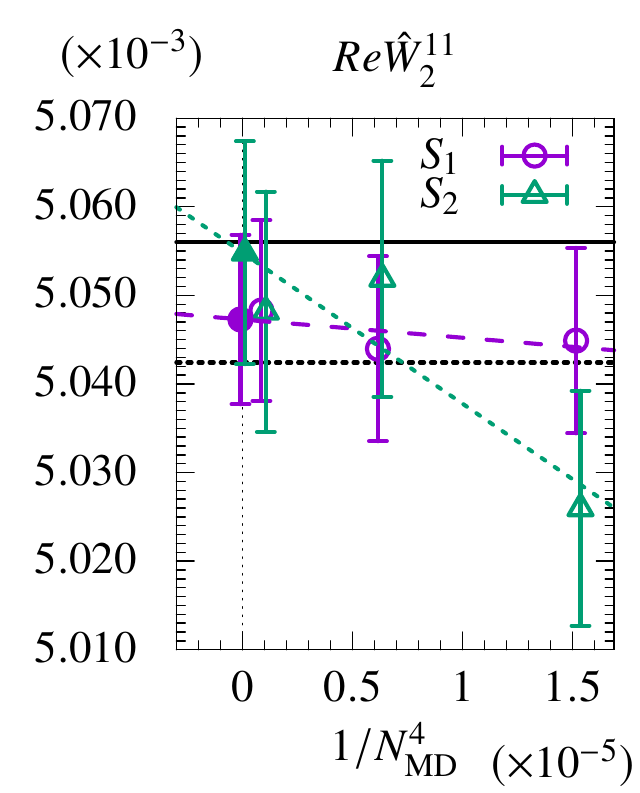}
\includegraphics[scale=\figscale,clip]{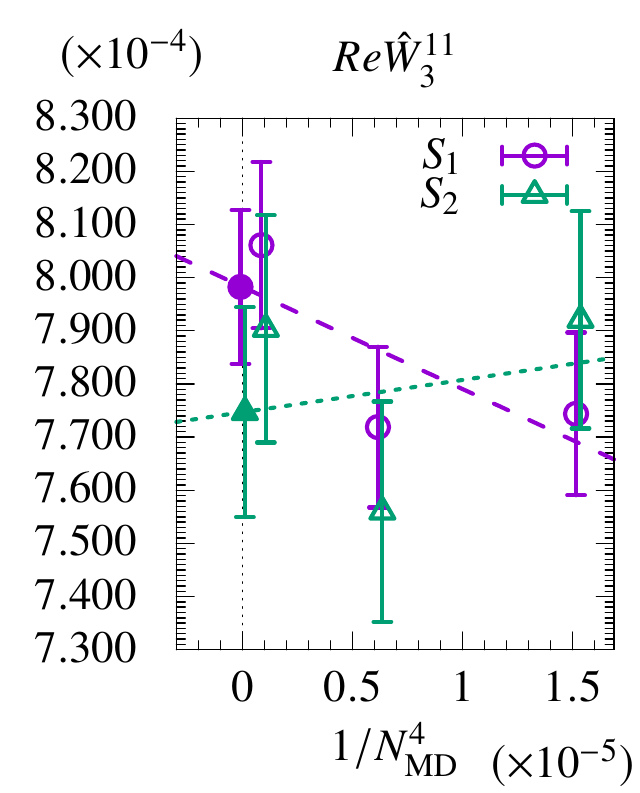}
\includegraphics[scale=\figscale,clip]{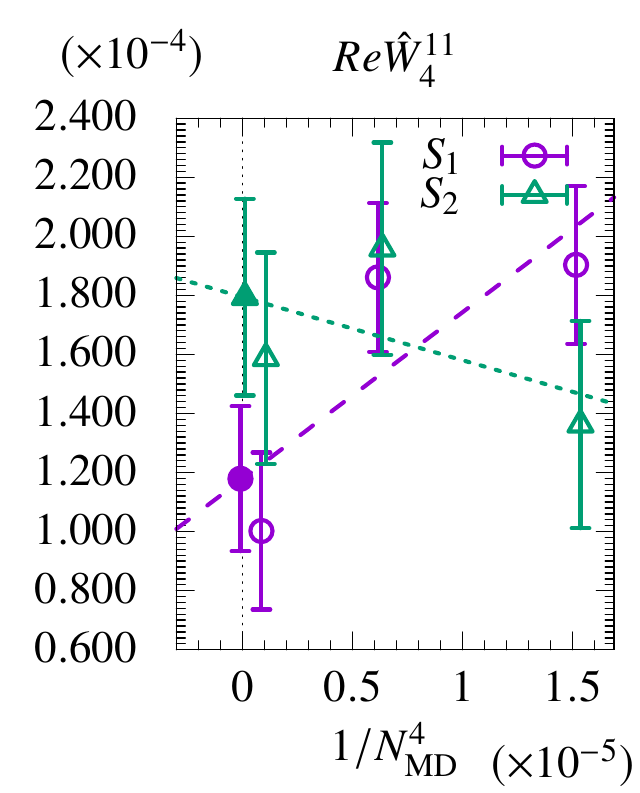}
\includegraphics[scale=\figscale,clip]{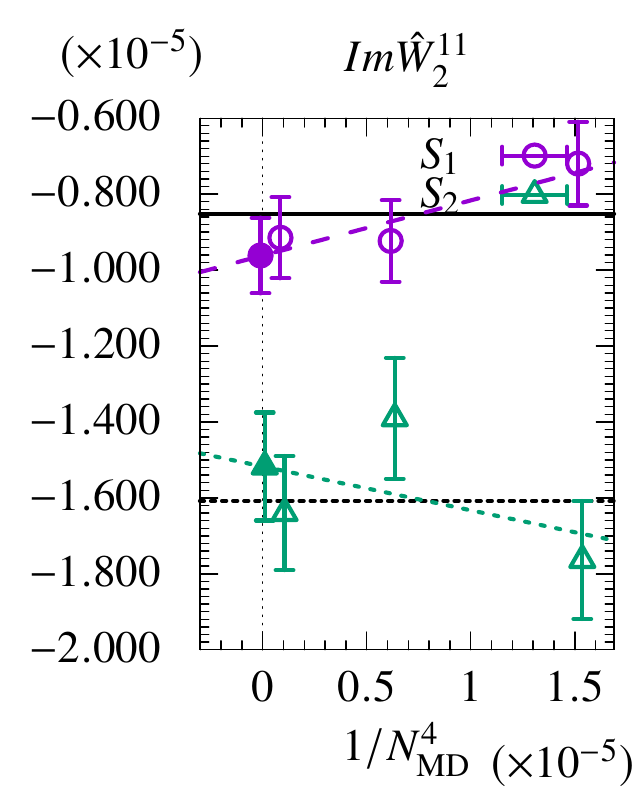}
\includegraphics[scale=\figscale,clip]{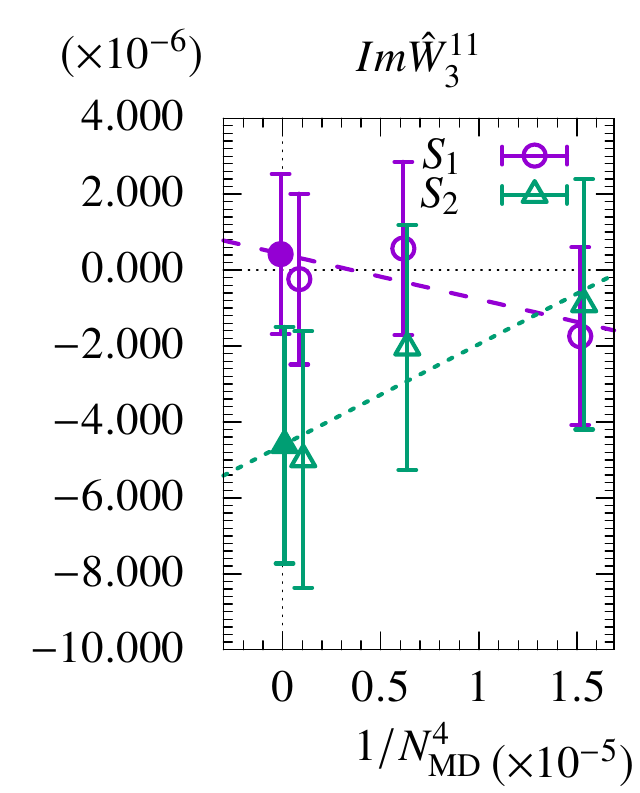}
\includegraphics[scale=\figscale,clip]{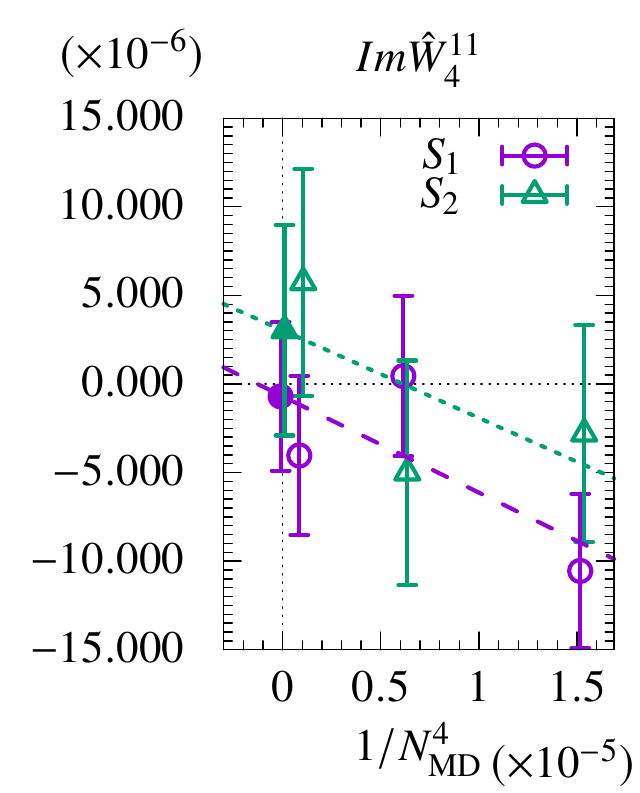}
    \caption{MD step size dependence and extrapolation for $\RE \hat{W}^{11}_\ell$ (upper) 
             and $\IM \hat{W}^{11}_\ell$ (lower) at $(\Nc,k)=(16,1)$.  }
    \label{fig:ReW11ImW11MDDepSU16K1}
\end{figure}

\begin{figure}[t]
    \centering
\includegraphics[scale=\figscale,clip]{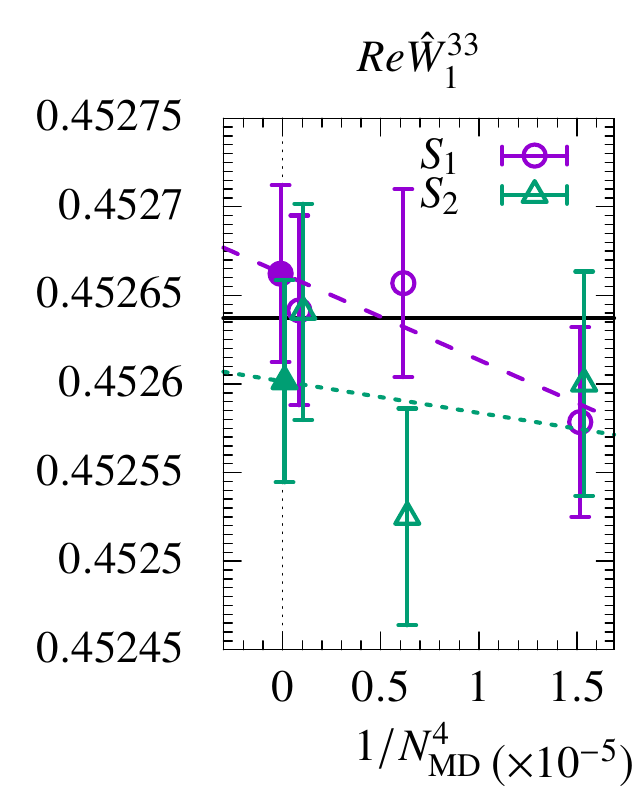}
\includegraphics[scale=\figscale,clip]{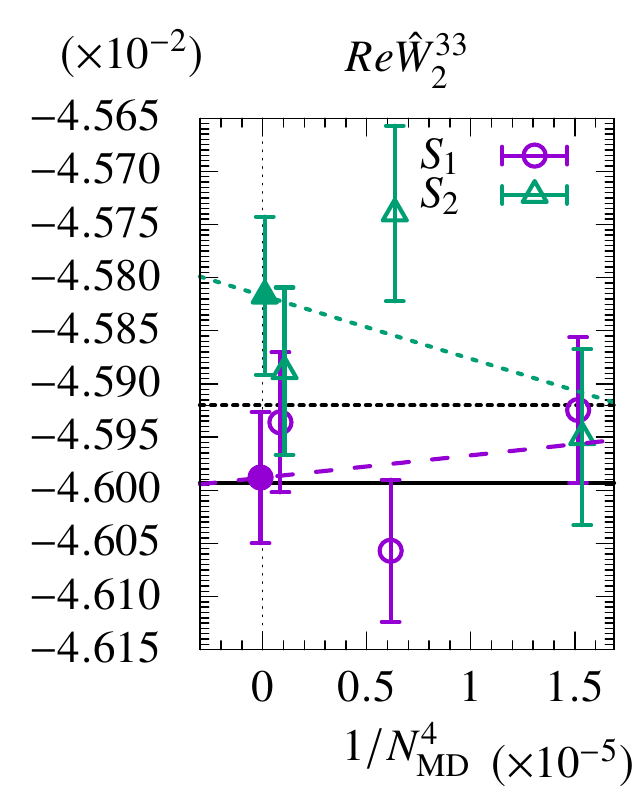}
\includegraphics[scale=\figscale,clip]{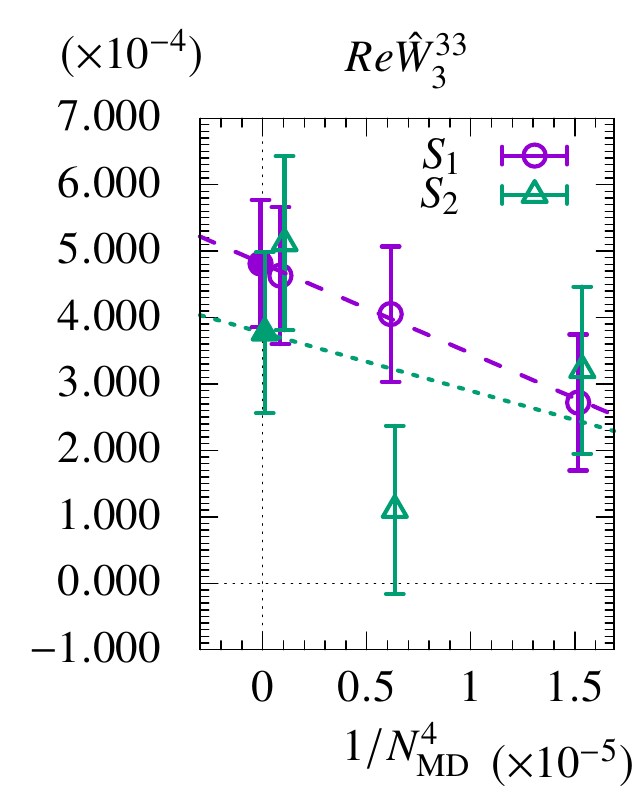}
\includegraphics[scale=\figscale,clip]{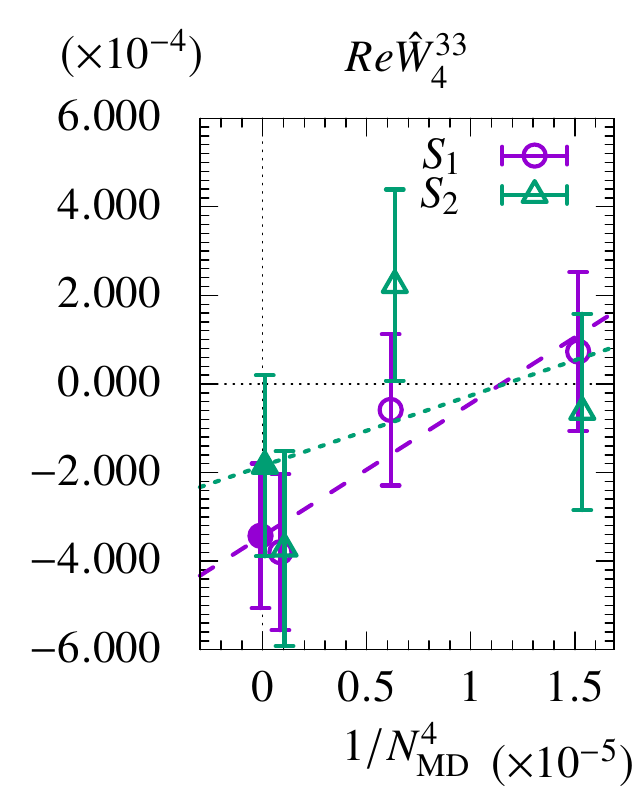}
\includegraphics[scale=\figscale,clip]{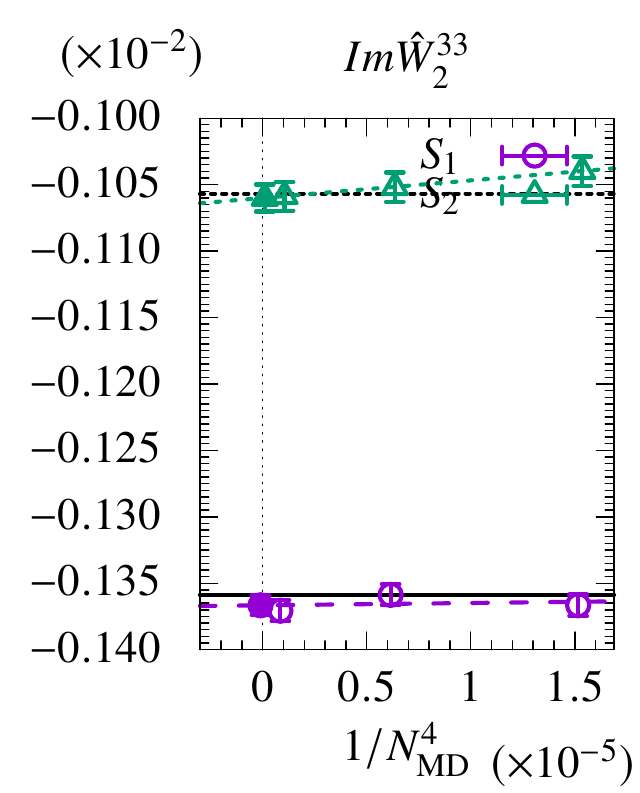}
\includegraphics[scale=\figscale,clip]{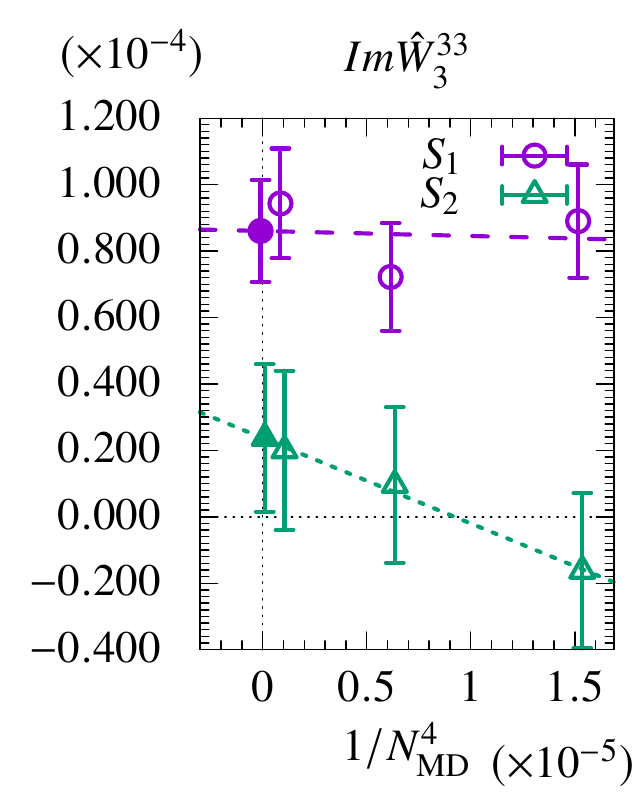}
\includegraphics[scale=\figscale,clip]{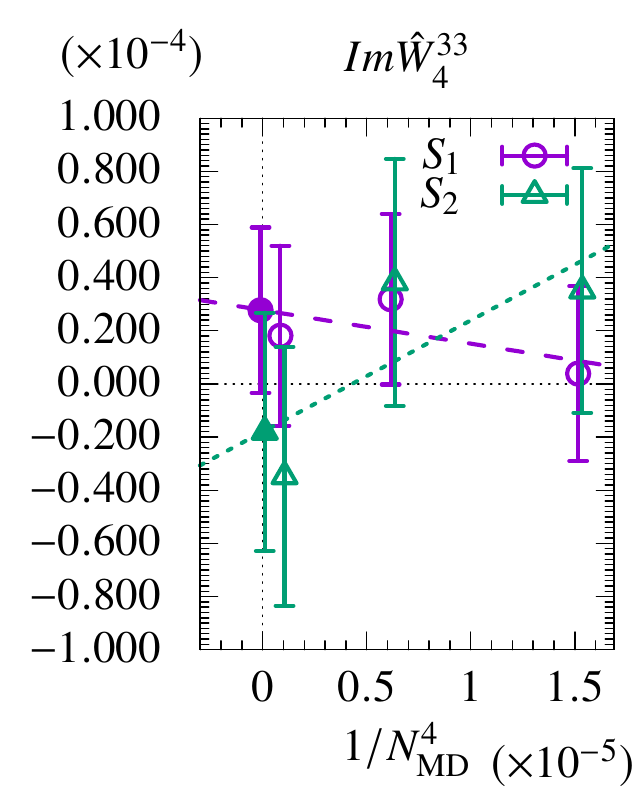}
    \caption{MD step size dependence and extrapolation for $\RE
    \hat{W}^{33}_\ell$ (upper) and $\IM \hat{W}^{33}_\ell$ (lower) at $(\Nc,k)=(16,1)$.}
    \label{fig:ReW33ImW33MDDepSU16K1}
\end{figure}

\subsection{Comparison with  analytic calculations}
As mentioned earlier the presence of the twist breaks part of the
symmetries of the standard lattice symmetries with periodic boundary
conditions. In particular, part of the cubic  group is
broken. This translates into a dependence of the coefficients of the
Wilson loop perturbative expansion on the plane in which the Wilson
loop lies. Due to the remaining symmetry for our choice of twist, there
are two sets of planes  which we label as $S_1 = \{(\mu,\nu):(1,2),(2,3),(3,4),(4,1)\}$
and $S_2 = \{(\mu,\nu):(1,3),(2,4)\}$~\cite{Perez:2017jyq}, such that the
results for all $\mu-\nu$ planes contained in each $S_i$ should be the same.
The results for all planes in $S_1$ need not be equal to those  contained in $S_2$. To increase the
statistics we can average all the planes within each set. 

Our results  for $\hat{W}^{11}_\ell$ and $\hat{W}^{33}_\ell$ at
$(\Nc,k)=(16,1)$ are shown  in  
figures~\ref{fig:ReW11ImW11MDDepSU16K1} and
\ref{fig:ReW33ImW33MDDepSU16K1}, respectively. The mean values are plotted
as a function of $1/\NMD^4$ and extrapolated linearly to vanishing
step size. The linear dependence in this variable is the expectation
for our 4th order leapfrog scheme for the MD integrator, and our data
are consistent with this expectation. The data for the $S_1$ and $S_2$
planes are plotted and extrapolated separately (purple dashed : $S_1$, green dotted : $S_2$). 
The analytic results, depicted by horizontal lines in the same plots,
only predict  differences among planes for $\hat{W}_2$.
At all orders, the cubic symmetry breaking in the real parts is found to be small and 
of the order of the errors of our calculation. 
As mentioned earlier the imaginary parts beyond the leading order 
are non-zero and different for the two families of planes. Our results
reproduce both features and match nicely with the analytic results for
the first two coefficients.

\begin{table}[t]
    \centering
{\footnotesize
\sisetup{tight-spacing=true}
    \begin{tabular}{c
      S[table-format=1.8]
      S[table-format=2.10]
      S[table-format=1.4]
      S[table-format=1.8]
      S[table-format=2.10]
      S[table-format=1.4]}
\toprule
& \multicolumn{3}{c}{{$S_1$}} 
& \multicolumn{3}{c}{{$S_2$}} \\
              & {Analytic} & {NSPT} & {$\chi^2/\mathrm{DoF}$} 
              & {Analytic} & {NSPT} & {$\chi^2/\mathrm{DoF}$} \\
\midrule
 $Re \hat{W}_{1}^{11}$  & 0.12451172 &   0.1245223 \pm  0.0000093 &          1.0  & {$=S_1$} &  0.124512 \pm  0.000012 &  0.068 \\
 $Re \hat{W}_{1}^{22}$  & 0.32812500 &   0.328138  \pm  0.000037  &         0.60  & {$=S_1$} &  0.328096 \pm  0.000045 &    4.4 \\
 $Re \hat{W}_{1}^{33}$  & 0.45263672 &   0.452662  \pm  0.000050  &         0.35  & {$=S_1$} &  0.452601 \pm  0.000057 &    1.8 \\
\midrule
 $Re \hat{W}_{2}^{11}$  &  0.00505599 &   0.0050473 \pm  0.0000096 &                 0.057  &  0.00504242  &  0.005055 \pm  0.000013 &   0.52 \\
 $Re \hat{W}_{2}^{22}$  & -0.01467313 &  -0.014691  \pm  0.000043  &                   1.0  & -0.01473699  & -0.014698 \pm  0.000056 &    4.1 \\
 $Re \hat{W}_{2}^{33}$  & -0.04599308 &  -0.045988  \pm  0.000062  &                   2.3  & -0.04592018  & -0.045817 \pm  0.000074 &    2.9 \\
 $Re \hat{W}_{2}^{44}$  &  0.15838433 &   0.158400  \pm  0.000045  & {9.4 $\times 10^{-8}$} &  0.15789605  &  0.157894 \pm  0.000045 &  0.023 \\
\midrule
 $Re \hat{W}_{3}^{11}$  &&   0.000798 \pm   0.000014 &          1.4  &&  0.000775 \pm  0.000020 &    1.8 \\
 $Re \hat{W}_{3}^{22}$  &&   0.000114 \pm   0.000069 &         0.38  &&  0.000051 \pm  0.000089 &    2.3 \\
 $Re \hat{W}_{3}^{33}$  &&   0.000481 \pm   0.000096 &        0.010  &&   0.00038 \pm   0.00012 &    4.4 \\
 $Re \hat{W}_{3}^{44}$  &&  -0.023093 \pm   0.000053 &        0.050  && -0.022910 \pm  0.000056 &    1.2 \\
\midrule
 $Re \hat{W}_{4}^{11}$  &&   0.000118 \pm   0.000025 &         2.7   &&  0.000179 \pm  0.000033 &    1.1 \\
 $Re \hat{W}_{4}^{22}$  &&  -0.00013  \pm   0.00012  &         2.5   &&   0.00010 \pm   0.00015 &    1.2 \\
 $Re \hat{W}_{4}^{33}$  &&  -0.00034  \pm   0.00016  &         0.51  &&  -0.00018 \pm   0.00020 &    3.2 \\
 $Re \hat{W}_{4}^{44}$  &&   0.000837 \pm   0.000083 &         0.72  &&  0.000775 \pm  0.000093 &   0.83 \\
\midrule
 $|Im \hat{W}_{2}^{11}|$ &  0.00000851  &  0.00000962 \pm  0.00000100 &   0.37  &  0.00001608  &  0.0000152 \pm  0.0000014 &      2.3 \\
 $|Im \hat{W}_{2}^{22}|$ &  0.00020833  &  0.0002084  \pm  0.0000036  &   0.90  &  0.00004485  &  0.0000507 \pm  0.0000051 &      1.4 \\
 $|Im \hat{W}_{2}^{33}|$ &  0.00135865  &  0.0013666  \pm  0.0000074  &   1.1   &  0.00105671  &  0.001060  \pm  0.000010  &  0.00016 \\
\midrule
 $|Im \hat{W}_{3}^{11}|$ &&   0.0000004 \pm   0.0000021 &    0.23  &&  0.0000046 \pm  0.0000031 &     0.13 \\
 $|Im \hat{W}_{3}^{22}|$ &&   0.0000023 \pm   0.0000071 &    1.7   &&   0.000016 \pm   0.000010 &    0.075 \\
 $|Im \hat{W}_{3}^{33}|$ &&   0.000086  \pm   0.000015  &    0.99  &&   0.000024 \pm   0.000022 &    0.011 \\
 $|Im \hat{W}_{3}^{44}|$ &&   0.0000007 \pm   0.0000012 &    0.18  &&  0.0000012 \pm  0.0000017 &    0.035 \\
\midrule
 $|Im \hat{W}_{4}^{11}|$ &&   0.0000007 \pm   0.0000042 &    1.5   &&  0.0000030 \pm  0.0000059 &     0.94 \\
 $|Im \hat{W}_{4}^{22}|$ &&   0.000004  \pm   0.000013  &    0.24  &&   0.000017 \pm   0.000019 &  0.00084 \\
 $|Im \hat{W}_{4}^{33}|$ &&   0.000028  \pm   0.000031  &    0.22  &&   0.000018 \pm   0.000045 &     0.65 \\
 $|Im \hat{W}_{4}^{44}|$ &&   0.0000052 \pm   0.0000036 &    0.20  &&  0.0000028 \pm  0.0000051 &      2.6 \\
\bottomrule        
    \end{tabular}
}
    \caption{Perturbative coefficients for Wilson loops on $S_1,S_2$ planes. (SU(16),$k=1$)}
    \label{tab:wloopsS1S2SU16K1}
\end{table}

\begin{table}[t]
    \centering
{\footnotesize
\sisetup{tight-spacing=true}
    \begin{tabular}{c
      S[table-format=1.8]
      S[table-format=2.10]
      S[table-format=1.4]
      S[table-format=1.8]
      S[table-format=2.10]
      S[table-format=1.4]}
\toprule
& \multicolumn{3}{c}{{$S_1$}} 
& \multicolumn{3}{c}{{$S_2$}} \\
              & {Analytic} & {NSPT} & {$\chi^2/\mathrm{DoF}$} 
              & {Analytic} & {NSPT} & {$\chi^2/\mathrm{DoF}$} \\
\midrule
 $Re \hat{W}_{1}^{11}$ & 0.12494794  &   0.1249542 \pm  0.0000053 &  0.030  & {$=S_1$} &  0.1249579 \pm  0.0000065 &   0.038 \\
 $Re \hat{W}_{1}^{22}$ & 0.34134568  &   0.341372  \pm  0.000021  &  2.7    & {$=S_1$} &  0.341395  \pm   0.000025 &    0.78 \\
 $Re \hat{W}_{1}^{33}$ & 0.57010177  &   0.570171  \pm  0.000046  &  0.34   & {$=S_1$} &  0.570128  \pm   0.000054 &     1.6 \\
 $Re \hat{W}_{1}^{44}$ & 0.78949299  &   0.789596  \pm  0.000071  &  0.44   & {$=S_1$} &  0.789543  \pm   0.000082 &     1.1 \\
\midrule
 $Re \hat{W}_{2}^{11}$ &  0.00510103 &   0.0051037 \pm  0.0000061 &  7.6    &  0.00510026  &   0.0051012 \pm  0.0000077 &    0.22 \\
 $Re \hat{W}_{2}^{22}$ & -0.01667806 &  -0.016682  \pm  0.000026  &  0.013  & -0.01668076  &  -0.016677  \pm   0.000033 &     1.0 \\
 $Re \hat{W}_{2}^{33}$ & -0.08822803 &  -0.088249  \pm  0.000061  &  0.0043 & -0.08823509  &  -0.088183  \pm   0.000074 &   0.018 \\
 $Re \hat{W}_{2}^{44}$ & -0.20618250 &  -0.20618   \pm  0.00010   &  0.12   & -0.20624012  &  -0.20613   \pm    0.00012 &   0.034 \\
\midrule
 $Re \hat{W}_{3}^{11}$ &&   0.0007741 \pm  0.0000096 &  1.1    &&   0.000784 \pm   0.000013 &     1.3 \\
 $Re \hat{W}_{3}^{22}$ &&  -0.000029  \pm  0.000043  &  0.021  &&  -0.000107 \pm   0.000054 &     2.6 \\
 $Re \hat{W}_{3}^{33}$ &&   0.002795  \pm  0.000095  &  0.068  &&   0.00264  \pm    0.00012 &   0.064 \\
 $Re \hat{W}_{3}^{44}$ &&   0.01948   \pm  0.00016   &  0.0025 &&   0.01924  \pm    0.00019 &  0.0023 \\
\midrule
 $Re \hat{W}_{4}^{11}$ &&    0.000177 \pm   0.000018 &   0.48  &&   0.000171 \pm   0.000024 &    0.64 \\
 $Re \hat{W}_{4}^{22}$ &&    0.000128 \pm   0.000077 &   0.044 &&   0.000254 \pm   0.000097 &    0.22 \\
 $Re \hat{W}_{4}^{33}$ &&    0.00021  \pm   0.00017  &   0.019 &&   0.00055  \pm    0.00021 &    0.43 \\
 $Re \hat{W}_{4}^{44}$ &&   -0.00017  \pm   0.00027  &   0.045 &&   0.00038  \pm    0.00034 &    0.28 \\
\midrule
 $|Im \hat{W}_{2}^{11}|$ & 0.00000029 &   0.00000055 \pm  0.00000059 &  0.31 & 0.00000102  &  0.00000001 \pm  0.00000085 &    1.7 \\
 $|Im \hat{W}_{2}^{22}|$ & 0.00001424 &   0.0000135  \pm  0.0000027  &  0.32 & 0.00000199  &  0.0000041  \pm   0.0000038 &   0.35 \\
 $|Im \hat{W}_{2}^{33}|$ & 0.00001824 &   0.0000164  \pm  0.0000063  &  1.3  & 0.00006497  &  0.0000544  \pm   0.0000089 &  0.097 \\
 $|Im \hat{W}_{2}^{44}|$ & 0.00038314 &   0.000386   \pm  0.000011   &  1.2  & 0.00044756  &  0.000438   \pm    0.000017 &   0.14 \\
\midrule
 $|Im \hat{W}_{3}^{11}|$ &&    0.0000001 \pm   0.0000013 &  0.0017 &&  0.0000022 \pm   0.0000020 &    1.5 \\
 $|Im \hat{W}_{3}^{22}|$ &&    0.0000059 \pm   0.0000056 &  3.9    &&  0.0000044 \pm   0.0000076 &   0.75 \\
 $|Im \hat{W}_{3}^{33}|$ &&    0.000018  \pm   0.000014  &  1.4    &&  0.000031  \pm    0.000018 &   0.67 \\
 $|Im \hat{W}_{3}^{44}|$ &&    0.000102  \pm   0.000025  &  0.61   &&  0.000088  \pm    0.000037 &  0.053 \\
\midrule
 $|Im \hat{W}_{4}^{11}|$ &&    0.0000008 \pm   0.0000026 &   0.044 &&  0.0000035 \pm   0.0000039 &   0.22 \\
 $|Im \hat{W}_{4}^{22}|$ &&    0.000003  \pm   0.000011  &   2.2   &&  0.000008  \pm    0.000015 &   0.39 \\
 $|Im \hat{W}_{4}^{33}|$ &&    0.000056  \pm   0.000029  &   0.25  &&  0.000068  \pm    0.000039 &    1.8 \\
 $|Im \hat{W}_{4}^{44}|$ &&    0.000108  \pm   0.000053  &   0.86  &&  0.000021  \pm    0.000079 &   0.15 \\
\bottomrule        
    \end{tabular}
}
    \caption{Perturbative coefficients for Wilson loops on $S_1,S_2$ planes. (SU(49),$k=2$)}
    \label{tab:wloopsS1S2SU49K2}
\end{table}

All our results for $(\Nc,k)=(16,1)$ and  $(49,2)$, extrapolated
linearly in $1/\NMD^4$ to zero, are collected in Tables
\ref{tab:wloopsS1S2SU16K1} and \ref{tab:wloopsS1S2SU49K2}.
We also tabulate the analytic values from ref.~\cite{Perez:2017jyq}
and $\chi^2/\mathrm{DoF}$ from the extrapolating fit.
The violation of CP and cubic invariance due to the twist is well seen
for the $\Nc=16$ case at the two loop level
and the differences between $S_1$ and $S_2$ averages are consistent
with the analytic values. 
For the  $(\Nc,k)=(49,2)$ case, the results of our  NSPT analysis are
also  consistent with analytic values. However,
the violation of CP and cubic invariance is too small to be seen in
comparison with the statistical errors.  
Since the violation of CP and cubic invariance disappears in the large
$\Nc$ limit, we conclude that in practice one would not be able see
any effect of this symmetry breaking for even larger values of $\Nc$.
Hence, hereafter, we show the coefficients averaged over all
$\mu$--$\nu$ planes ($S=S_1+S_2$) in order to study
the $k$ dependence and $\Nc$ dependence. We emphasize the small errors
of our coefficient determinations, typically of order $10^{-5}$. This
gives about 3 to 5 significant digits in some determinations.

\renewcommand{\figscale}{0.58}
\graphicspath{{FIGS/}}
\begin{figure}[t]
    \centering
\includegraphics[scale=\figscale,clip]{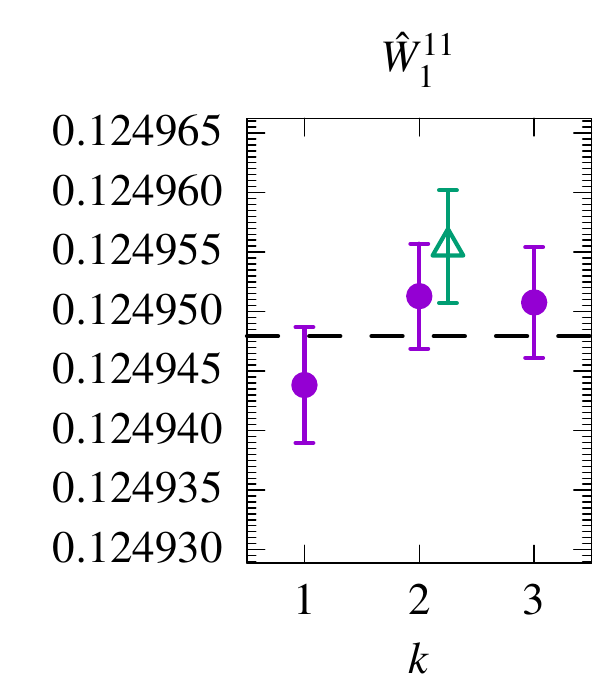}
\includegraphics[scale=\figscale,clip]{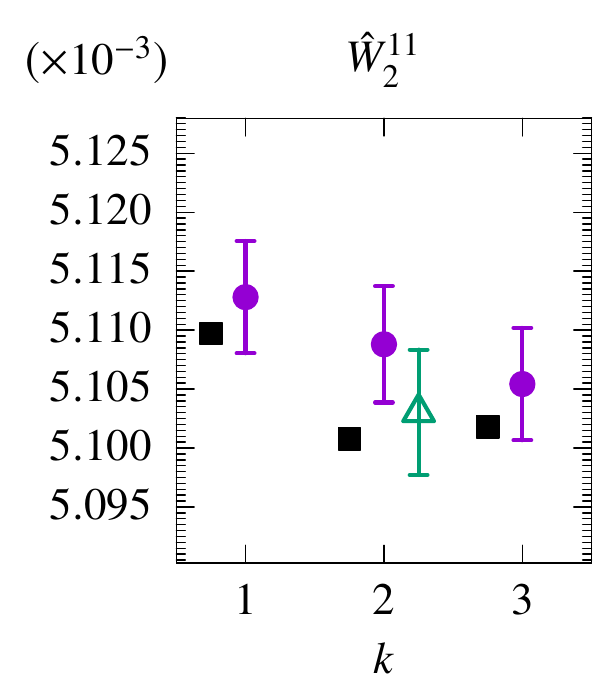}
\includegraphics[scale=\figscale,clip]{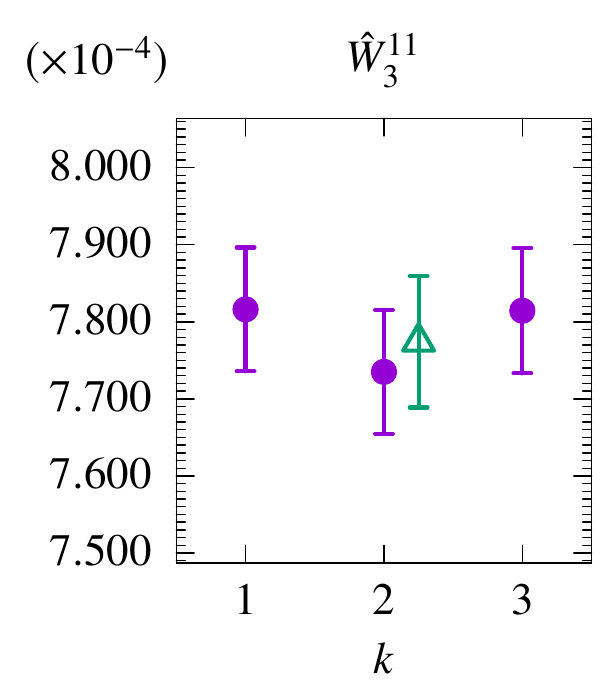}
\includegraphics[scale=\figscale,clip]{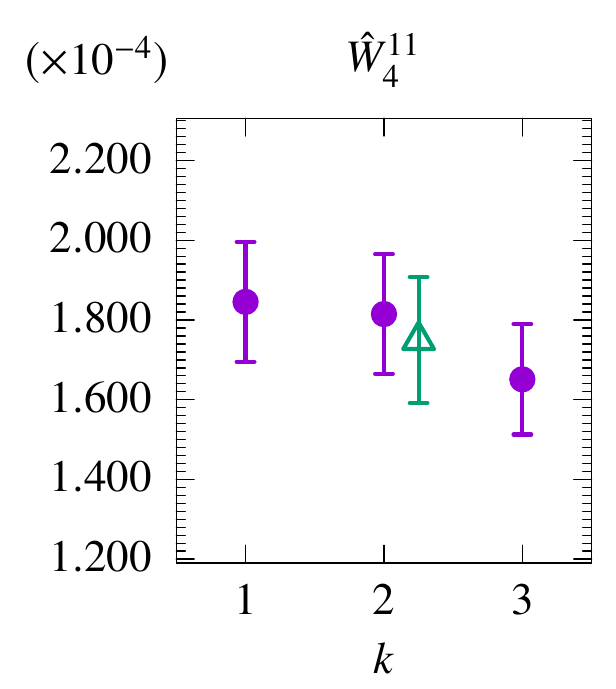}
\includegraphics[scale=\figscale,clip]{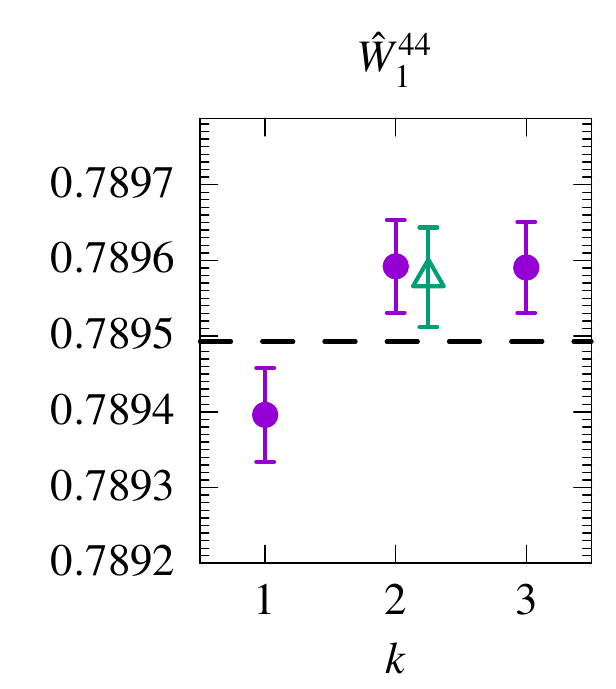}
\includegraphics[scale=\figscale,clip]{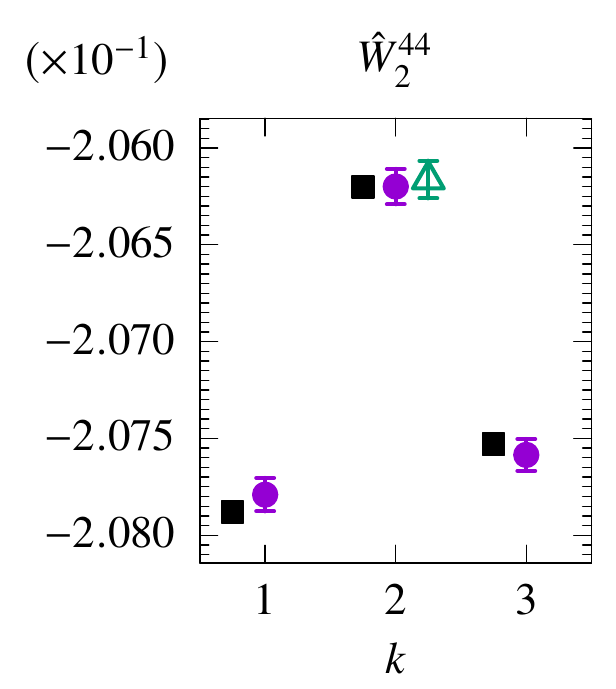}
\includegraphics[scale=\figscale,clip]{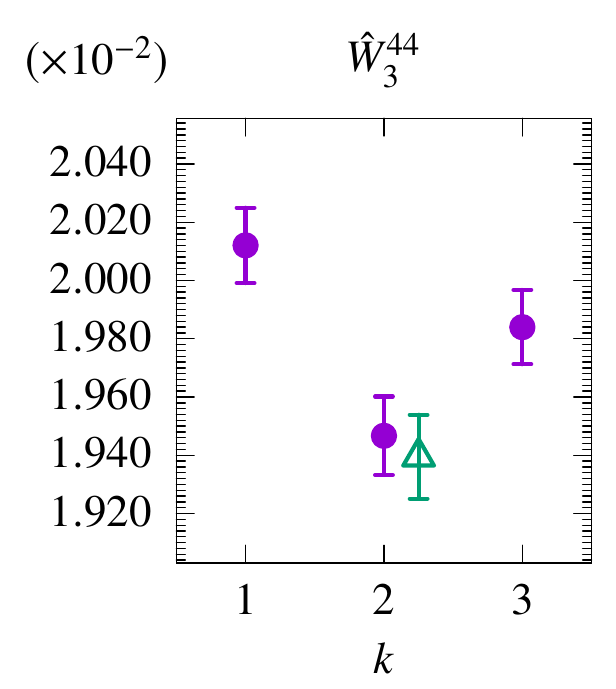}
\includegraphics[scale=\figscale,clip]{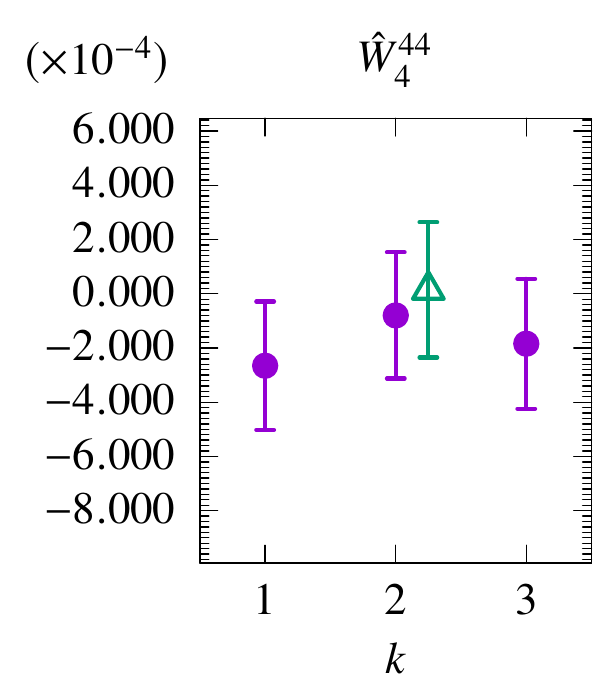}
    \caption{$k$ dependence for $\hat{W}_{\ell}^{11}$ and $\hat{W}_{\ell}^{44}$ at $\Nc=49$. 
    Open triangles show the results extrapolated to $1/\NMD\to 0 $ and
    the other values are at $\NMD=32$.
    Dashed line in $\hat{W}^{RR}_1$ and filled squares in $\hat{W}^{RR}_2$ are the analytic values~\cite{Perez:2017jyq,privcomm}.
    }
    \label{fig:KDEPW11W44SU49}
\end{figure}

We next investigate the dependence on the flux parameter $k$  at $\Nc=49$.
For that purpose we studied the $k=1$ and $k=3$ results at fixed value
of $\NMD=32$. The results for  the coefficients for the plaquette and
$4\times 4$ Wilson loop are displayed in figure~\ref{fig:KDEPW11W44SU49}.
The results for $k=2$ are given both at $\NMD=32$ as well as
extrapolated $1/\NMD\to 0$ (open triangles). From the data we conclude that  $\NMD=32$ is 
sufficient for the current analysis as the statistical error and the systematic 
error from the finite MD time step are comparable. The horizontal dash
lines in $\hat{W}^{RR}_1$ 
and the filled squares in $\hat{W}^{RR}_2$ correspond to the analytic values.
For the plaquette, the observed $k$-dependence is compatible with
statistical errors. As expected, the differences are small but well
beyond statistical errors  for the $4\times 4$ loop. This is also the case for the
third order coefficient $\hat{W}^{RR}_3$ for which we have no analytic
results. Anyhow, no dramatic $k$ dependence is seen for the third and
fourth order coefficients.

Our  results averaged over all planes are displayed in
table~\ref{tab:WLOOPSALLFINAL}. The choice of values
$(\Nc,k)=(16,1),(49,2),(121,3)$ are selected so as to keep $k/\Lbox$
at $\sim 0.27$  ($k/\Lbox = 1/4, 2/7, 3/11$ for each data set). The
results at the one and two-loop level are consistent with the analytic
values within  two standard deviations. We also
display (NPT fit) an estimate of the  three-loop coefficients for $\Nc=16$ and $49$ 
obtained by fitting nonperturbative expectation values to a third order polynomial in
$\lambda$ with the first  coefficients fixed to the analytic values~\cite{Perez:2017jyq}.
The results are roughly consistent with our NSPT determinations, which
are expected  to be more reliable.

\begin{table}[t]
    \centering
{\footnotesize
    \begin{tabular}{c
      S[table-format=2.10]
      S[table-format=2.10]
      S[table-format=2.9]
      S[table-format=2.9]}
\toprule
 {SU(16), $k=1$} & {$\ell=1$} & {$\ell=2$} & {$\ell=3$} & {$\ell=4$} \\
\midrule
$\hat{W}^{11}_{\ell}$ & 0.1245188 \pm  0.0000086 &  0.0050497 \pm  0.0000084 &  0.000791 \pm  0.000012 &  0.000138 \pm  0.000021 \\
$\hat{W}^{22}_{\ell}$ & 0.328123  \pm  0.000034  & -0.014695  \pm  0.000038  &  0.000094 \pm  0.000061 & -0.00005  \pm  0.00011  \\
$\hat{W}^{33}_{\ell}$ & 0.452642  \pm  0.000047  & -0.045930  \pm  0.000056  &  0.000448 \pm  0.000088 & -0.00029  \pm  0.00015  \\
$\hat{W}^{44}_{\ell}$ & 0.0                      &  0.158232  \pm  0.000044  & -0.023032 \pm  0.000051 &  0.000816 \pm  0.000079 \\
\midrule
 & {PT Analytic} & {PT Analytic} & {NPT Fit}  & \\
\midrule
$\hat{W}^{11}_{\ell}$ & 0.12451172 &   0.00505146  &  0.000826(36) &  \\
$\hat{W}^{22}_{\ell}$ & 0.32812500 &  -0.01469441  &  0.00004 (12) &  \\
$\hat{W}^{33}_{\ell}$ & 0.45263672 &  -0.04596878  &  0.00034 (16) &  \\
$\hat{W}^{44}_{\ell}$ & 0.0        &   0.15822157  & -0.02288 (6)  &  \\
\midrule        
{SU(49), $k=2$} & {$\ell=1$} & {$\ell=2$} & {$\ell=3$} & {$\ell=4$}  \\
\midrule
$\hat{W}^{11}_{\ell}$ &  0.1249555 \pm  0.0000047 &  0.0051030 \pm  0.0000053 &  0.0007774 \pm  0.0000085 &  0.000175 \pm  0.000016  \\
$\hat{W}^{22}_{\ell}$ &  0.341380  \pm  0.000019  & -0.016680  \pm  0.000024  & -0.000055  \pm  0.000040  &  0.000171 \pm  0.000071  \\
$\hat{W}^{33}_{\ell}$ &  0.570156  \pm  0.000042  & -0.088227  \pm  0.000055  &  0.002743  \pm  0.000084  &  0.00033  \pm  0.00015   \\
$\hat{W}^{44}_{\ell}$ &  0.789578  \pm  0.000066  & -0.206163  \pm  0.000097  &  0.01940   \pm  0.00014   &  0.00001  \pm  0.00025   \\
\midrule
 & {PT Analytic} & {PT Analytic} & {NPT Fit} &   \\
\midrule
$\hat{W}^{11}_{\ell}$ & 0.12494794 &   0.00510077 & 0.000883(9)   \\
$\hat{W}^{22}_{\ell}$ & 0.34134568 &  -0.01667896 & 0.000086(36)  \\
$\hat{W}^{33}_{\ell}$ & 0.57010177 &  -0.08823039 & 0.00293(5)    \\
$\hat{W}^{44}_{\ell}$ & 0.78949299 &  -0.20620171 & 0.01959(7)    \\
\midrule        
{SU(121), $k=3$} & {$\ell=1$} & {$\ell=2$} & {$\ell=3$} & {$\ell=4$}  \\
\midrule
$\hat{W}^{11}_{\ell}$ &  0.1249887 \pm  0.0000037 &   0.0051036 \pm  0.0000041 &   0.0007884 \pm  0.0000069 &   0.000156 \pm  0.000012  \\
$\hat{W}^{22}_{\ell}$ &  0.342179  \pm  0.000015  &  -0.016760  \pm  0.000019  &  -0.000030  \pm  0.000032  &   0.000020 \pm  0.000055  \\
$\hat{W}^{33}_{\ell}$ &  0.575489  \pm  0.000034  &  -0.090700  \pm  0.000043  &   0.002995  \pm  0.000068  &   0.00003  \pm  0.00012   \\
$\hat{W}^{44}_{\ell}$ &  0.812537  \pm  0.000056  &  -0.220758  \pm  0.000080  &   0.02206   \pm  0.00012   &  -0.00039  \pm  0.00021   \\
$\hat{W}^{55}_{\ell}$ &  1.049316  \pm  0.000088  &  -0.40728   \pm  0.00014   &   0.07062   \pm  0.00021   &  -0.00459  \pm  0.00038   \\
$\hat{W}^{66}_{\ell}$ &  1.28172   \pm  0.00012   &  -0.64507   \pm  0.00021   &   0.16012   \pm  0.00033   &  -0.01933  \pm  0.00057   \\
\midrule
 & {PT Analytic} & {PT Analytic~\cite{privcomm}} &  &   \\
\midrule
$\hat{W}^{11}_{\ell}$ &  0.12499146  &   0.00510592  &   \\
$\hat{W}^{22}_{\ell}$ &  0.34218332  &  -0.01679772  &   \\
$\hat{W}^{33}_{\ell}$ &  0.57549981  &  -0.09073426  &   \\
$\hat{W}^{44}_{\ell}$ &  0.81255000  &  -0.22085102  &   \\
$\hat{W}^{55}_{\ell}$ &  1.04933625  &  -0.40740949  &   \\
$\hat{W}^{66}_{\ell}$ &  1.28175142  &  -0.64531188  &   \\
\bottomrule        
    \end{tabular}
}
    \caption{Perturbative coefficients for Wilson loops ($1/\NMD\to 0$ extrapolated).
     Data in PT Analytic are obtained analytically and 
     data in NPT Fit are obtained by fitting the non-perturbative data with a polynomial 
             of $\lambda$~\cite{Perez:2017jyq}.}
    \label{tab:WLOOPSALLFINAL}
\end{table}

Since the  main interest of the TEK model is its large $\Nc$ limit,
which coincides with Yang-Mills  at infinite $\Nc$ and infinite
volume, we give the infinite $\Nc$ extrapolation of our results. As mentioned
previously finite $\Nc$ corrections are expected to grow for
large values of $R/\Lbox$. Thus, we exclude the data of $(\Nc,k)=(16,1)$ and base our
analysis on $(\Nc,k)=(121,3)$ and $(49,2)$ only and assuming a
$1/\Nc^2$ dependence of the coefficients. The results are displayed  
 in figure~\ref{fig:NcINFW11W44} and the numerical values of the
 extrapolated coefficients  are tabulated in table~\ref{tab:WLOOPSSUNcInf}.
In the figure we only show  the plaquette and $4\times 4$ Wilson loop
up to three-loop order. The error includes only the statistical one.
The values extrapolated to $\Nc \to \infty$  are shown as filled
circles, while the horizontal solid lines are the corresponding analytic 
values~\cite{DiGiacomo:1981lcx,Alles:1993dn,Alles:1998is,Perez:2017jyq,privcomm}.
We see a reasonable good agreement within errors. The deviations for
the $4\times 4$ loop might include also a systematic error due to the
relatively large values of $4/\Lbox$. For a more precise control of
the systematic errors in the extrapolated results it would be necessary to 
have at least 3 sufficiently large values of $\Nc$, which is left out
of this exploratory work.

\begin{table}[t]
    \centering
{
    \begin{tabular}{c
        S[table-format=2.12]
        S[table-format=2.12]
        S[table-format=2.12]}
\toprule
{$\Nc\to \infty$} & {$\ell=1$} & {$\ell=2$} & {$\ell=3$}   \\
\midrule
$\hat{W}^{11}_{\ell}$ &  0.1249999  \pm  0.0000054  &   0.0051134  \pm  0.0000059  &   0.0007905  \pm  0.0000099  \\
$\hat{W}^{22}_{\ell}$ &  0.342336   \pm  0.000021   &  -0.016775   \pm  0.000027   &  -0.000025   \pm  0.000046   \\
$\hat{W}^{33}_{\ell}$ &  0.576534   \pm  0.000049   &  -0.091185   \pm  0.000063   &   0.003044   \pm  0.000098   \\
$\hat{W}^{44}_{\ell}$ &  0.817004   \pm  0.000080   &  -0.22362    \pm  0.00011    &   0.02259    \pm  0.00017    \\
\midrule
 & {PT Analytic} & {PT Analytic} & {PT Analytic}    \\
\midrule
$\hat{W}^{11}_{\ell}$ &  0.12500000                 &   0.0051069297 & 0.000794223 \pm 0.000000019 \\
$\hat{W}^{22}_{\ell}$ &  0.34232788                 &  -0.016814     &  \\
$\hat{W}^{33}_{\ell}$ &  0.57629827                 &  -0.09107      &  \\
$\hat{W}^{44}_{\ell}$ &  0.81537098                 &  -0.22287      &  \\
\bottomrule        
    \end{tabular}
}
    \caption{Perturbative coefficients for $\hat{W}^{RR}_\ell$ in $\Nc \to \infty$. 
             Values in PT Analytic for $\hat{W}^{11}_{\ell}$ are $\Nc \to \infty$ limit of results
             from lattice $\SUNc$ gauge theory with Wilson gauge action~\cite{Alles:1998is}.}
    \label{tab:WLOOPSSUNcInf}
\end{table}

\renewcommand{\figscale}{0.70}
\graphicspath{{FIGS/}}

\begin{figure}[t]
    \centering
\includegraphics[scale=\figscale,clip]{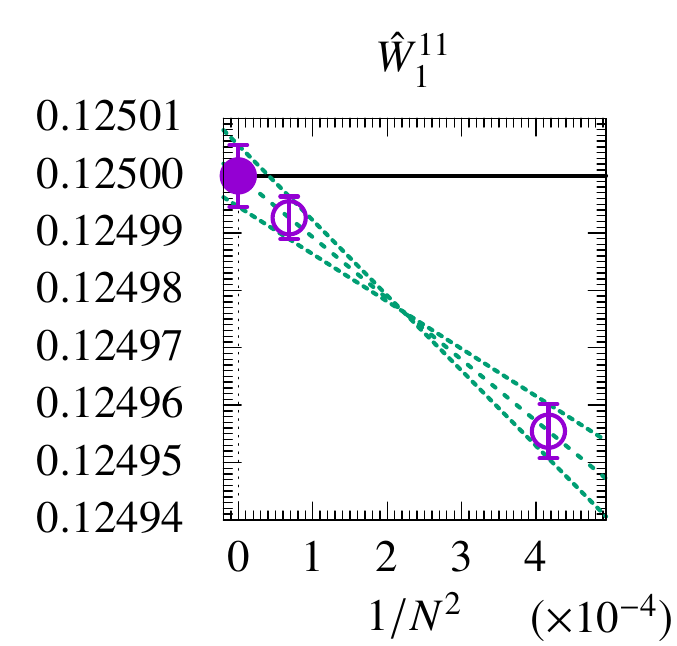}
\includegraphics[scale=\figscale,clip]{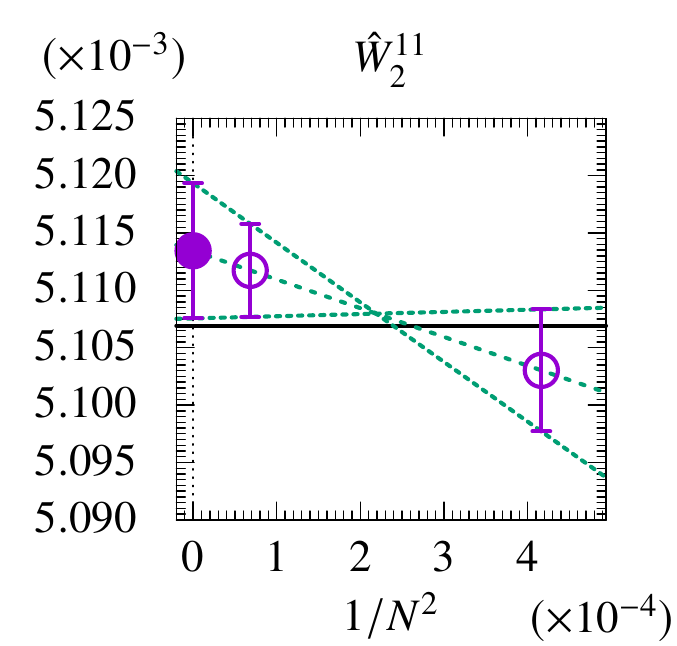}
\includegraphics[scale=\figscale,clip]{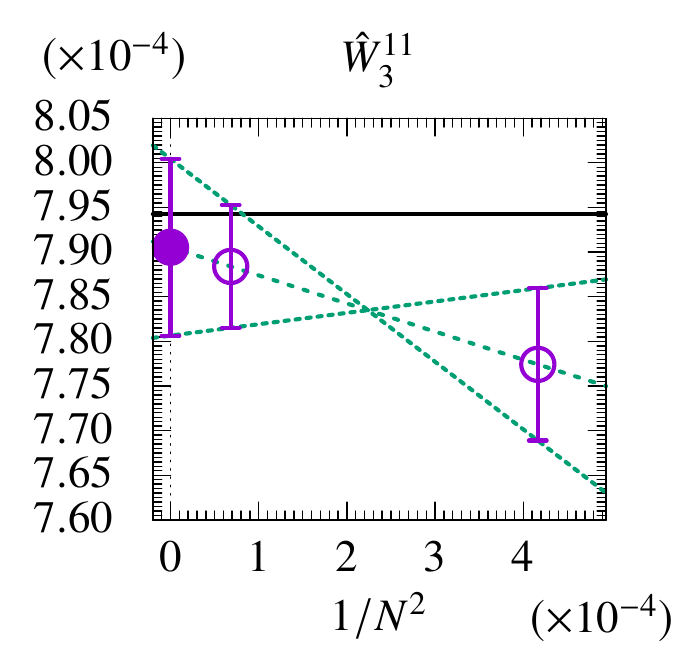}
\includegraphics[scale=\figscale,clip]{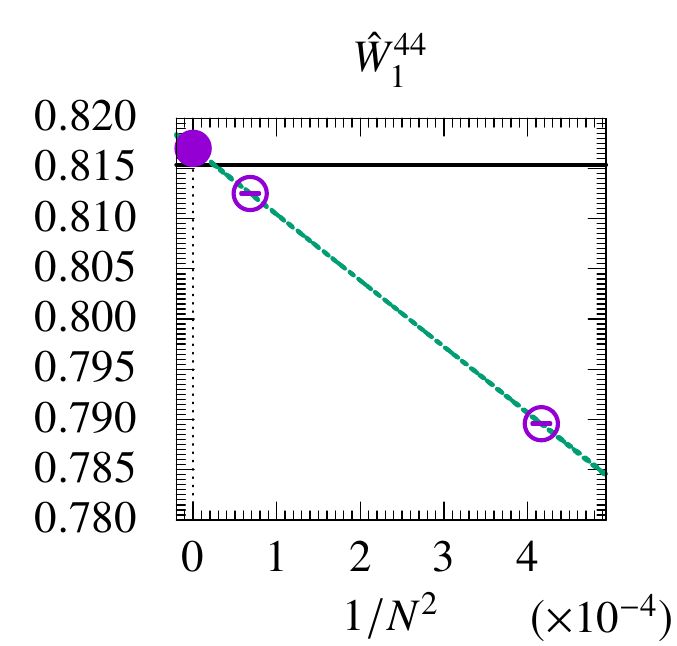}
\includegraphics[scale=\figscale,clip]{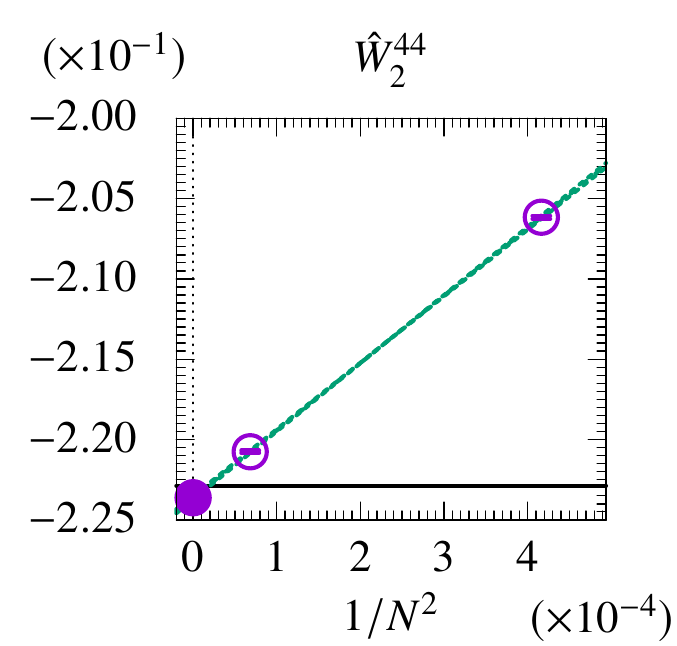}
\includegraphics[scale=\figscale,clip]{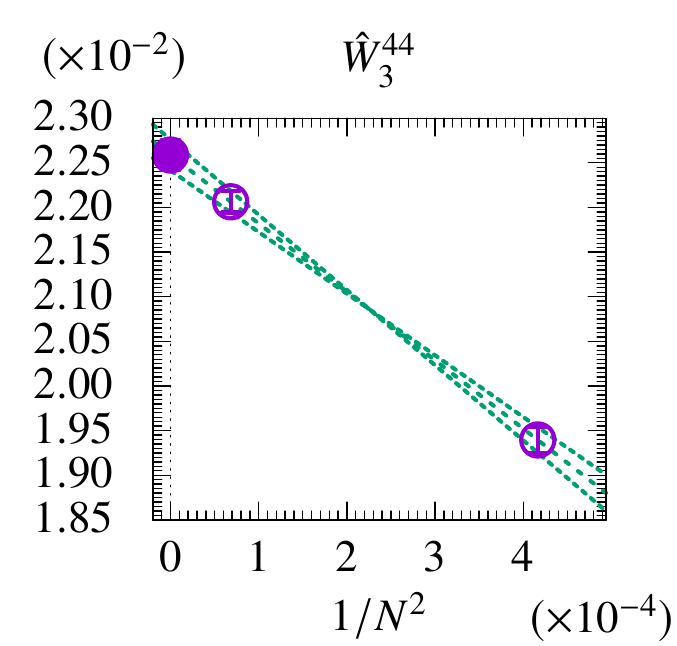}
    \caption{$\Nc \to \infty$ limit of perturbative coefficients 
    using data from $(N,k)=(49,2)$ and $(121,3)$ (upper : plaquette, lower : $4\times 4$ Wilson loop).}
    \label{fig:NcINFW11W44}
\end{figure}

\renewcommand{\figscale}{0.421}
\graphicspath{{FIGS/}}

\begin{figure}[t]
    \centering
\includegraphics[scale=\figscale,clip]{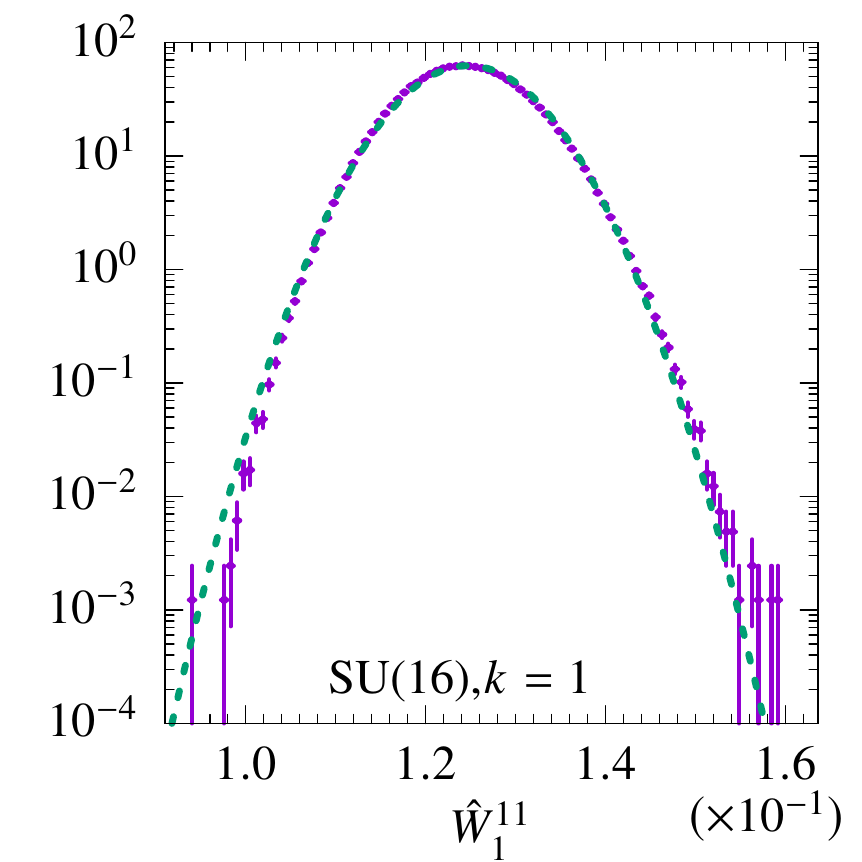}
\includegraphics[scale=\figscale,clip]{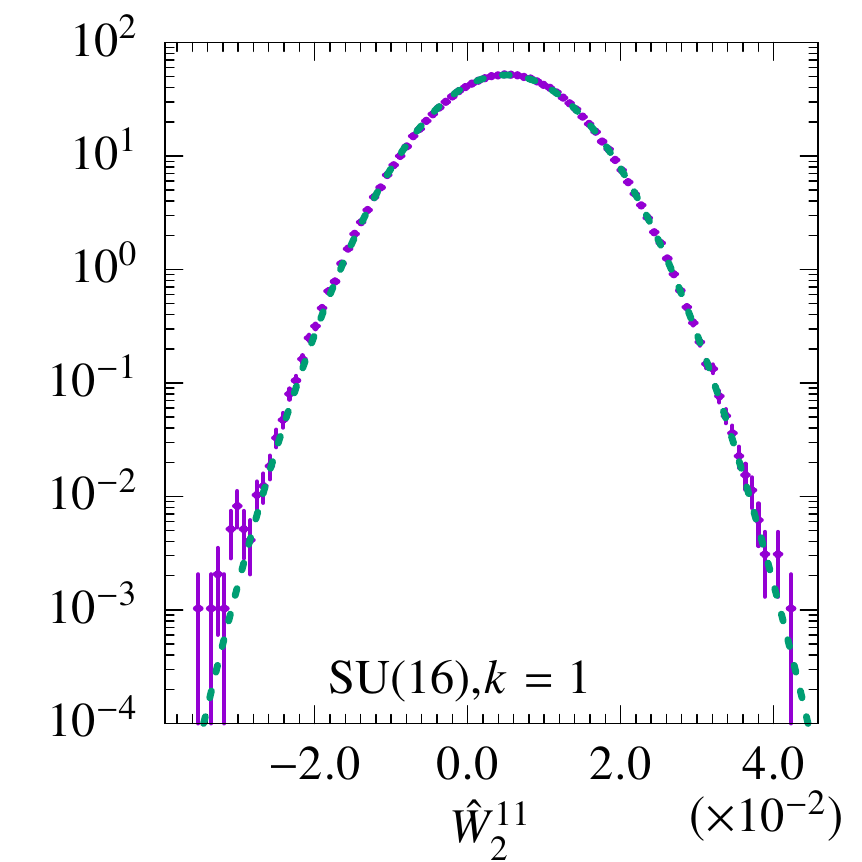}
\includegraphics[scale=\figscale,clip]{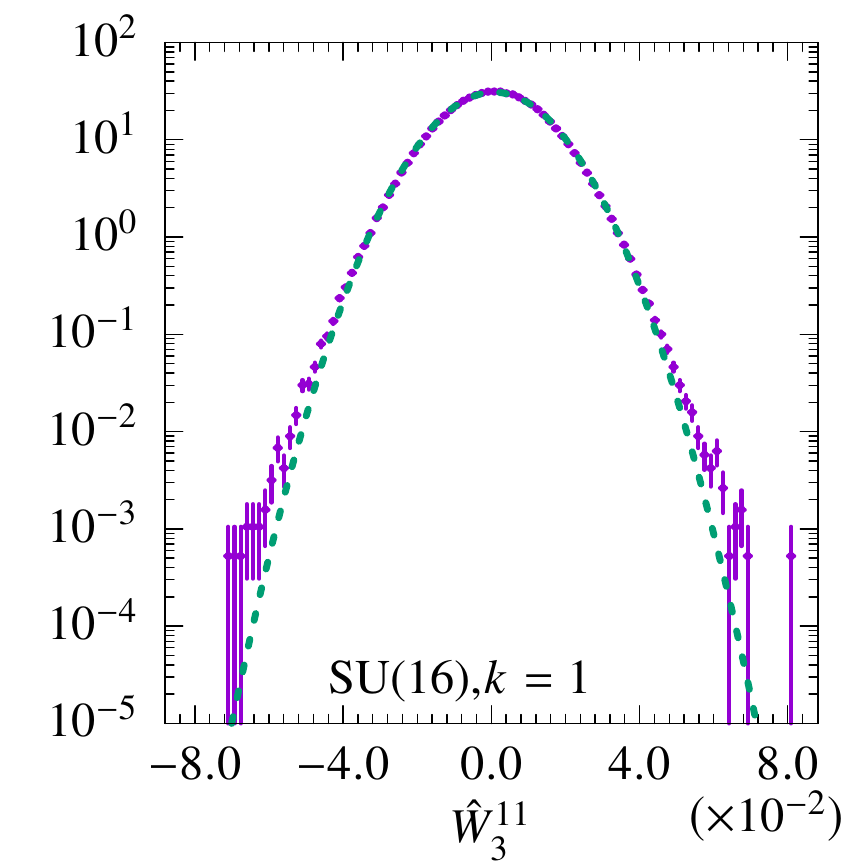}
\includegraphics[scale=\figscale,clip]{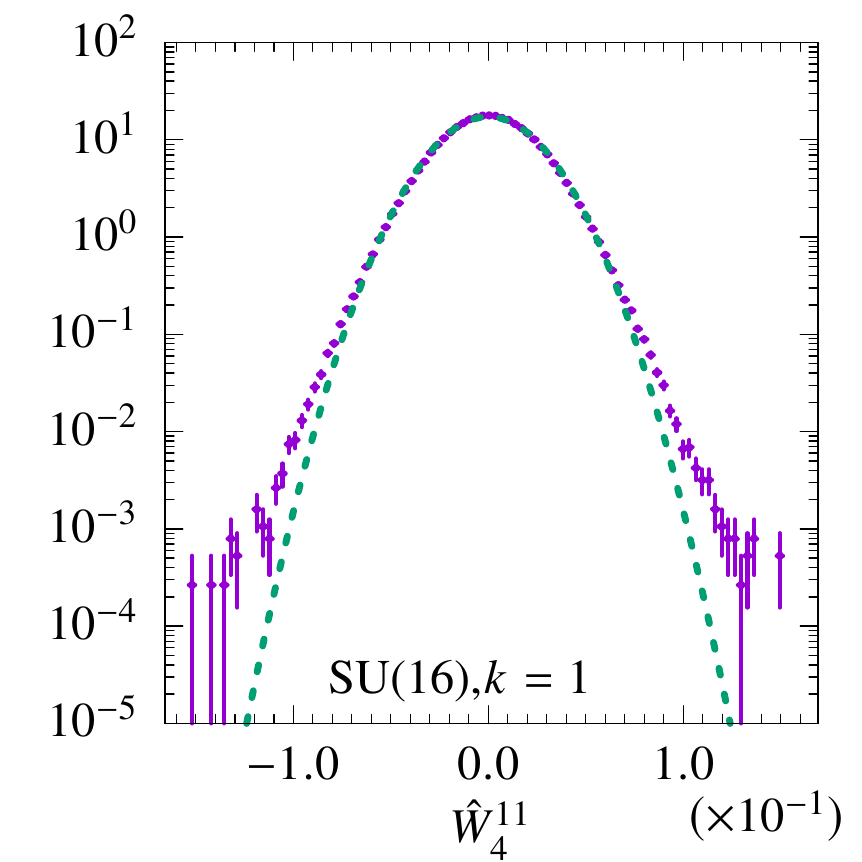}

\includegraphics[scale=\figscale,clip]{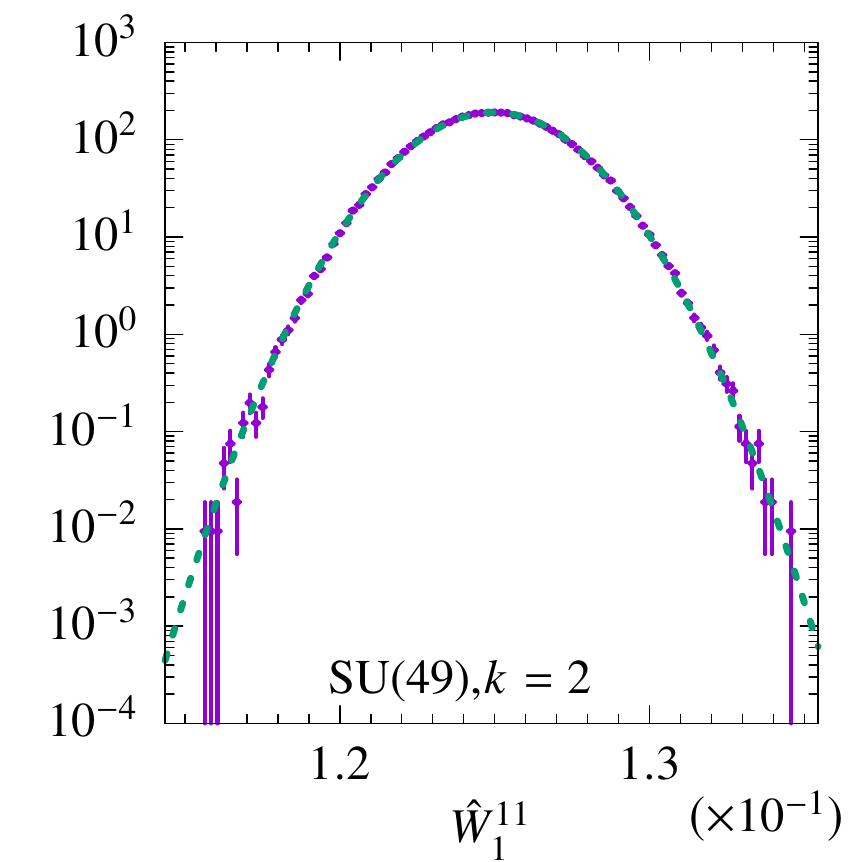}
\includegraphics[scale=\figscale,clip]{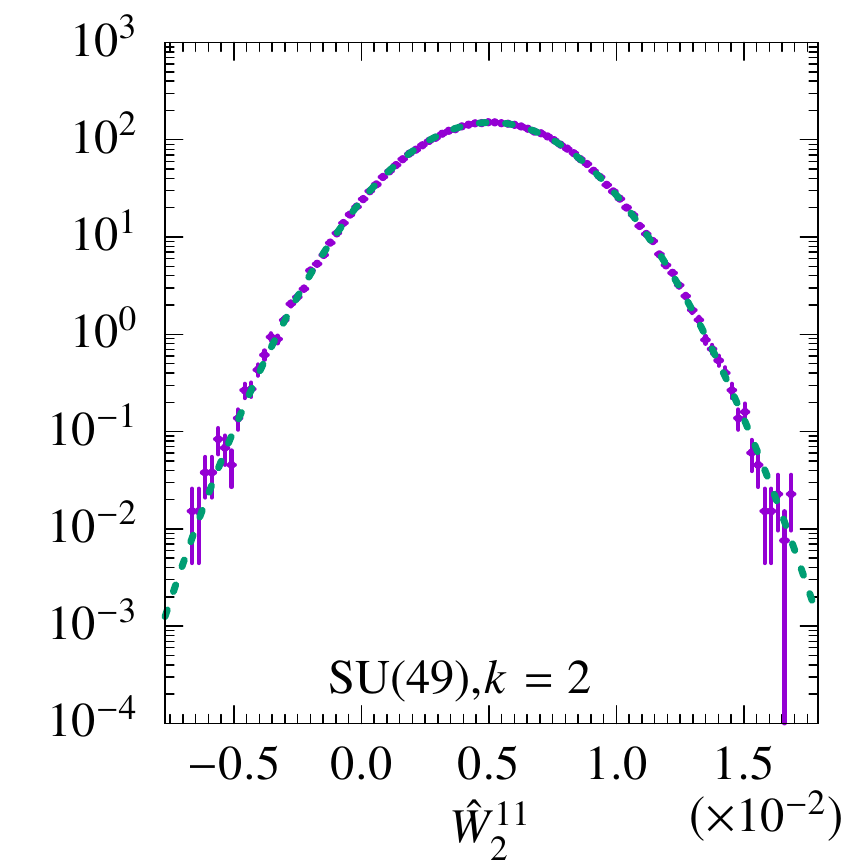}
\includegraphics[scale=\figscale,clip]{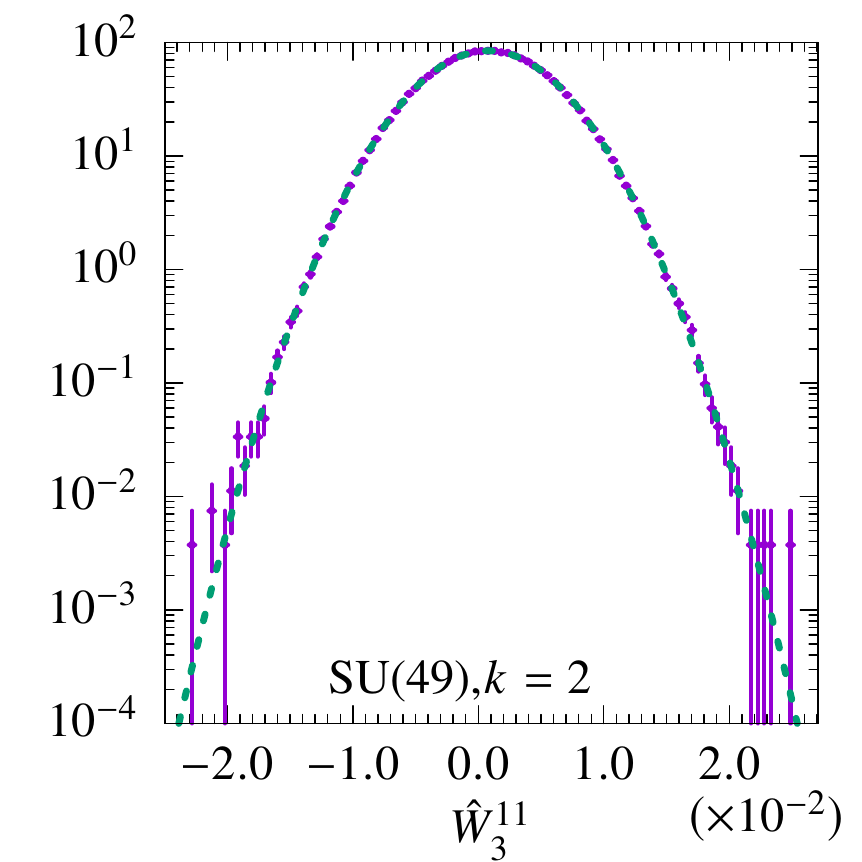}
\includegraphics[scale=\figscale,clip]{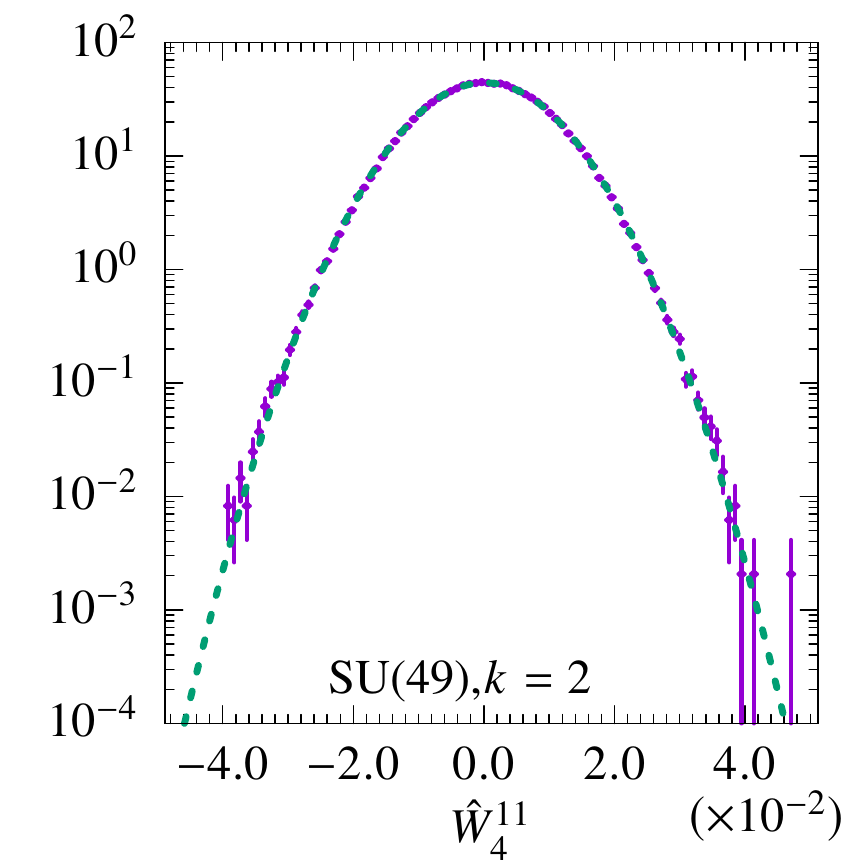}

\includegraphics[scale=\figscale,clip]{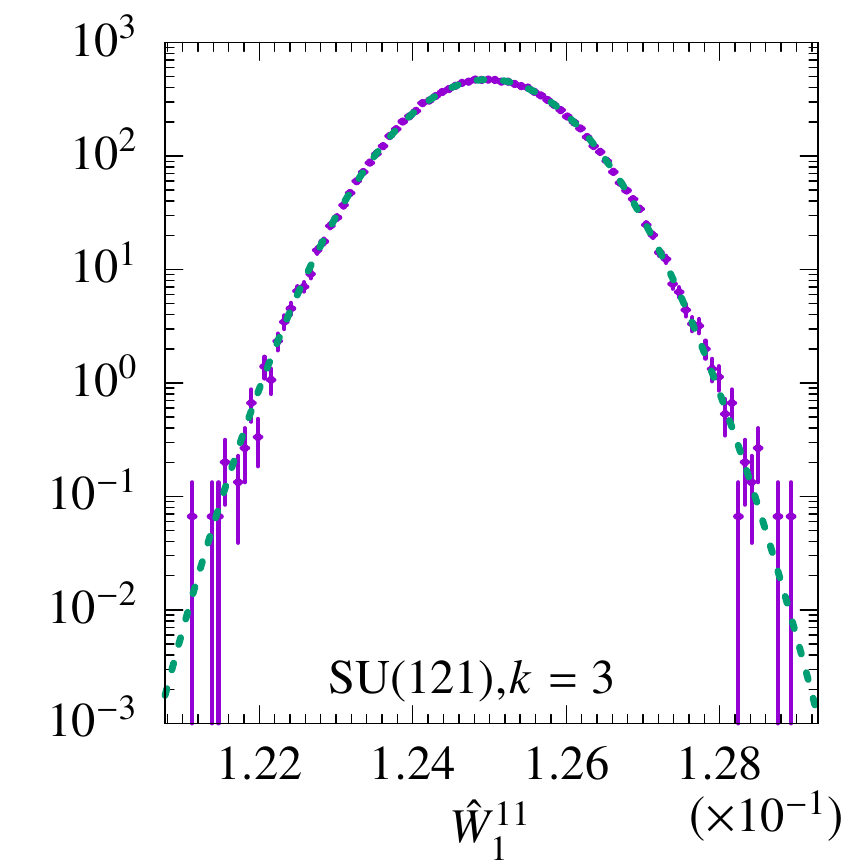}
\includegraphics[scale=\figscale,clip]{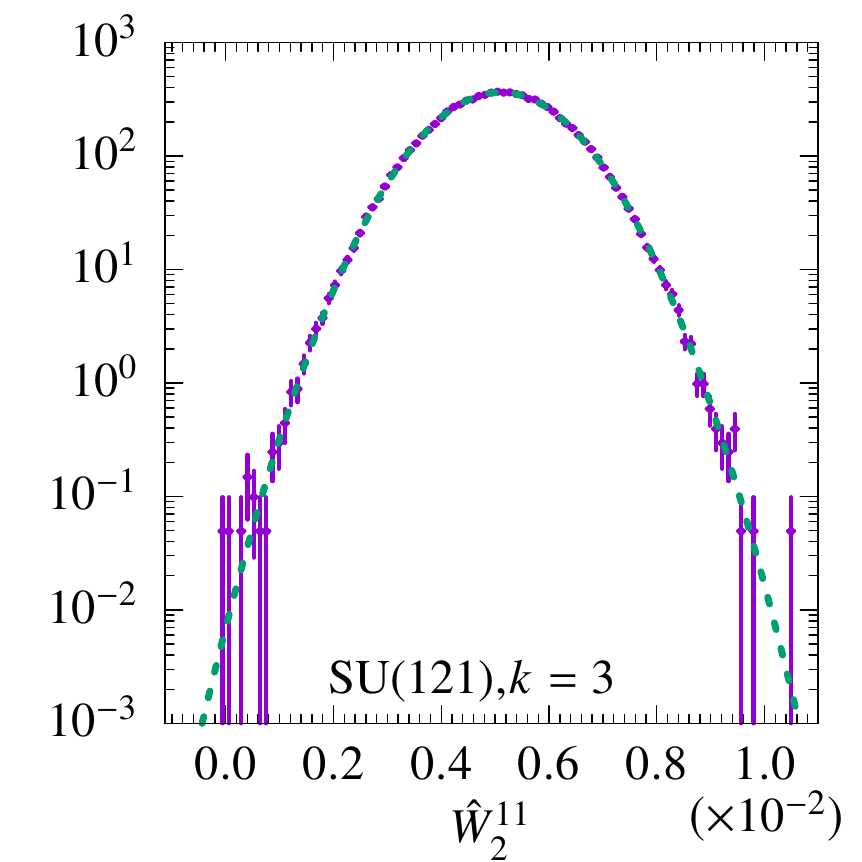}
\includegraphics[scale=\figscale,clip]{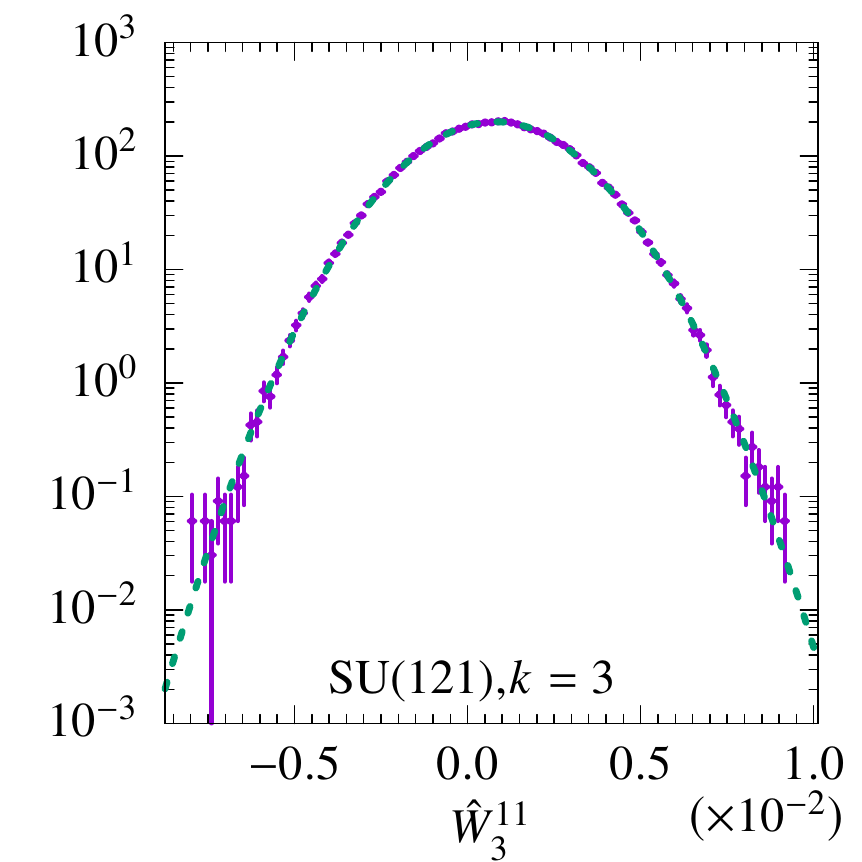}
\includegraphics[scale=\figscale,clip]{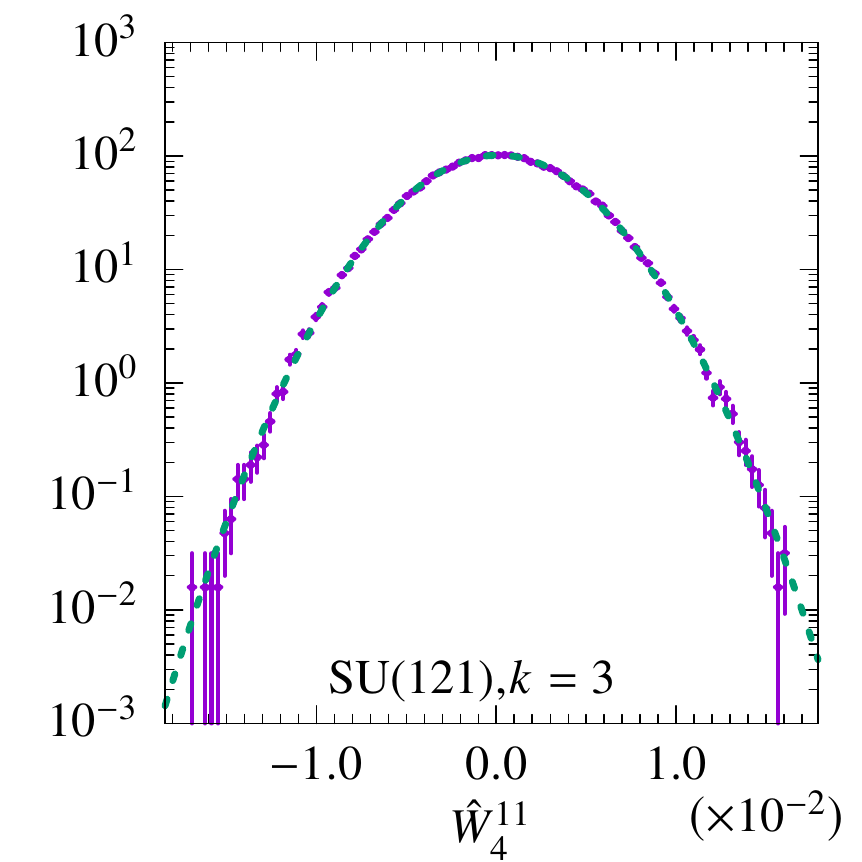}
    \caption{Histogram of $\hat{W}_{\ell}^{11}$ at $\ell=1,2,3,4$ from left to right (top row: $\Nc=16$, middle: $\Nc=49$, bottom : $\Nc=121$). Results from $\NMD=32$.}
    \label{fig:COMPHISTW11}
\end{figure}
\begin{figure}[t]
    \centering
\includegraphics[scale=\figscale,clip]{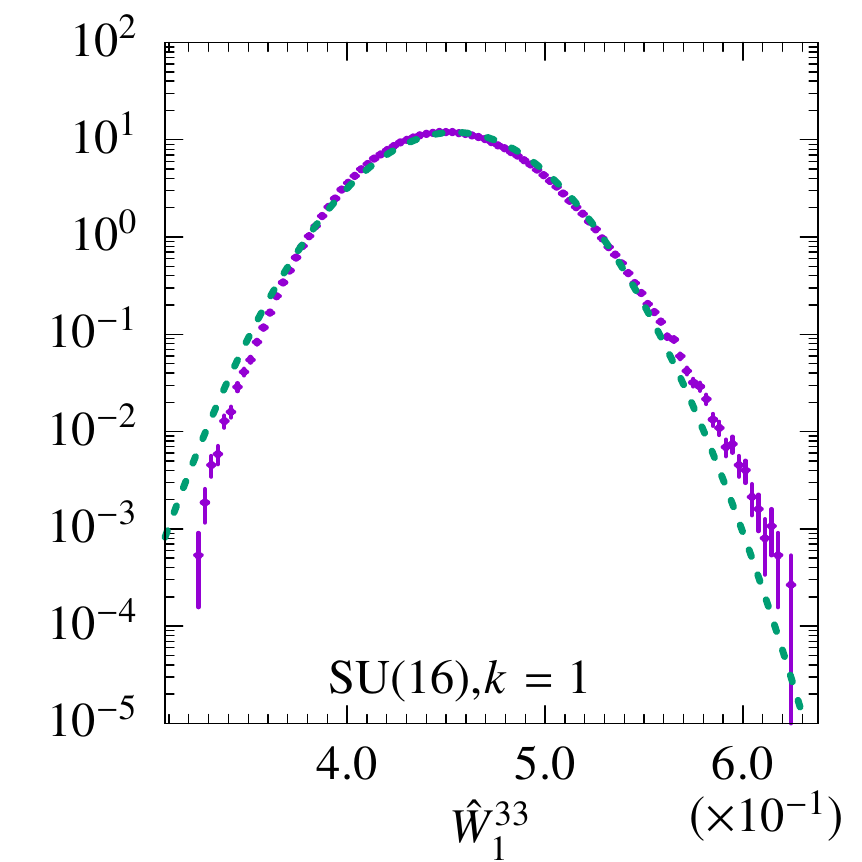}
\includegraphics[scale=\figscale,clip]{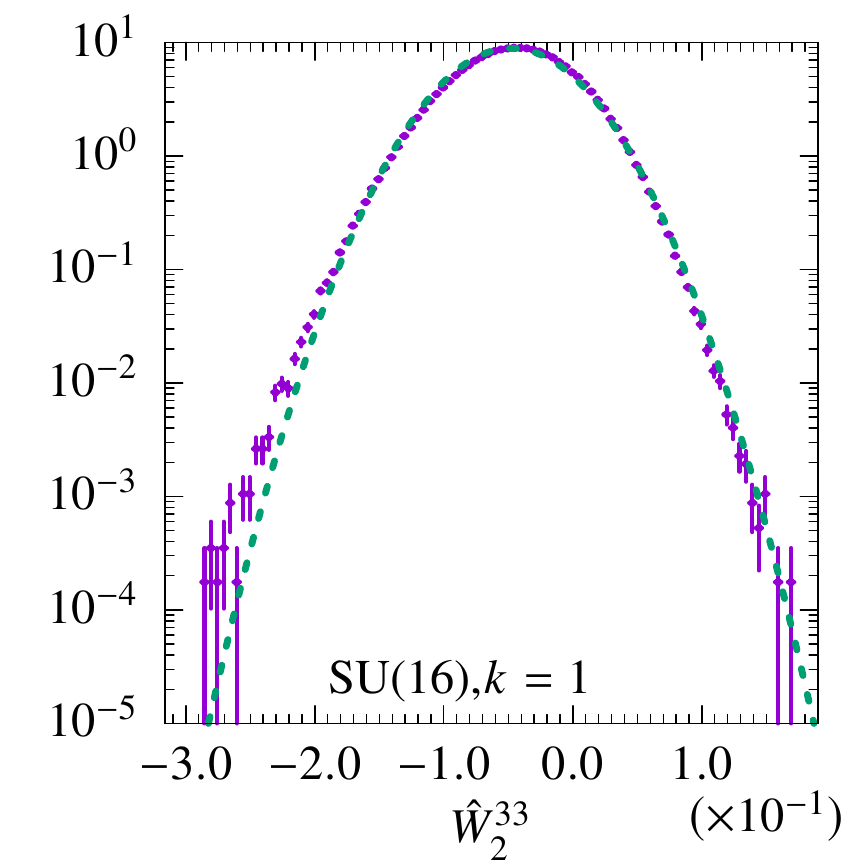}
\includegraphics[scale=\figscale,clip]{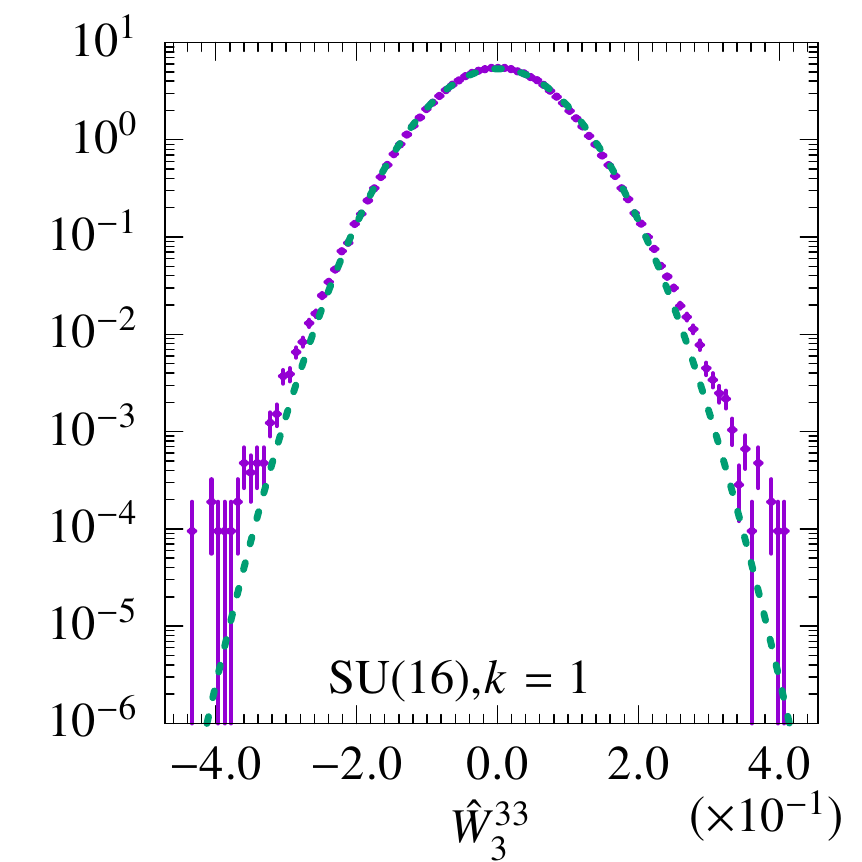}
\includegraphics[scale=\figscale,clip]{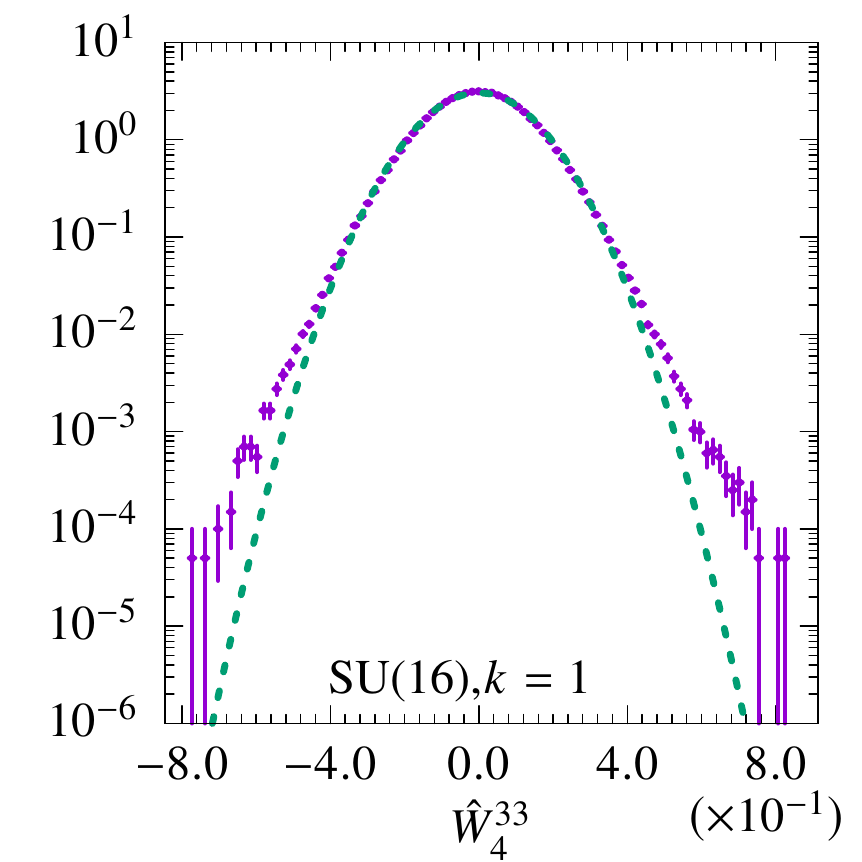}

\includegraphics[scale=\figscale,clip]{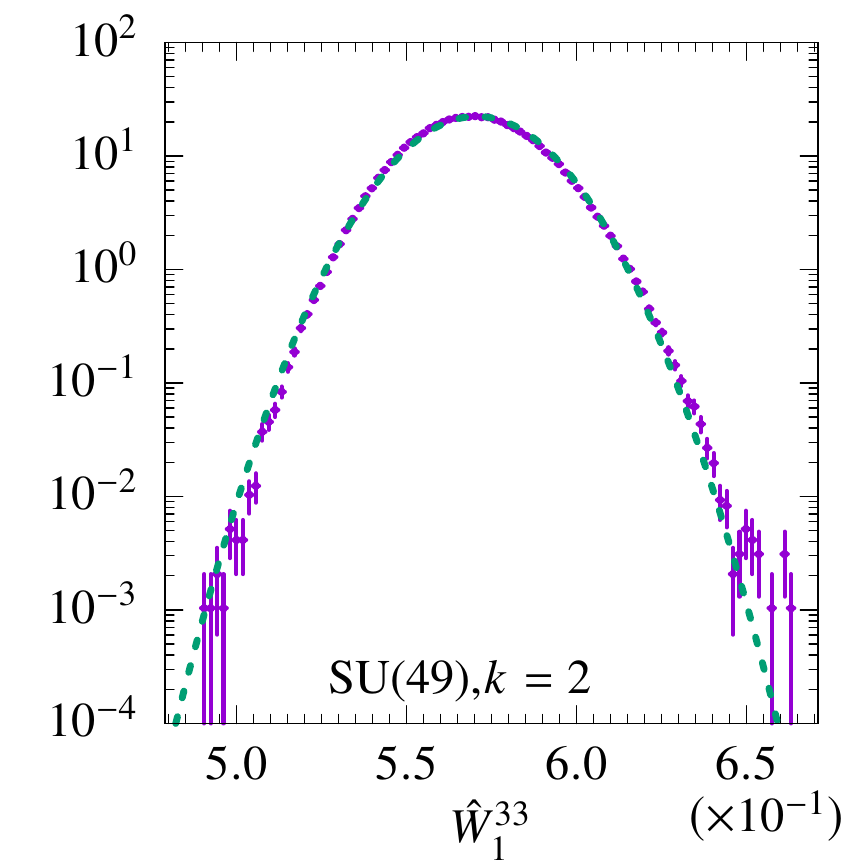}
\includegraphics[scale=\figscale,clip]{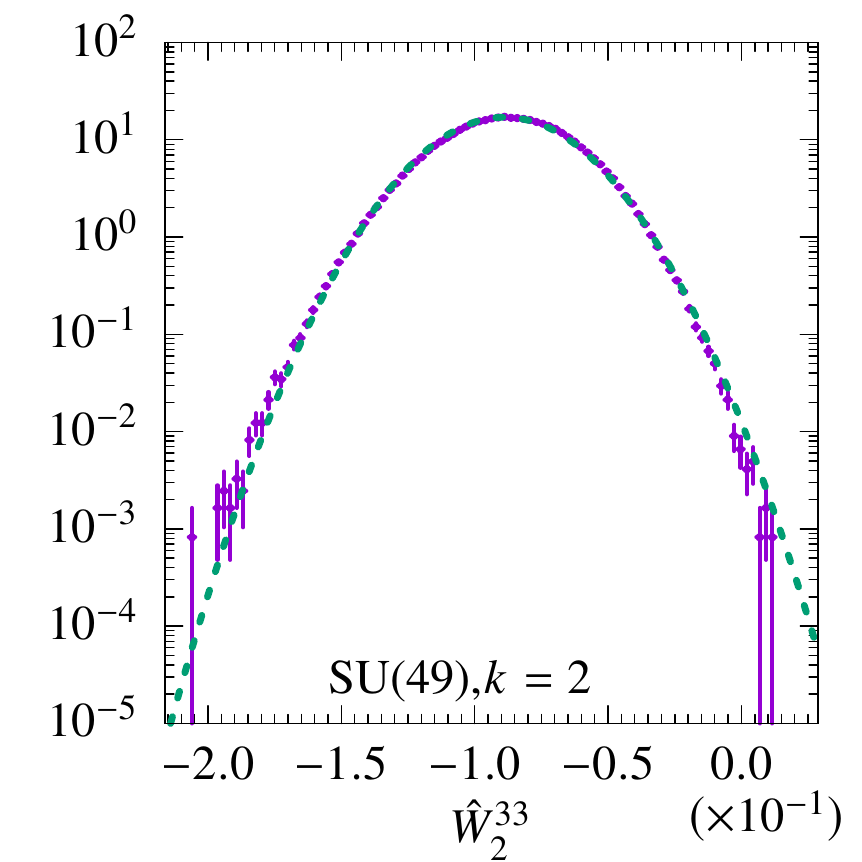}
\includegraphics[scale=\figscale,clip]{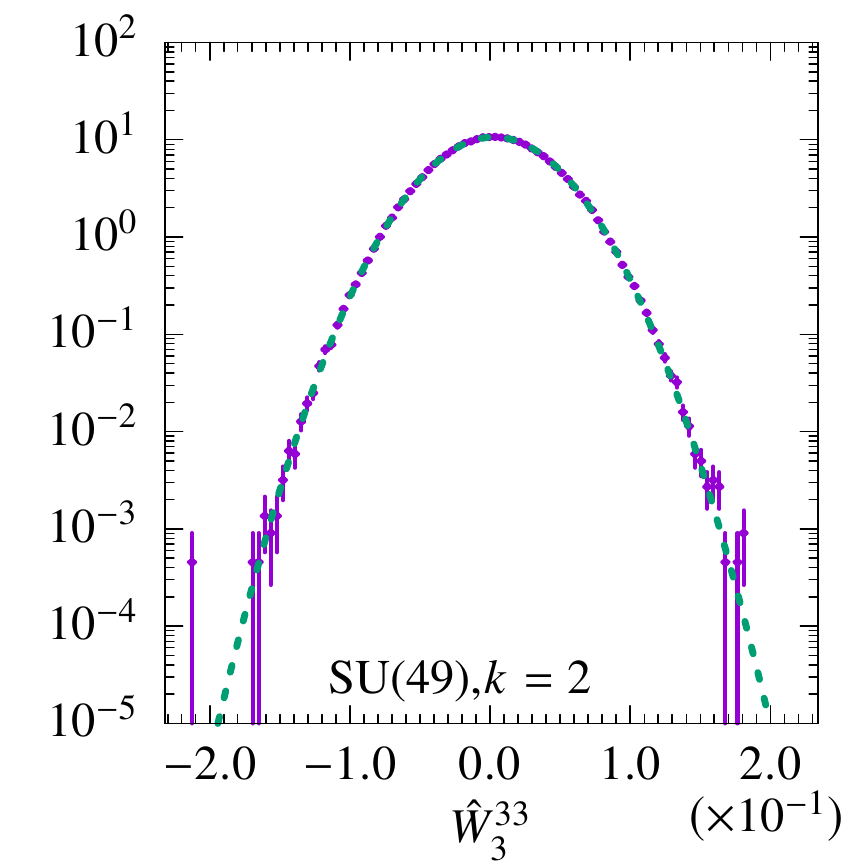}
\includegraphics[scale=\figscale,clip]{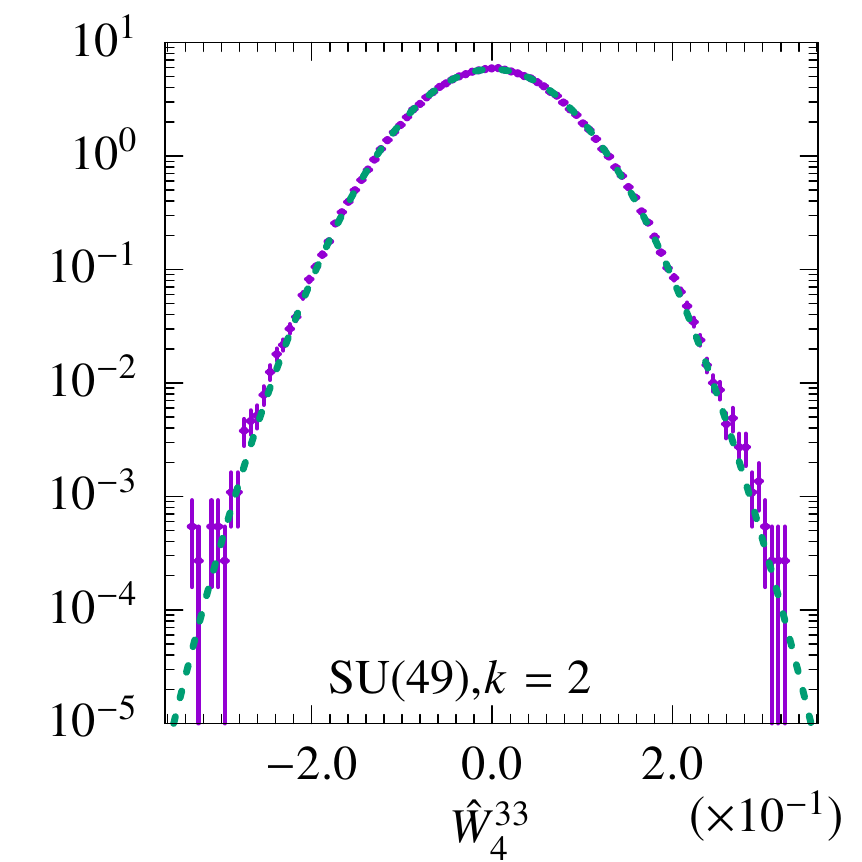}

\includegraphics[scale=\figscale,clip]{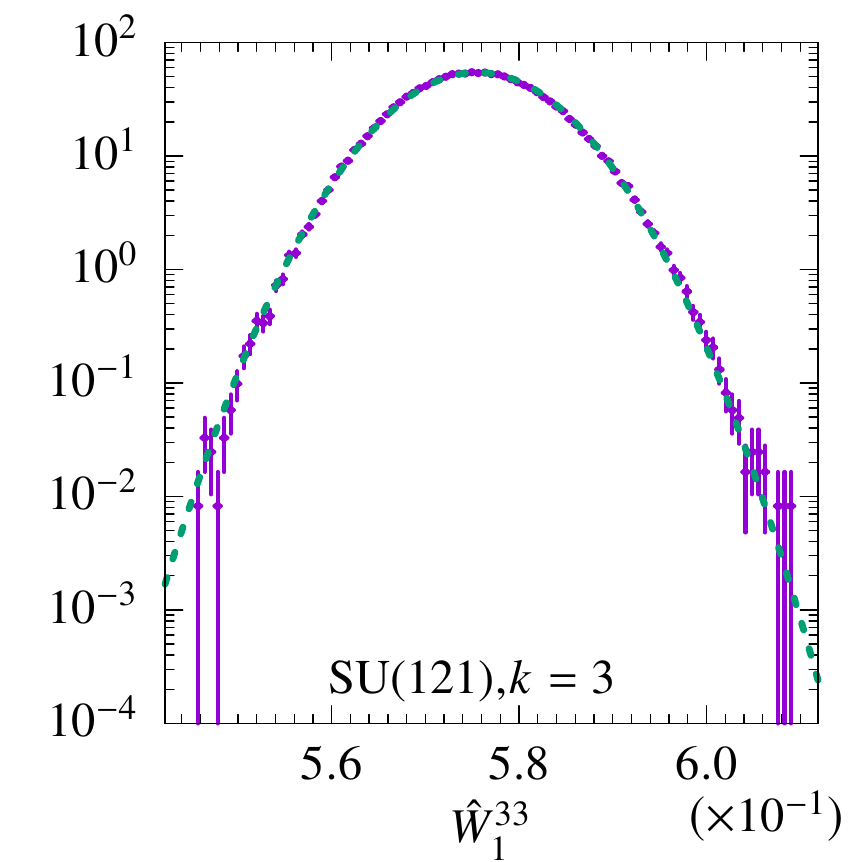}
\includegraphics[scale=\figscale,clip]{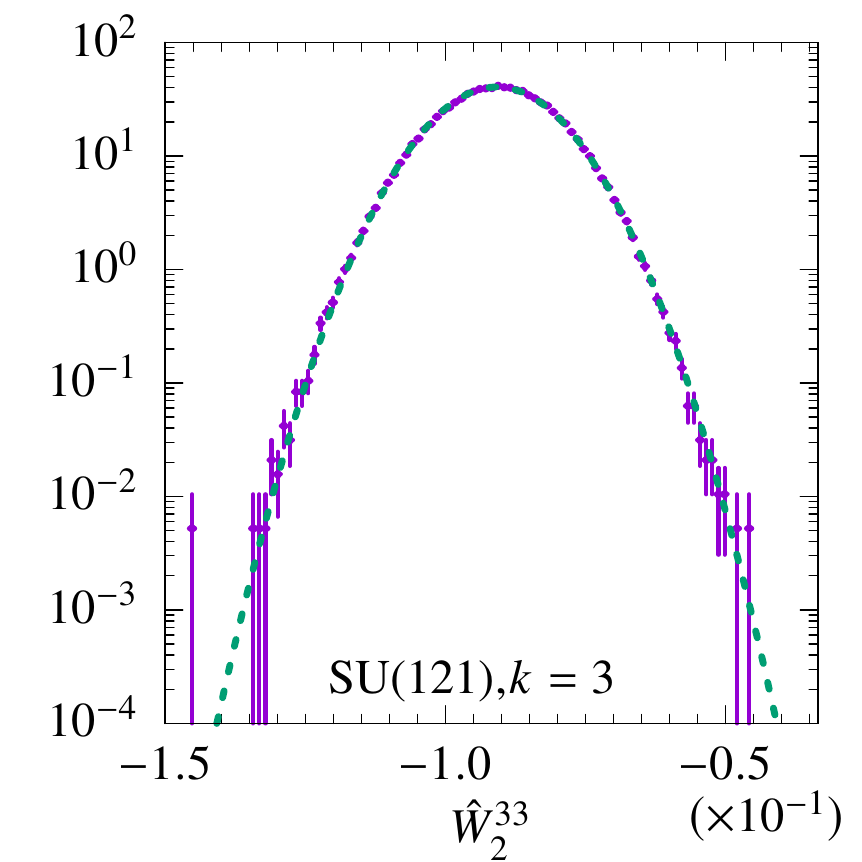}
\includegraphics[scale=\figscale,clip]{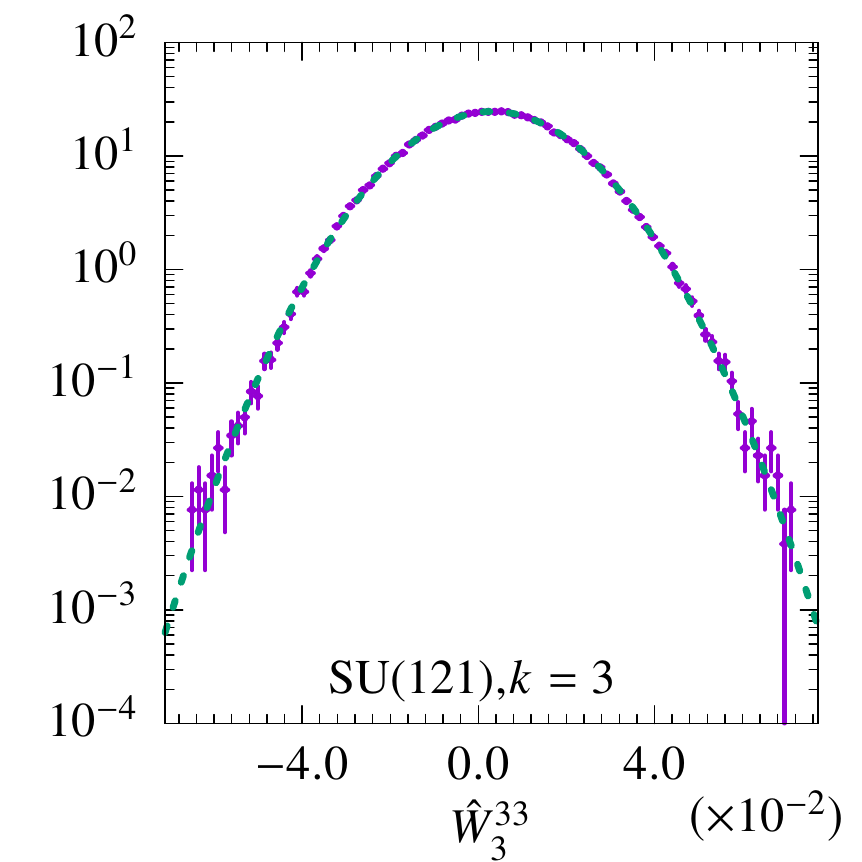}
\includegraphics[scale=\figscale,clip]{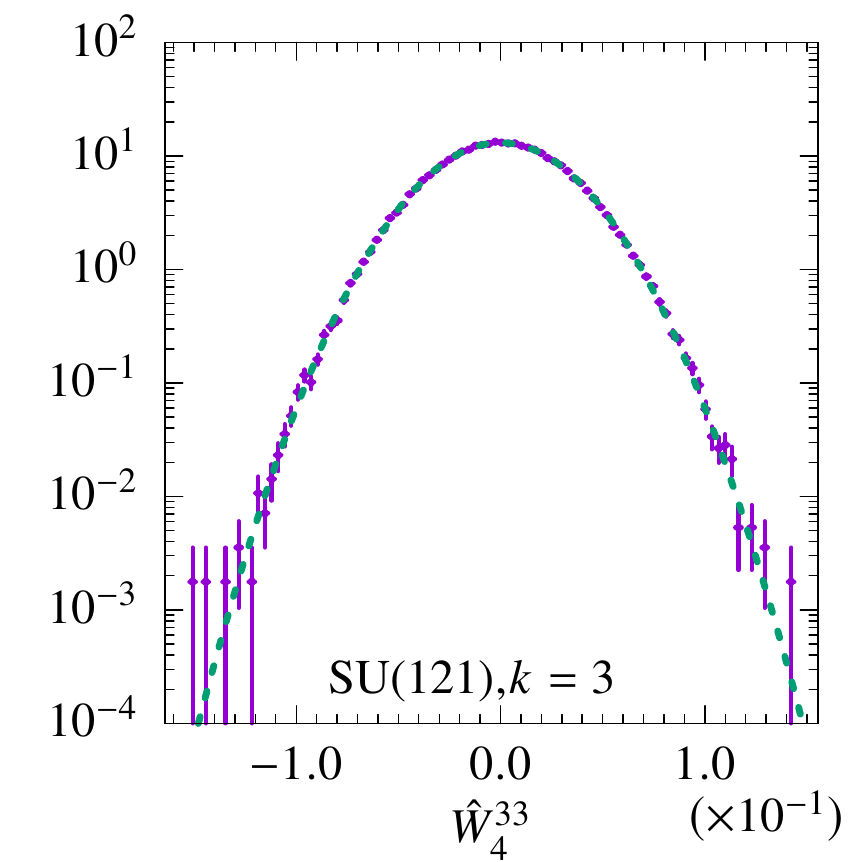}
    \caption{Same as figure~\ref{fig:COMPHISTW11}, but for $\hat{W}_{\ell}^{33}$.}
    \label{fig:COMPHISTW33}
\end{figure}

\subsection{Analysis of the distribution}

For the purpose of determining the statistical requirements involved
in a precise determination of the perturbative coefficients we
performed an
analysis of the probability distribution of the corresponding NSPT
estimates. As a quantitative measure we computed the cumulants 
 up to the 5th cumulant. 
The $\Nc$ dependence of the variance (second cumulant) is important since the statistical error at fixed statistics 
is proportional to the square root of this variance.
The non-zero value of higher order cumulants measures  the deviation from the normal distribution.
This is of practical importance since, as seen in ref.~\cite{Alfieri:2000ce}, the distribution for perturbative coefficients 
at higher order in NSPT could show strong deviations from a  normal
distribution, the so-called ``Pepe effect''.
We show the histograms for $\hat{W}^{11}_{\ell}$ and $\hat{W}^{33}_{\ell}$
in figures~\ref{fig:COMPHISTW11} and \ref{fig:COMPHISTW33},  respectively.
We have fixed $\NMD=32$ for better comparison. The distributions are
plotted in logarithmic scale for which the normal distribution
corresponds to a parabola.  The dotted lines are the best fit to a normal distribution. 
One can see deviations in the tail for some plots for $\Nc=16$, but
it disappears for larger $\Nc$. Hence, the problem of ``Pepe effect''
is absent in the TEK model for sufficiently large $\Nc$, 
similarly to what happens for  pure SU(3) lattice gauge theory.
Since the 3rd--5th cumulants have minor effect or are not detectable on the distribution,
we focus on the variance only in the following analysis.

We have studied the dependence of the variance of $\hat{W}_{\ell}^{RR}$ on $R$,
$\ell$ and $\Nc$. Notice that we can also use the results for
half-integer $\ell$ in this analysis. The $R$ dependence is illustrated in
figure~\ref{fig:LSIZEDEPVARIANCE}, and is seen to go linearly with
$R^4$. On the other hand the dependence on $j=2\ell$, displayed  in
figure~\ref{fig:PTDEPVARIANCE} shows an exponential behaviour for
sufficiently large $j$. Altogether, an ansatz of the form 
\begin{align}
  \sigma^2(\Nc,j) &=
  \dfrac{c R^4}{\Nc^2}\left(2-\dfrac{1}{\sqrt{\Nc}}\right)^{2 \ell},
  \label{eq:VARIANCEANSATZ}
  \end{align}
with $c=0.0013$ (solid lines in these figures) describes our data
quite well. For large $\Nc$ this corresponds to a dependence on the 
ratio  of the loop size with respect to the effective size of the
box $R/\Lbox$ to the fourth power, similar to the observed subleading  
corrections in  perturbation theory. 

The $1/\Nc^2$ dependence of the variance follows from the perturbative
realization of factorization $\langle W^2\rangle = \langle W\rangle^2
+O(1/\Nc^2)$. Similar arguments lead to the prediction that the 
$p$-th cumulant scales as $O(1/\Nc^{2p-2})$. This by itself explains
why the distribution tends to normal at large $\Nc$. We cannot check
this behaviour of the higher cumulants since their values are too small.

The  exponential dependence  of the variance on the order  $4^{\ell}$
coincides precisely with the characteristic growth in the number of planar
diagrams~\cite{Brezin:1977sv,Koplik:1977pf,Marino:2012zq}, given by
the Catalan number. In the
absence of renormalons, a similar growth
is also expected for the mean values $\hat{W}_{\ell}$, but it is hard to
check this behaviour with our results reaching only up to order $\ell=4$.
Indeed, as observed in ref.~\cite{Bali:2014fea}, extremely large order
perturbative coefficients are needed  to detect
the expected factorial behavior for the pure SU(3) lattice gauge theory.
Hence, it would be interesting to extend our results to higher order to
explore this point. The computational requirements will be studied in
the next subsection.

\renewcommand{\figscale}{0.56}

\begin{figure}[t]
    \centering
\includegraphics[scale=\figscale,clip]{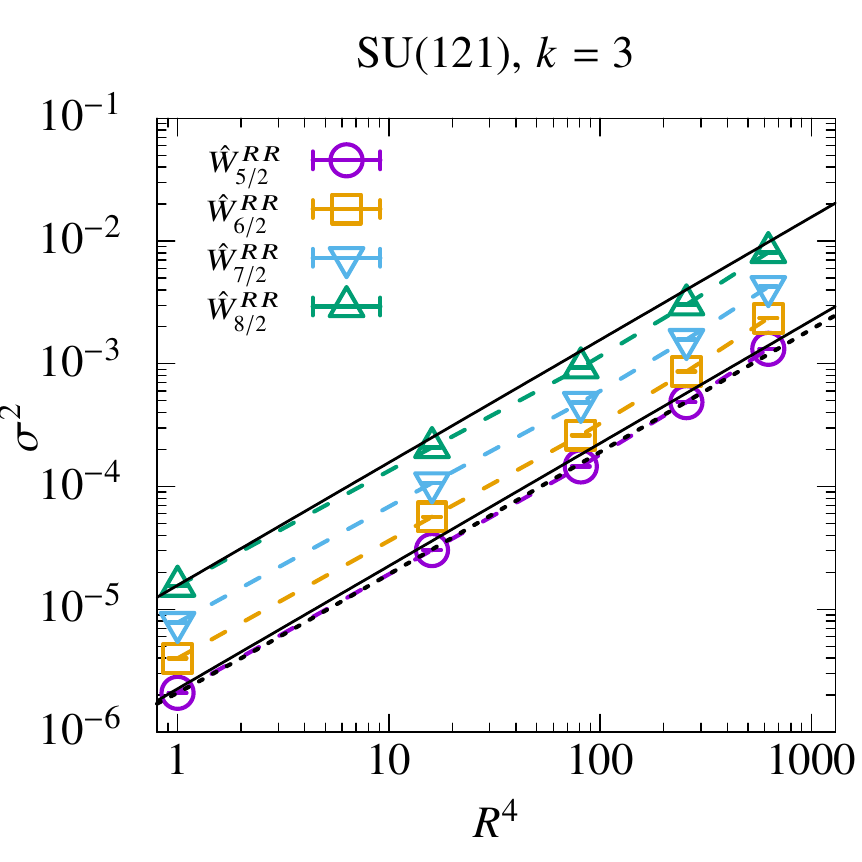}
    \caption{The loop size dependence of the variance $\sigma^2$ for
    $\hat{W}_{j/2}^{RR}$ ($\NMD=32$, $j \ge 5$ and SU(121), $k=3$).
             The dotted line shows the  result of a fit. 
             Solid lines represent the ansatz eq.~\eqref{eq:VARIANCEANSATZ} with $\ell=5/2$ and $8/2$.}
    \label{fig:LSIZEDEPVARIANCE}
\end{figure}

\begin{figure}[t]
    \centering
\includegraphics[scale=\figscale,clip]{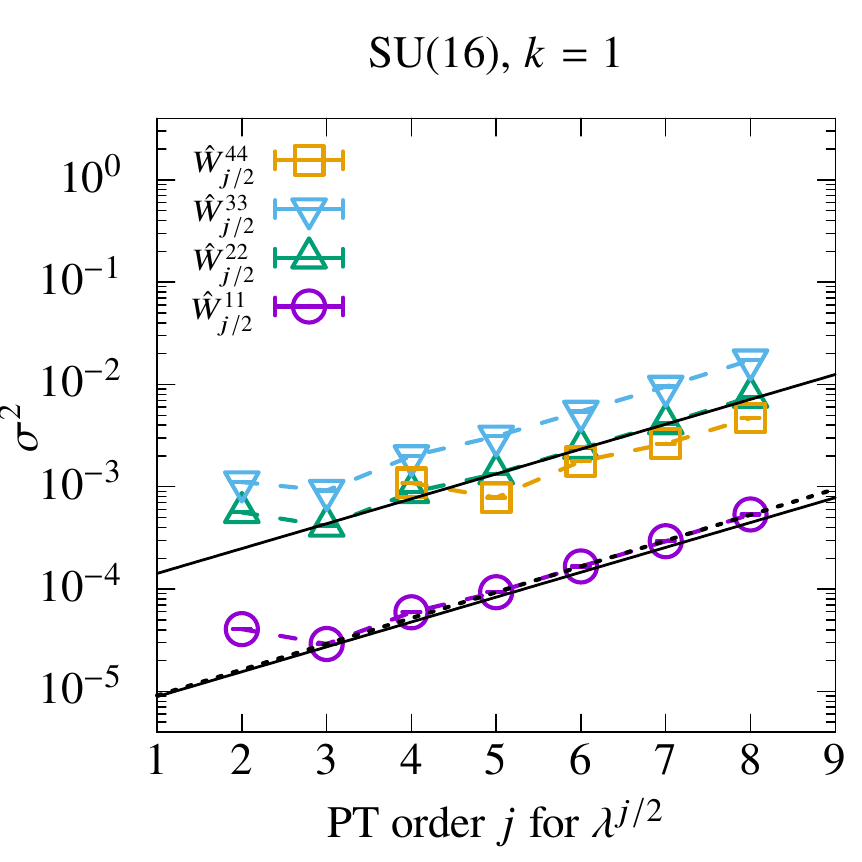}
\includegraphics[scale=\figscale,clip]{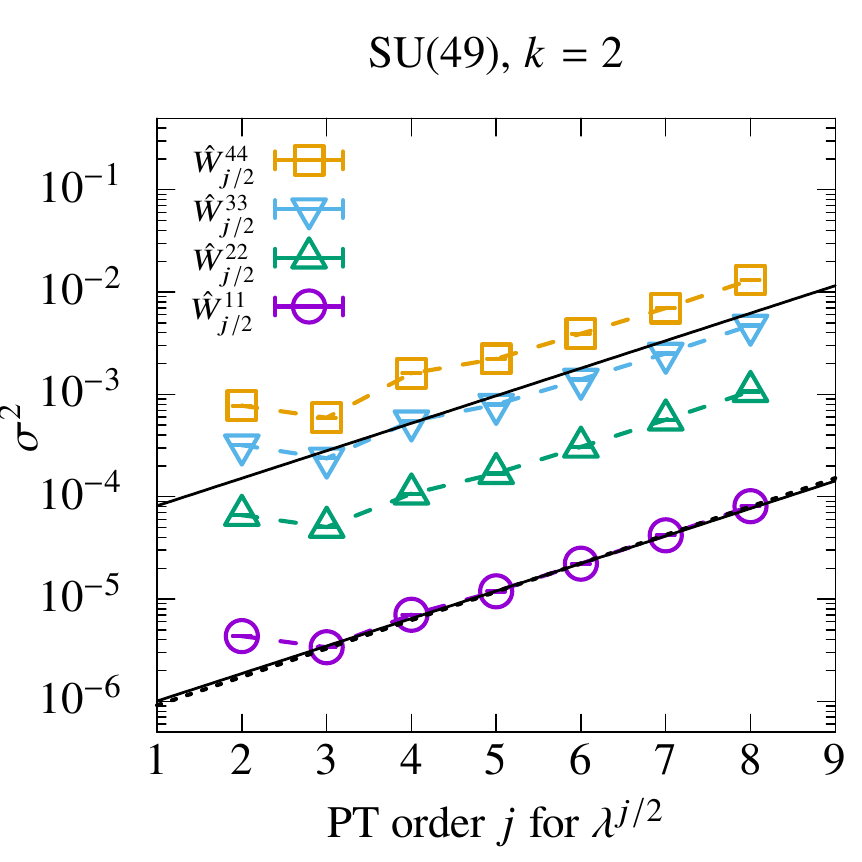}
\includegraphics[scale=\figscale,clip]{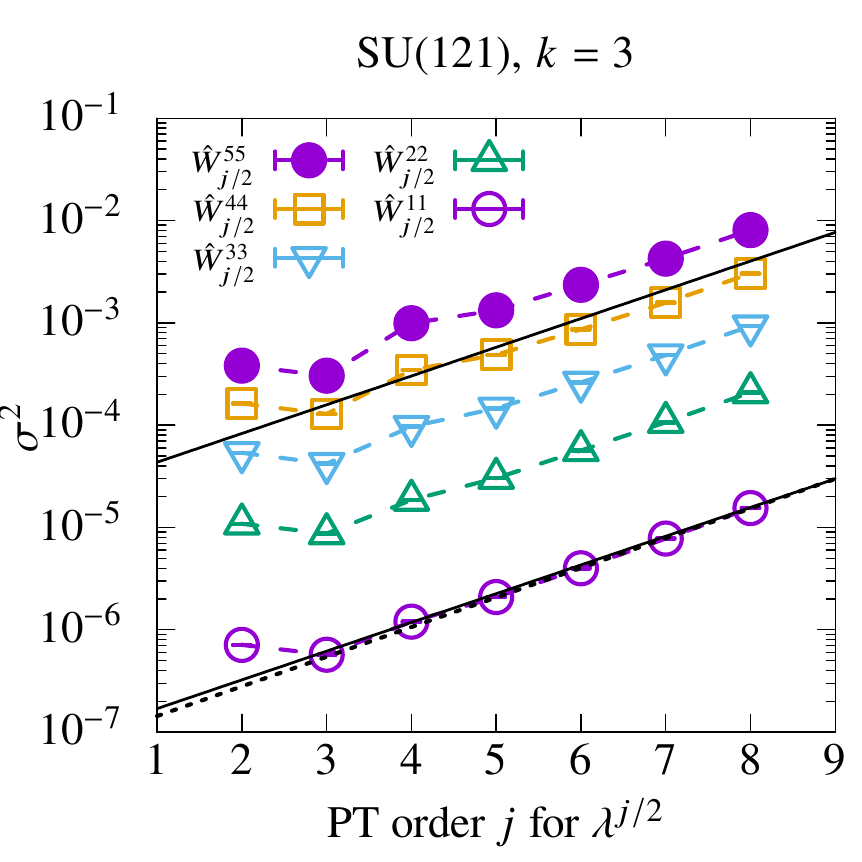}
    \caption{Variance $\sigma^2$ for $\hat{W}_{j/2}^{RR}$ at $\NMD=32$.
             Dotted lines are the  results of a fit to the  $j\ge 5$ data.  
             Solid lines represent the ansatz eq.~\eqref{eq:VARIANCEANSATZ} with $R=1$ for all $\SUNc$
             and with $R = 2, 3, 4$ for $\Nc = 16, 49, 121$, respectively.}
    \label{fig:PTDEPVARIANCE}
\end{figure}

\section{Estimate of the computational cost at high order and large $\Nc$}
\label{sec:costestimate}
 
Here we will evaluate the computer requirements  to extend our calculation $\hat{W}_{\ell}$ up to order
$\ell=\Nptmax/2$. The precision in the determination is proportional to
the standard deviation of the corresponding probability distribution.
Hence, if we want to keep this precision fixed our data sample should
grow with the square, the variance. We previously estimated that for
each sample point the computational cost goes as  $O(\Nc^{7/2}
\Nptmax^3)$ for the MD part and as $O(\Nc^3\Nptmax^4)$ for the reunitarization part. 
Using eq.~\eqref{eq:VARIANCEANSATZ}, we can then give an estimate of
the total computational cost as follows:
\begin{align}
\mathrm{[Cost]}& \propto (\Nc^{7/2} \Nptmax^3)\times \sigma^2(\Nc,\Nptmax) \propto \Nc^{3/2} \Nptmax^3 \left(2-\dfrac{1}{\sqrt{\Nc}}\right)^{\Nptmax},
\label{eq:COSTMD}
\end{align}
or
\begin{align}
\mathrm{[Cost]}& \propto (\Nc^{3} \Nptmax^4)\times \sigma^2(\Nc,\Nptmax) \propto \Nc \Nptmax^4 \left(2-\dfrac{1}{\sqrt{\Nc}}\right)^{\Nptmax}.
\label{eq:COSTREUNIT}
\end{align}

This estimate can be reduced  by improving the algorithm. To obtain
perturbative coefficients at very higher order,
the exponential scaling behaviour, $2^{\Nptmax}$, has to be relaxed by reducing the variance using a method 
such as a kind of reweighting method in NSPT. Furthermore, 
the factor $\Nptmax^4$ comes from our implementation of the reunitarization algorithm, 
for which a better algorithm could be applicable. One possibility is
to use the perturbative Gram-Schmidt algorithm 
for the reorthogonalization followed by the perturbative subtraction of the U(1) phase. 
The estimation of the U(1) phase still requires the perturbative
trace-log computation whose cost is of $O(\Nptmax^3)$.
At least one  can reduce the factor from $\Nptmax^4$ to $\Nptmax^3$.
In that case eq.~\eqref{eq:COSTMD} would dominate the total cost.

The cost of $O(\Nptmax^2)$ in the convolutional product of two series could be 
ameliorated  by using the fast Fourier transformation (FFT) algorithm,
which reduces the cost to  \linebreak $O(\Nptmax\log\Nptmax)$.
Improvement in this direction  can be used in future studies.

\section{Summary}
In this paper we have developed and applied an HMD-based  NSPT
algorithm to compute the perturbative expansion of Wilson loops in the
TEK model. At large $\Nc$ these coefficients coincide with those of
large volume Yang-Mills theory, hence their interest. The low order coefficients are very precise 
and match with the values obtained by the analytic calculation of
ref.~\cite{Perez:2017jyq}. We have been able to extend the
perturbative series two more orders, up to $O(\lambda^4)$.

We also studied the statistical properties of the coefficients in NSPT. 
We found that their distribution tends
to normal at large $\Nc$ and there is no \emph{Pepe effect}.
Furthermore, we have determined the dependence of the corresponding
variance on $\Nc$ and the order of perturbation theory. This has
allowed us to estimate the computational requirements necessary to
extend our results to higher orders. This would allow extending
studies of the type described in ref.~\cite{Marino:2012zq} to
Yang-Mills theory in the limit of infinite number of colours.

\acknowledgments

We thank Margarita Garc\'{i}a P\'{e}rez for discussions and for providing us with the
analytic values of some perturbative coefficients,
and Alberto Ramos for his encouraging comments on the draft.
A.G.-A. acknowledges financial support from the MINECO/FEDER grant
FPA2015-68541-P  and the MINECO Centro de Excelencia Severo Ochoa Program  SEV-2016-0597.
K.-I.I. and M.O. are supported by JSPS KAKENHI Grant Numbers 16K05326 and 17K05417, respectively.
K.-I.I. and I.K. acknowledge for financial support by Priority Issue 9 to be tackled by using Post K Computer.
The numerical computation for this paper were performed on the subsystem A of the ITO supercomputer system
at Research Institute for Information Technology, Kyushu University, and workstations at 
INSAM (Institute for Nonlinear Sciences and Applied Mathematics),
Hiroshima University.



\appendix
\section{Perturbative matrix functions}
\label{sec:AppendixA}

In this appendix we describe the details of evaluating the perturbative 
coefficients of a matrix function with a perturbative series as the argument.
After introducing the generic algorithm for perturbative matrix functions, 
we explain the cases with matrix exponential and matrix logarithm explicitly.

We consider a matrix function $f(F)$ whose argument $F$ is a matrix.  
We assume that $f(F)$ has the following Taylor expansion form:
\begin{align}
    f(F) &= \sum_{k=0}^{\infty} c_k F^k.
\end{align}
We also assume that the perturbative expansion of $F$ is
\begin{align}
    F&=\sum_{k=1}^{\infty} g^k F^{(k)},
\label{eq:PTexF}
\end{align}
where $g$ is the coupling constant and $F^{(k)}$ are the coefficient matrices. Note that the leading term of $F$ is $O(g)$.
We truncate the perturbative expansion at $\Nptmax$-th order.
\begin{align}
    F&=\sum_{k=1}^{\Nptmax} g^k F^{(k)}.
\label{eq:ARGPT}
\end{align}

The perturbative expansion of $f(F)$ with eq.~\eqref{eq:ARGPT} is also truncated at $\Nptmax$ as
\begin{align}
    f(\sum_{k=1}^{\Nptmax} g^k F^{(k)})&= c_0 I + \sum^{\Nptmax}_{k=1} g^{k} \left[f(F)\right]^{(k)}.
\end{align}
We would like to know the coefficient $\left[f(F)\right]^{(k)}$ in terms of $c_k$'s and $F^{(k)}$'s.
A recursion relation to evaluate the coefficient $\left[f(F)\right]^{(k)}$ can be obtained by expanding 
the Horner's method for polynomials as follows.

As $F$ is truncated at $\Nptmax$-th order, the Taylor expansion is also truncated at $\Nptmax$-th order,
\begin{align}
    f(F) & \simeq G = \sum_{k=0}^{\Nptmax} c_k F^k,
\end{align}
where the matrix $G$ is introduced for the truncated form.
$G$ can be evaluated via the Horner's method:
\begin{align}
G = f(F) &= c_0 + c_1 F + c_2 F^2 + \cdots + c_{\Nptmax} F^{\Nptmax}
\notag\\                                                                                                                    
&= c_0 \left[ I + d_1 F \left[ I + d_2 F \left[ I + \cdots \left[ I + d_{\Nptmax} F\right] \cdots\right]\right]\right],             
\\                                                                                                                          
d_j &\equiv  c_j/c_{j-1}.
\end{align}
This form leads to the following recursion relation:
\begin{align}
    G_N & = I + d_{\Nptmax} F,\\
    G_j & = I + d_j F G_{j+1},\quad \mbox{for $j=\Nptmax-1,\Nptmax-2,\dots,2,1$},\\
    G   & = c_0 G_1,
\end{align}
where $G_j$ are working area.
As $F$ has the series form eq.~\eqref{eq:PTexF}, 
$G_j$ also has the series form:
\begin{align}
    G_j &= I + \sum_{k=1}^{\Nptmax}g^{k}G_j^{(k)}.
\label{eq:WORKPT}
\end{align}
Substituting the series form 
\eqref{eq:ARGPT} for $F$
and \eqref{eq:WORKPT} for $G_j$ into the recursion relation, we obtain
\begin{align}
    G_{j} &= I + \sum_{k=1}^{\Nptmax}g^k G_{j}^{(k)} 
\notag\\
&=      
I + d_{j}
 \left(    \sum_{k=1}^{N} g^k F^{(k)}      \right) 
 \left(I + \sum_{k=1}^{N} g^k G_{j+1}^{(k)}\right) 
\notag\\
&=
I + d_{j}
\left[ 
  \sum_{k=1}^{\Nptmax}                        g^k F^{(k)} 
+ \sum_{k=1}^{\Nptmax}\sum_{\ell=1}^{\Nptmax} g^{k+\ell} F^{(k)} G_{j+1}^{(\ell)}
\right] 
\notag\\
&=
I + d_{j}
\left[
  \sum_{k=1}^{\Nptmax}g^{k} F^{(k)}
+ \sum_{k=2}^{\Nptmax}g^{k} \sum_{\ell=1}^{k-1} F^{(k-\ell)} G_{j+1}^{(\ell)}
\right].
\end{align}
Here we discard higher order terms $O(g^{k})$ with $k>\Nptmax$.

Consequently the perturbative coefficient $[f(F)]^{(k)}$ of $f(F)$ can 
be obtained as algorithm~\ref{eq:RECURSIONF}. The computational cost scales with $\Nptmax^3$.

{

\begin{algorithm}[t]
\caption{Recursion algorithm for the perturbative expansion of a matrix function $f(F)$. 
The maximum order $\Nptmax$ of expansion is fixed.
$G^{(k)}$ is the perturbative coefficient $[f(F)]^{(k)}$.}
\label{eq:RECURSIONF}
\begin{algorithmic}[1]
\FOR{$k=1$ to $\Nptmax$}
\STATE $G_{\Nptmax}^{(k)} = d_{\Nptmax} F^{(k)}$
\ENDFOR
\FOR{$j=\Nptmax-1$ to $1$}
\FOR{$k=\Nptmax$ to $1$}
\STATE $G_{j}^{(k)} = d_{j} \left[  F^{(k)} + \sum_{\ell=1}^{k-1} F^{(k-\ell)}G_{j+1}^{(\ell)}\right]$
\ENDFOR
\ENDFOR
\FOR{$k=1$ to $\Nptmax$}
\STATE $G^{(k)} = c_0 G_{1}^{(k)}$
\ENDFOR
\end{algorithmic}
\end{algorithm}

}

\subsection{Matrix exponential for updating $U$}
\label{subsec:matexp}
Here we consider the case of $f(F)=\exp(F)=\exp(i P \Delta \tau)$ with 
$P=\sum_{k=1}^{\infty} \beta^{-k/2} P^{(k)}$.
The Taylor expansion for $f(F)$ is
\begin{align}
    f(F)=\exp(F)&= I + \sum_{k=1}^{\infty} \dfrac{1}{k!}F^{k}
\notag\\
\exp(iP\Delta \tau)&= I +\sum_{k=1}^{\infty} \dfrac{(i\Delta \tau)^k}{k!}P^{k}.
\end{align}
Thus the coefficients are $c_{k}=(i\Delta\tau)^{k}/{k!}$. 
We identify $F=P$, $F^{(k)}=P^{(k)}$, $d_j=i \Delta \tau/j$, and 
$[\exp(iP\Delta \tau)]^{(k)}=G_1^{(k)}$ with $g=\beta^{-1/2}$
for the recursion algorithm~\ref{eq:RECURSIONF} (omitting the last step lines 9--11 as $c_0=1$).

\subsection{Matrix logarithm for reunitarization}
\label{subsec:matlog}
Next we consider the case of $f(F)=\ln(I+F)=\ln(U)$ with 
$U=I+\sum_{k=1}^{\infty} \beta^{-k/2} U^{(k)}=I+F$.
The Taylor expansion for $f(F)$ is
\begin{align}
    f(F)=\ln(I+F)&= \sum_{\ell=1}^{\infty} \dfrac{(-1)^{\ell+1}}{\ell}F^{\ell}.
\end{align}
Thus the coefficients are $c_0=0$, $c_{\ell}=(-1)^{\ell+1}/\ell$. 
As $F$ is expanded as $F=\sum_{k=1}^{\infty} \beta^{-k/2} U^{(k)}$, 
we can identify $F^{(k)}=U^{(k)}$, and
$[\ln(U)]^{(k)}=G_1^{(k)}$ with $g=\beta^{-1/2}$ and $d_j=-(j-1)/j$
for the recursion algorithm~\ref{eq:RECURSIONF}.

\section{Perturbative reunitarization}
\label{sec:AppendixB}

Reunitarization of $U$ is applied perturbatively on $U^{(k)}$ via the perturbative expansion of the matrix logarithm:
\begin{align}
    A\equiv \ln[U] & \Leftrightarrow U = \exp[A],
\end{align}
where $A$ should be anti-Hermitian and traceless for $U$ to be $\SUNc$.
From the perturbative expansion of the matrix logarithm and $A=\sum_{k=1}^{\infty}g^k A^{(k)}$, we have
\begin{align}
\left[\ln[U]\right]^{(k)} = A^{(k)},
\end{align}
where $\left[\ln[U]\right]^{(k)}$ is described in subsection~\ref{subsec:matlog}.
The $\SUNc$ condition on the perturbative coefficients $A^{(k)}$ is
\begin{align}
   {A^{(k)}}^{\dag}&=-A^{(k)},\qquad\Tr[A^{(k)}]=0.
\end{align}
By inspecting algorithm~\ref{eq:RECURSIONF}, we can find that the coefficient 
$\left[\ln[U]\right]^{(k)}=A^{(k)}$ has the following dependency on $U^{(k)}$:
\begin{align}
 A^{(1)} &= U^{(1)},\\
 A^{(k)} &= U^{(k)} + X[U^{(1)},U^{(2)},\dots,U^{(k-1)}],\quad\mbox{for $1<k$},
\end{align}
where $X[U^{(1)},U^{(2)},\dots,U^{(k-1)}]\equiv \left[\ln[U]\right]^{(k)}-U^{(k)}$.
Thus the $\SUNc$ condition on $A^{(k)}$ can be guaranteed 
by applying algorithm~\ref{alg:REUNIT} on $U^{(k)}$. 
The computational cost of algorithm~\ref{alg:REUNIT} is $O(\Nptmax^4)$.

In order to apply algorithm~\ref{alg:REUNIT} to the TEK model, we have to reunitarize 
the matrix $V^{(k)}_{\mu}=U^{(k)}_{\mu}\Gamma^{\dag}_{\mu}$
as the perturbative vacuum is $U^{(0)}_{\mu}=\Gamma_{\mu}$. 
After reunitarizing $V_{\mu}^{(k)}$, $U^{(k)}_{\mu}= \Gamma_{\mu}V^{(k)}_{\mu}$ is computed.

{

\begin{algorithm}[t]
\caption{Perturbative reunitarization algorithm for a $\SUNc$ matrix. $A$ and $B$ are working area.}
\label{alg:REUNIT}
\begin{algorithmic}[1]
\FOR{$k=1$ to $\Nptmax$}
\IF{$k=1$}
\STATE $A = U^{(1)}$
\ELSE
\STATE $A = U^{(k)}+X[U^{(1)},U^{(2)},\dots,U^{(k-1)}]$
\ENDIF
\STATE $B=\left(A-A^{\dag}\right)/2$
\STATE $A=B-\Tr[B]/\Nc$
\IF{$k=1$}
\STATE $U^{(1)}=A^{(1)}$
\ELSE
\STATE $U^{(k)}=A^{(k)}-X[U^{(1)},U^{(2)},\dots,U^{(k-1)}]$
\ENDIF
\ENDFOR
\end{algorithmic}
\end{algorithm}

}



\begin{thebibliography}{10}

\bibitem{tHooft:1973alw}
G.~'t~Hooft, \emph{{A Planar Diagram Theory for Strong Interactions}},
  \href{https://doi.org/10.1016/0550-3213(74)90154-0}{\emph{Nucl. Phys.}
  {\bfseries B72} (1974) 461}.

\bibitem{Maldacena:1997re}
J.~M. Maldacena, \emph{{The Large N limit of superconformal field theories and
  supergravity}}, \href{https://doi.org/10.1023/A:1026654312961,
  10.4310/ATMP.1998.v2.n2.a1}{\emph{Int. J. Theor. Phys.} {\bfseries 38} (1999)
  1113} [\href{https://arxiv.org/abs/hep-th/9711200}{{\ttfamily
  hep-th/9711200}}].

\bibitem{Wilson:1974sk}
K.~G. Wilson, \emph{{Confinement of Quarks}},
  \href{https://doi.org/10.1103/PhysRevD.10.2445}{\emph{Phys. Rev.} {\bfseries
  D10} (1974) 2445}.

\bibitem{Lucini:2012gg}
B.~Lucini and M.~Panero, \emph{{SU(N) gauge theories at large N}},
  \href{https://doi.org/10.1016/j.physrep.2013.01.001}{\emph{Phys. Rept.}
  {\bfseries 526} (2013) 93} [\href{https://arxiv.org/abs/1210.4997}{{\ttfamily
  1210.4997}}].

\bibitem{Lucini:2003zr}
B.~Lucini, M.~Teper and U.~Wenger, \emph{{The High temperature phase transition
  in SU(N) gauge theories}},
  \href{https://doi.org/10.1088/1126-6708/2004/01/061}{\emph{JHEP} {\bfseries
  01} (2004) 061} [\href{https://arxiv.org/abs/hep-lat/0307017}{{\ttfamily
  hep-lat/0307017}}].

\bibitem{Lucini:2004my}
B.~Lucini, M.~Teper and U.~Wenger, \emph{{Glueballs and k-strings in SU(N)
  gauge theories: Calculations with improved operators}},
  \href{https://doi.org/10.1088/1126-6708/2004/06/012}{\emph{JHEP} {\bfseries
  06} (2004) 012} [\href{https://arxiv.org/abs/hep-lat/0404008}{{\ttfamily
  hep-lat/0404008}}].

\bibitem{Lucini:2005vg}
B.~Lucini, M.~Teper and U.~Wenger, \emph{{Properties of the deconfining phase
  transition in SU(N) gauge theories}},
  \href{https://doi.org/10.1088/1126-6708/2005/02/033}{\emph{JHEP} {\bfseries
  02} (2005) 033} [\href{https://arxiv.org/abs/hep-lat/0502003}{{\ttfamily
  hep-lat/0502003}}].

\bibitem{Eguchi:1982nm}
T.~Eguchi and H.~Kawai, \emph{{Reduction of Dynamical Degrees of Freedom in the
  Large N Gauge Theory}},
  \href{https://doi.org/10.1103/PhysRevLett.48.1063}{\emph{Phys. Rev. Lett.}
  {\bfseries 48} (1982) 1063}.

\bibitem{Bhanot:1982sh}
G.~Bhanot, U.~M. Heller and H.~Neuberger, \emph{{The Quenched Eguchi-Kawai
  Model}}, \href{https://doi.org/10.1016/0370-2693(82)90106-X}{\emph{Phys.
  Lett.} {\bfseries 113B} (1982) 47}.

\bibitem{Gross:1982at}
D.~J. Gross and Y.~Kitazawa, \emph{{A Quenched Momentum Prescription for Large
  N Theories}}, \href{https://doi.org/10.1016/0550-3213(82)90278-4}{\emph{Nucl.
  Phys.} {\bfseries B206} (1982) 440}.

\bibitem{Narayanan:2003fc}
R.~Narayanan and H.~Neuberger, \emph{{Large N reduction in continuum}},
  \href{https://doi.org/10.1103/PhysRevLett.91.081601}{\emph{Phys. Rev. Lett.}
  {\bfseries 91} (2003) 081601}
  [\href{https://arxiv.org/abs/hep-lat/0303023}{{\ttfamily hep-lat/0303023}}].

\bibitem{Kovtun:2007py}
P.~Kovtun, M.~{\"{U}}nsal and L.~G. Yaffe, \emph{{Volume independence in large
  N(c) QCD-like gauge theories}},
  \href{https://doi.org/10.1088/1126-6708/2007/06/019}{\emph{JHEP} {\bfseries
  06} (2007) 019} [\href{https://arxiv.org/abs/hep-th/0702021}{{\ttfamily
  hep-th/0702021}}].

\bibitem{Unsal:2008ch}
M.~{\"{U}}nsal and L.~G. Yaffe, \emph{{Center-stabilized Yang-Mills theory:
  Confinement and large N volume independence}},
  \href{https://doi.org/10.1103/PhysRevD.78.065035}{\emph{Phys. Rev.}
  {\bfseries D78} (2008) 065035}
  [\href{https://arxiv.org/abs/0803.0344}{{\ttfamily 0803.0344}}].

\bibitem{GonzalezArroyo:1982ub}
A.~Gonz{\'{a}}lez-Arroyo and M.~Okawa, \emph{{A Twisted Model for Large $N$
  Lattice Gauge Theory}},
  \href{https://doi.org/10.1016/0370-2693(83)90647-0}{\emph{Phys. Lett.}
  {\bfseries 120B} (1983) 174}.

\bibitem{GonzalezArroyo:1982hz}
A.~Gonz{\'{a}}lez-Arroyo and M.~Okawa, \emph{{The Twisted Eguchi-Kawai Model: A
  Reduced Model for Large N Lattice Gauge Theory}},
  \href{https://doi.org/10.1103/PhysRevD.27.2397}{\emph{Phys. Rev.} {\bfseries
  D27} (1983) 2397}.

\bibitem{GonzalezArroyo:1983pw}
A.~Gonz{\'{a}}lez-Arroyo and M.~Okawa, \emph{{String Tension for Large $N$
  Gauge Theory}},
  \href{https://doi.org/10.1016/0370-2693(83)90818-3}{\emph{Phys. Lett.}
  {\bfseries 133B} (1983) 415}.

\bibitem{Fabricius:1984un}
K.~Fabricius and O.~Haan, \emph{{Numerical Analysis of the Twisted Eguchi-kawai
  Model in Four-dimensions}},
  \href{https://doi.org/10.1016/0370-2693(84)91083-9}{\emph{Phys. Lett.}
  {\bfseries 139B} (1984) 293}.

\bibitem{GonzalezArroyo:2012fx}
A.~Gonz{\'{a}}lez-Arroyo and M.~Okawa, \emph{{The string tension from smeared
  Wilson loops at large N}},
  \href{https://doi.org/10.1016/j.physletb.2012.12.027}{\emph{Phys. Lett.}
  {\bfseries B718} (2013) 1524}
  [\href{https://arxiv.org/abs/1206.0049}{{\ttfamily 1206.0049}}].

\bibitem{Gonzalez-Arroyo:2014dua}
A.~Gonz{\'{a}}lez-Arroyo and M.~Okawa, \emph{{Testing volume independence of
  SU(N) pure gauge theories at large N}},
  \href{https://doi.org/10.1007/JHEP12(2014)106}{\emph{JHEP} {\bfseries 12}
  (2014) 106} [\href{https://arxiv.org/abs/1410.6405}{{\ttfamily 1410.6405}}].

\bibitem{Douglas:2001ba}
M.~R. Douglas and N.~A. Nekrasov, \emph{{Noncommutative field theory}},
  \href{https://doi.org/10.1103/RevModPhys.73.977}{\emph{Rev. Mod. Phys.}
  {\bfseries 73} (2001) 977}
  [\href{https://arxiv.org/abs/hep-th/0106048}{{\ttfamily hep-th/0106048}}].

\bibitem{GonzalezArroyo:1983ac}
A.~Gonz{\'{a}}lez-Arroyo and C.~P. Korthals~Altes, \emph{{Reduced Model for
  Large $N$ Continuum Field Theories}},
  \href{https://doi.org/10.1016/0370-2693(83)90526-9}{\emph{Phys. Lett.}
  {\bfseries 131B} (1983) 396}.

\bibitem{Connes:1987ue}
A.~Connes and M.~A. Rieffel, \emph{{Yang-Mills for noncommutative two-tori}},
  {\emph{Contemp. Math.} {\bfseries 62} (1987) 237}.

\bibitem{tHooft:1979rtg}
G.~'t~Hooft, \emph{{A Property of Electric and Magnetic Flux in Nonabelian
  Gauge Theories}},
  \href{https://doi.org/10.1016/0550-3213(79)90595-9}{\emph{Nucl. Phys.}
  {\bfseries B153} (1979) 141}.

\bibitem{Perez:2017jyq}
M.~Garc{\'{i}a}~P{\'{e}}rez, A.~Gonz{\'{a}}lez-Arroyo and M.~Okawa,
  \emph{{Perturbative contributions to Wilson loops in twisted lattice boxes
  and reduced models}},
  \href{https://doi.org/10.1007/JHEP10(2017)150}{\emph{JHEP} {\bfseries 10}
  (2017) 150} [\href{https://arxiv.org/abs/1708.00841}{{\ttfamily
  1708.00841}}].

\bibitem{Dunne:2015eaa}
G.~V. Dunne and M.~{\"{U}}nsal, \emph{{What is QFT? Resurgent trans-series,
  Lefschetz thimbles, and new exact saddles}},
  \href{https://doi.org/10.22323/1.251.0010}{\emph{PoS} {\bfseries LATTICE2015}
  (2016) 010} [\href{https://arxiv.org/abs/1511.05977}{{\ttfamily
  1511.05977}}].

\bibitem{Cherman:2014ofa}
A.~Cherman, D.~Dorigoni and M.~{\"{U}}nsal, \emph{{Decoding perturbation theory
  using resurgence: Stokes phenomena, new saddle points and Lefschetz
  thimbles}}, \href{https://doi.org/10.1007/JHEP10(2015)056}{\emph{JHEP}
  {\bfseries 10} (2015) 056} [\href{https://arxiv.org/abs/1403.1277}{{\ttfamily
  1403.1277}}].

\bibitem{Marino:2012zq}
M.~Mari{\~{n}}o, \emph{{Lectures on non-perturbative effects in large $N$ gauge
  theories, matrix models and strings}},
  \href{https://doi.org/10.1002/prop.201400005}{\emph{Fortsch. Phys.}
  {\bfseries 62} (2014) 455} [\href{https://arxiv.org/abs/1206.6272}{{\ttfamily
  1206.6272}}].

\bibitem{Gross:1980he}
D.~J. Gross and E.~Witten, \emph{{Possible Third Order Phase Transition in the
  Large N Lattice Gauge Theory}},
  \href{https://doi.org/10.1103/PhysRevD.21.446}{\emph{Phys. Rev.} {\bfseries
  D21} (1980) 446}.

\bibitem{Wadia:2012fr}
S.~R. Wadia, \emph{{A Study of U(N) Lattice Gauge Theory in 2-dimensions}},
  \href{https://arxiv.org/abs/1212.2906}{{\ttfamily 1212.2906}}.

\bibitem{Buividovich:2015oju}
P.~V. Buividovich, G.~V. Dunne and S.~N. Valgushev, \emph{{Complex Path
  Integrals and Saddles in Two-Dimensional Gauge Theory}},
  \href{https://doi.org/10.1103/PhysRevLett.116.132001}{\emph{Phys. Rev. Lett.}
  {\bfseries 116} (2016) 132001}
  [\href{https://arxiv.org/abs/1512.09021}{{\ttfamily 1512.09021}}].

\bibitem{tHooft:1977xjm}
G.~'t~Hooft, \emph{{Can We Make Sense Out of Quantum Chromodynamics?}},
  {\emph{Subnucl. Ser.} {\bfseries 15} (1979) 943}.

\bibitem{Beneke:1998ui}
M.~Beneke, \emph{{Renormalons}},
  \href{https://doi.org/10.1016/S0370-1573(98)00130-6}{\emph{Phys. Rept.}
  {\bfseries 317} (1999) 1}
  [\href{https://arxiv.org/abs/hep-ph/9807443}{{\ttfamily hep-ph/9807443}}].

\bibitem{DiRenzo:1993hs}
F.~Di~Renzo, G.~Marchesini, P.~Marenzoni and E.~Onofri, \emph{{Lattice
  perturbation theory by Langevin dynamics}},
  \href{https://arxiv.org/abs/hep-lat/9308006}{{\ttfamily hep-lat/9308006}}.

\bibitem{DiRenzo:1994av}
F.~Di~Renzo, G.~Marchesini, P.~Marenzoni and E.~Onofri, \emph{{Lattice
  perturbation theory on the computer}},
  \href{https://doi.org/10.1016/0920-5632(94)90517-7}{\emph{Nucl. Phys. Proc.
  Suppl.} {\bfseries 34} (1994) 795}.

\bibitem{DiRenzo:1994sy}
F.~Di~Renzo, E.~Onofri, G.~Marchesini and P.~Marenzoni, \emph{{Four loop result
  in SU(3) lattice gauge theory by a stochastic method: Lattice correction to
  the condensate}},
  \href{https://doi.org/10.1016/0550-3213(94)90026-4}{\emph{Nucl. Phys.}
  {\bfseries B426} (1994) 675}
  [\href{https://arxiv.org/abs/hep-lat/9405019}{{\ttfamily hep-lat/9405019}}].

\bibitem{DiRenzo:1995qc}
F.~Di~Renzo, E.~Onofri and G.~Marchesini, \emph{{Renormalons from eight loop
  expansion of the gluon condensate in lattice gauge theory}},
  \href{https://doi.org/10.1016/0550-3213(95)00525-0}{\emph{Nucl. Phys.}
  {\bfseries B457} (1995) 202}
  [\href{https://arxiv.org/abs/hep-th/9502095}{{\ttfamily hep-th/9502095}}].

\bibitem{Burgio:1997hc}
G.~Burgio, F.~Di~Renzo, G.~Marchesini and E.~Onofri, \emph{{$\Lambda^2$
  contribution to the condensate in lattice gauge theory}},
  \href{https://doi.org/10.1016/S0370-2693(98)00057-4}{\emph{Phys. Lett.}
  {\bfseries B422} (1998) 219}
  [\href{https://arxiv.org/abs/hep-ph/9706209}{{\ttfamily hep-ph/9706209}}].

\bibitem{DiRenzo:2004hhl}
F.~Di~Renzo and L.~Scorzato, \emph{{Numerical stochastic perturbation theory
  for full QCD}},
  \href{https://doi.org/10.1088/1126-6708/2004/10/073}{\emph{JHEP} {\bfseries
  10} (2004) 073} [\href{https://arxiv.org/abs/hep-lat/0410010}{{\ttfamily
  hep-lat/0410010}}].

\bibitem{Parisi:1980ys}
G.~Parisi and Y.-s. Wu, \emph{{Perturbation Theory Without Gauge Fixing}},
  {\emph{Sci. Sin.} {\bfseries 24} (1981) 483}.

\bibitem{Damgaard:1987rr}
P.~H. Damgaard and H.~H{\"{u}}ffel, \emph{{Stochastic Quantization}},
  \href{https://doi.org/10.1016/0370-1573(87)90144-X}{\emph{Phys. Rept.}
  {\bfseries 152} (1987) 227}.

\bibitem{BOOK_SQ1}
P.~H. Damgaard and H.~H{\"{u}}ffel, eds., \emph{Stochastic Quantization}. World
  Scientific Publishing Company, 1988,
  \href{https://doi.org/10.1142/0375}{10.1142/0375}.

\bibitem{BOOK_SQ2}
M.~Namiki, \emph{Stochastic Quantization}, Lecture Notes in Physics Monographs
  9. Springer-Verlag Berlin Heidelberg, 1992,
  \href{https://doi.org/10.1007/978-3-540-47217-9}{10.1007/978-3-540-47217-9}.

\bibitem{DiRenzo:2000ua}
F.~Di~Renzo and L.~Scorzato, \emph{{A Consistency check for renormalons in
  lattice gauge theory: $\beta^{-10}$ contributions to the SU(3) plaquette}},
  \href{https://doi.org/10.1088/1126-6708/2001/10/038}{\emph{JHEP} {\bfseries
  10} (2001) 038} [\href{https://arxiv.org/abs/hep-lat/0011067}{{\ttfamily
  hep-lat/0011067}}].

\bibitem{Rakow:2005yn}
P.~E.~L. Rakow, \emph{{Stochastic perturbation theory and the gluon
  condensate}}, \href{https://doi.org/10.22323/1.020.0284}{\emph{PoS}
  {\bfseries LAT2005} (2006) 284}
  [\href{https://arxiv.org/abs/hep-lat/0510046}{{\ttfamily hep-lat/0510046}}].

\bibitem{Bauer:2011ws}
C.~Bauer, G.~S. Bali and A.~Pineda, \emph{{Compelling Evidence of Renormalons
  in QCD from High Order Perturbative Expansions}},
  \href{https://doi.org/10.1103/PhysRevLett.108.242002}{\emph{Phys. Rev. Lett.}
  {\bfseries 108} (2012) 242002}
  [\href{https://arxiv.org/abs/1111.3946}{{\ttfamily 1111.3946}}].

\bibitem{Horsley:2012ra}
R.~Horsley, G.~Hotzel, E.~M. Ilgenfritz, R.~Millo, H.~Perlt, P.~E.~L. Rakow
  et~al., \emph{{Wilson loops to 20th order numerical stochastic perturbation
  theory}}, \href{https://doi.org/10.1103/PhysRevD.86.054502}{\emph{Phys. Rev.}
  {\bfseries D86} (2012) 054502}
  [\href{https://arxiv.org/abs/1205.1659}{{\ttfamily 1205.1659}}].

\bibitem{Bali:2013pla}
G.~S. Bali, C.~Bauer, A.~Pineda and C.~Torrero, \emph{{Perturbative expansion
  of the energy of static sources at large orders in four-dimensional SU(3)
  gauge theory}}, \href{https://doi.org/10.1103/PhysRevD.87.094517}{\emph{Phys.
  Rev.} {\bfseries D87} (2013) 094517}
  [\href{https://arxiv.org/abs/1303.3279}{{\ttfamily 1303.3279}}].

\bibitem{Horsley:2013pra}
R.~Horsley, H.~Perlt, P.~E.~L. Rakow, G.~Schierholz and A.~Schiller, \emph{{The
  SU(3) Beta Function from Numerical Stochastic Perturbation Theory}},
  \href{https://doi.org/10.1016/j.physletb.2013.11.012}{\emph{Phys. Lett.}
  {\bfseries B728} (2014) 1} [\href{https://arxiv.org/abs/1309.4311}{{\ttfamily
  1309.4311}}].

\bibitem{Bali:2014fea}
G.~S. Bali, C.~Bauer and A.~Pineda, \emph{{Perturbative expansion of the
  plaquette to ${\cal O}(\alpha^{35})$ in four-dimensional SU(3) gauge
  theory}}, \href{https://doi.org/10.1103/PhysRevD.89.054505}{\emph{Phys. Rev.}
  {\bfseries D89} (2014) 054505}
  [\href{https://arxiv.org/abs/1401.7999}{{\ttfamily 1401.7999}}].

\bibitem{DelDebbio:2018ftu}
L.~Del~Debbio, F.~Di~Renzo and G.~Filaci, \emph{{Large-order NSPT for lattice
  gauge theories with fermions: the plaquette in massless QCD}},
  \href{https://doi.org/10.1140/epjc/s10052-018-6458-9}{\emph{Eur. Phys. J.}
  {\bfseries C78} (2018) 974}
  [\href{https://arxiv.org/abs/1807.09518}{{\ttfamily 1807.09518}}].

\bibitem{DelDebbio:2018vhr}
L.~Del~Debbio, F.~Di~Renzo and G.~Filaci, \emph{{Non perturbative physics from
  NSPT: renormalons, the gluon condensate and all that}},  in \emph{{36th
  International Symposium on Lattice Field Theory (Lattice 2018) East Lansing,
  MI, United States, July 22-28, 2018}}, 2018,
  \href{https://arxiv.org/abs/1811.05427}{{\ttfamily 1811.05427}}.

\bibitem{DallaBrida:2017pex}
M.~Dalla~Brida, M.~Garofalo and A.~D. Kennedy, \emph{{Investigation of New
  Methods for Numerical Stochastic Perturbation Theory in $\varphi^4$ Theory}},
  \href{https://doi.org/10.1103/PhysRevD.96.054502}{\emph{Phys. Rev.}
  {\bfseries D96} (2017) 054502}
  [\href{https://arxiv.org/abs/1703.04406}{{\ttfamily 1703.04406}}].

\bibitem{DallaBrida:2017tru}
M.~Dalla~Brida and M.~L{\"{u}}scher, \emph{{SMD-based numerical stochastic
  perturbation theory}},
  \href{https://doi.org/10.1140/epjc/s10052-017-4839-0}{\emph{Eur. Phys. J.}
  {\bfseries C77} (2017) 308}
  [\href{https://arxiv.org/abs/1703.04396}{{\ttfamily 1703.04396}}].

\bibitem{Zwanziger:1981kg}
D.~Zwanziger, \emph{{Covariant Quantization of Gauge Fields Without Gribov
  Ambiguity}}, \href{https://doi.org/10.1016/0550-3213(81)90202-9}{\emph{Nucl.
  Phys.} {\bfseries B192} (1981) 259}.

\bibitem{Kennedy:2000ju}
A.~D. Kennedy and B.~Pendleton, \emph{{Cost of the generalized hybrid Monte
  Carlo algorithm for free field theory}},
  \href{https://doi.org/10.1016/S0550-3213(01)00129-8}{\emph{Nucl. Phys.}
  {\bfseries B607} (2001) 456}
  [\href{https://arxiv.org/abs/hep-lat/0008020}{{\ttfamily hep-lat/0008020}}].

\bibitem{Rossi:1987hv}
P.~Rossi, C.~T.~H. Davies and G.~P. Lepage, \emph{{A Comparison of a Variety of
  Matrix Inversion Algorithms for Wilson Fermions on the Lattice}},
  \href{https://doi.org/10.1016/0550-3213(88)90021-1}{\emph{Nucl. Phys.}
  {\bfseries B297} (1988) 287}.

\bibitem{Davies:1987vs}
C.~T.~H. Davies, G.~G. Batrouni, G.~R. Katz, A.~S. Kronfeld, G.~P. Lepage,
  K.~G. Wilson et~al., \emph{{Fourier Acceleration in Lattice Gauge Theories.
  1. Landau Gauge Fixing}},
  \href{https://doi.org/10.1103/PhysRevD.37.1581}{\emph{Phys. Rev.} {\bfseries
  D37} (1988) 1581}.

\bibitem{OMF}
I.~Omelyan, I.~Mryglod and R.~Folk, \emph{Symplectic analytically integrable
  decomposition algorithms: classification, derivation, and application to
  molecular dynamics, quantum and celestial mechanics simulations},
  \href{https://doi.org/https://doi.org/10.1016/S0010-4655(02)00754-3}{\emph{Comput.
  Phys. Commun.} {\bfseries 151} (2003) 272 }.

\bibitem{Takaishi:2005tz}
T.~Takaishi and P.~de~Forcrand, \emph{{Testing and tuning new symplectic
  integrators for hybrid Monte Carlo algorithm in lattice QCD}},
  \href{https://doi.org/10.1103/PhysRevE.73.036706}{\emph{Phys. Rev.}
  {\bfseries E73} (2006) 036706}
  [\href{https://arxiv.org/abs/hep-lat/0505020}{{\ttfamily hep-lat/0505020}}].

\bibitem{GonzalezArroyo:2010ss}
A.~Gonz{\'{a}}lez-Arroyo and M.~Okawa, \emph{{Large $N$ reduction with the
  Twisted Eguchi-Kawai model}},
  \href{https://doi.org/10.1007/JHEP07(2010)043}{\emph{JHEP} {\bfseries 07}
  (2010) 043} [\href{https://arxiv.org/abs/1005.1981}{{\ttfamily 1005.1981}}].

\bibitem{privcomm}
M.~Garc{\'{i}}a~P{\'{e}}rez. private communication for analytic values for the
  two-loop coefficients not appeared in ref.~\cite{Perez:2017jyq}.

\bibitem{DiGiacomo:1981lcx}
A.~Di~Giacomo and G.~C. Rossi, \emph{{Extracting $\langle
  (\alpha/\pi)\Sigma_{a,\mu\nu} G^a_{\mu\nu}G^a_{\mu\nu}\rangle$ from Gauge
  Theories on a Lattice}},
  \href{https://doi.org/10.1016/0370-2693(81)90609-2}{\emph{Phys. Lett.}
  {\bfseries 100B} (1981) 481}.

\bibitem{Alles:1993dn}
B.~All{\'{e}}s, M.~Campostrini, A.~Feo and H.~Panagopoulos, \emph{{The
  three-loop lattice free energy}},
  \href{https://doi.org/10.1016/0370-2693(94)90218-6}{\emph{Phys. Lett.}
  {\bfseries B324} (1994) 433}
  [\href{https://arxiv.org/abs/hep-lat/9306001}{{\ttfamily hep-lat/9306001}}].

\bibitem{Alles:1998is}
B.~All{\'{e}}s, A.~Feo and H.~Panagopoulos, \emph{{Asymptotic scaling
  corrections in QCD with Wilson fermions from the 3-loop average plaquette}},
  \href{https://doi.org/10.1016/S0370-2693(98)00295-0}{\emph{Phys. Lett.}
  {\bfseries B426} (1998) 361}
  [\href{https://arxiv.org/abs/hep-lat/9801003}{{\ttfamily hep-lat/9801003}}].

\bibitem{Alfieri:2000ce}
R.~Alfieri, F.~Di~Renzo, E.~Onofri and L.~Scorzato, \emph{{Understanding
  stochastic perturbation theory: Toy models and statistical analysis}},
  \href{https://doi.org/10.1016/S0550-3213(00)00180-2}{\emph{Nucl. Phys.}
  {\bfseries B578} (2000) 383}
  [\href{https://arxiv.org/abs/hep-lat/0002018}{{\ttfamily hep-lat/0002018}}].

\bibitem{Brezin:1977sv}
E.~Brezin, C.~Itzykson, G.~Parisi and J.~B. Zuber, \emph{{Planar Diagrams}},
  \href{https://doi.org/10.1007/BF01614153}{\emph{Commun. Math. Phys.}
  {\bfseries 59} (1978) 35}.

\bibitem{Koplik:1977pf}
J.~Koplik, A.~Neveu and S.~Nussinov, \emph{{Some Aspects of the Planar
  Perturbation Series}},
  \href{https://doi.org/10.1016/0550-3213(77)90344-3}{\emph{Nucl. Phys.}
  {\bfseries B123} (1977) 109}.

\end{thebibliography}

\providecommand{\href}[2]{#2}\begingroup\raggedright\endgroup


\end{document}